\documentclass[11pt,a4paper,English,twoside]{article}

\usepackage{a4wide}
\usepackage{amssymb, amsmath, amsthm}
\usepackage{graphicx}
\usepackage{subcaption}
\usepackage[all]{xy}
\usepackage{enumerate}
\usepackage[pdftex,hyperref,svgnames]{xcolor}
\usepackage[pdftex,colorlinks=true,
pdfstartview=FitV,
pdfnewwindow=true,
linktoc = page,
linkcolor= blue,
citecolor= red,
urlcolor= blue,
hyperindex=true,
hyperfigures=false]{hyperref}
\hypersetup{linktocpage}
\usepackage{dsfont}
\usepackage{empheq}
\usepackage{cite}
\usepackage{float}
\usepackage{cancel}
\usepackage{relsize}
\usepackage{soul}
\usepackage{enumitem}
\usepackage{hhline}
\usepackage{multirow}

\usepackage{pdfpages}

\usepackage{setspace}

\newcommand{\beq}{\begin{equation}}
\newcommand{\eeq}{\end{equation}}
\def\bea#1\eea{\begin{align}#1\end{align}}
\def\beal#1\eeal{\begin{subequations}\begin{align}#1\end{align}\end{subequations}}
\newcommand{\nn}{\nonumber}
\newcommand{\w}{\wedge}
\newcommand{\R}{\mathcal{R}}
\renewcommand{\i}{\ensuremath{\textnormal{i}}}

\def\del {\partial}

\def\d {{\rm d}}

\def\N {\mathcal{N}}

\def\sin {\text{sin}}
\def\tan {\text{tan}}

\def\sinh {\text{sinh}}

\def\ln {\text{ln}}

\begin{document}
\numberwithin{equation}{section}

\begin{titlepage}
\begin{center}

\phantom{DRAFT}
\vspace{2.0cm}

{\LARGE \bf{Negative scalar potentials and the swampland:}\\\vspace{0.12in}
\bf{an Anti-Trans-Planckian Censorship Conjecture}}\\

\vspace{2.2 cm} {\Large David Andriot$^{1}$, Ludwig Horer$^{1,2}$, George Tringas$^{1}$}\\
\vspace{0.9 cm} {\small\slshape $^1$ Laboratoire d’Annecy-le-Vieux de Physique Th\'eorique (LAPTh)\\
CNRS, Universit\'e Savoie Mont Blanc (USMB), UMR 5108\\
9 Chemin de Bellevue, 74940 Annecy, France}\\
\vspace{0.2 cm} {\small\slshape $^2$ Institute for Theoretical Physics, TU Wien\\
Wiedner Hauptstrasse 8-10/136, A-1040 Vienna, Austria}\\
\vspace{0.5cm} {\upshape\ttfamily andriot@lapth.cnrs.fr; ludwig.horer@tuwien.ac.at; tringas@lapth.cnrs.fr}\\

\vspace{2.8cm}

{\bf Abstract}
\vspace{0.1cm}
\end{center}

\begin{quotation}
In this paper, we derive a characterisation of negative scalar potentials, $V<0$, in $d$-dimensional effective theories of quantum gravity. This is achieved thanks to an Anti-Trans-Planckian Censorship Conjecture (ATCC), inspired by a refined version of the TCC. The ATCC relies on the fact that in a contracting universe, modes that become sub-Planckian in length violate the validity of the effective theory. In the asymptotics of field space, we deduce that $-V'/V \geq c_0$ when $V' \geq 0$. The rate $c_0 = 2/\sqrt{(d-1)(d-2)}$ is successfully tested in several string compactifications for $d\geq 4$. In addition, a new asymptotic condition, $V''/V \geq c_0^2$, is derived. By extrapolation to anti-de Sitter solutions of radius $l$, we infer the existence of a scalar whose mass should obey $m^2 l^2 \lesssim -2$.
This property is verified in many supersymmetric examples.
\end{quotation}

\end{titlepage}

\tableofcontents

\section{Introduction and results summary}

Cosmological observations have reached an unprecedented level of precision in the last decades, putting tight constraints on cosmological models. Single field slow-roll inflation models are strongly constrained by the latest Planck data \cite{Planck:2018jri}, while quintessence models should soon get restricted by the Euclid mission. In spite of this progress, many such models remain in agreement with observations. In a $d$-dimensional spacetime, those are of the form
\beq
{\cal S}= \int \d^d x \sqrt{|g_d|} \left(\frac{M_p^2}{2} \R_d - \frac{1}{2} g_{ij} \del_{\mu}\varphi^i \del^{\mu}\varphi^j - V \right) \ ,\label{actionintro}
\eeq
with a minimal coupling of scalar fields $\varphi^i$ to gravity, and a scalar potential $V$. In this situation, additional theoretical input could be useful in order to distinguish between these models. It is commonly believed that a fundamental theory of quantum gravity should exist and provide an origin to such models \eqref{actionintro}, which then appear as effective theories. The swampland program \cite{Vafa:2005ui, Palti:2019pca} aims at characterising effective theories of quantum gravity, one example being the properties of the scalar potential $V$ in \eqref{actionintro}. Such a characterisation could help in selecting (cosmological) models which are compatible with quantum gravity. This has been the topic of many works in the swampland program, proposing criteria for admissible positive scalar potentials, $V>0$, under the name of de Sitter conjectures \cite{Obied:2018sgi, Andriot:2018wzk, Garg:2018reu, Ooguri:2018wrx, Andriot:2018mav, Rudelius:2019cfh, Bedroya:2019snp, Rudelius:2021oaz}. While the initial proposal was a strict condition on the ratio $\nabla V/V$, some refinements suggested to include the second derivative of the potential $\nabla \del V$ and others to restrict the characterisation of $V$ to the asymptotics of field space only. The latter was realised in particular in the Trans-Planckian Censorship Conjecture (TCC) \cite{Bedroya:2019snp}: it led to the following conditions on $V>0$, with $V'\leq0$, here for a single canonically normalized field
\beq
0 < V(\varphi)\, \leq\,  e^{ - c_0\, |\varphi - \varphi_i|} \ ,\quad \quad \left< -\frac{V'}{V} \right>_{\varphi \to \infty}  \  \geq \ c_0 \ ,\quad \quad c_0 = \frac{2}{\sqrt{(d-1)(d-2)}} \ ,\label{TCCVintro}
\eeq
in Planckian units $M_p=1$, for $3 \leq d \leq 10$.

In this work, we are interested in effective theories \eqref{actionintro} of quantum gravity with negative scalar potentials, $V<0$. Those may sound less relevant to cosmology, even though we can recall the ekpyrosis and bouncing cosmological models, which require negative potentials. Such scenarios seem able to reproduce, in a contracting phase, the early universe observations. Difficulties may occur later, at the bounce needed to catch up with a positive cosmological constant, where instabilities arise and are difficult to control. An incomplete list of related works includes \cite{Lehners:2010fy, Koehn:2013upa, Giambo:2015tja, Uzawa:2018sal, Lehners:2018vgi, Kist:2022mew, Agullo:2022klq, Antonini:2022fna} (see also \cite{Dutta:2018vmq, Visinelli:2019qqu, Calderon:2020hoc, Sen:2021wld} for cosmological models with a negative cosmological constant). Regardless of cosmology, negative scalar potentials include anti-de Sitter vacua: those are among the best understood quantum gravity backgrounds thanks to holography \cite{Maldacena:1997re, Gubser:1998bc, Witten:1998qj}. Characterising negative scalar potentials in quantum gravity effective theories of the form \eqref{actionintro} seems therefore a natural task in the swampland program.

De Sitter and anti-de Sitter solutions correspond to positive or negative critical points of the potential in \eqref{actionintro}. Those solutions have however very different status in string theory (see e.g.~\cite{Danielsson:2018ztv}). Nevertheless, positive and negative scalar potentials in effective theories are a priori formally identical: they would differ by values and signs of coefficients, depending e.g.~on details of a compactification, but the functional dependence on the fields is in principle the same. For this reason, it is reasonable to expect the characterisation of negative potentials to be similar to that of positive ones. Such an analogous characterisation has already been proposed in the literature, where criteria on admissible negative potentials were inspired by the refined de Sitter conjecture \cite{Ooguri:2018wrx}. Indeed, it was proposed in \cite{Lust:2019zwm} that $V<0$ should obey, in Planckian units
\beq
\frac{|\nabla V|}{|V|} \geq c \sim {\cal O}(1) \quad {\rm or} \quad \frac{{\rm min}\nabla \del V}{|V|} \leq c' \sim {\cal O}(1) \ ,\label{LPVcond}
\eeq
in line with the AdS moduli conjecture \cite{Gautason:2018gln}. In \eqref{LPVcond}, ${\rm min}\nabla \del V$ stands for the minimal eigenvalue of the mass matrix, of coefficients $g^{ik}\nabla_k \del_j V$. At an anti-de Sitter critical point, the first condition in \eqref{LPVcond} is violated, thus requiring the second condition to hold. The latter led to many recent discussions on scale separation. Let us mention in addition \cite{Bernardo:2021wnv} where the second condition in \eqref{LPVcond} was traded for ${\rm max}\nabla \del V / |V| \geq  -c' $, reminiscent of the Breitenlohner-Freedmann (BF) bound. Finally, the condition on the first derivative is also discussed and used in e.g.~\cite{Gonzalo:2021zsp}.

The refined de Sitter conjecture \cite{Ooguri:2018wrx}, very similar to \eqref{LPVcond}, got later weakened by the TCC. In the latter, the condition on the first derivative is present but is only valid in the asymptotics of field space, as indicated in \eqref{TCCVintro}. The condition on the second derivative got relaxed, and traded for a bound on the lifetime. In the present paper, we propose an analogous Anti-Trans-Planckian Censorship Conjecture (ATCC) which characterises negative scalar potentials. The same differences and weakening will occur with respect to \eqref{LPVcond}.\\

Prior to introducing our ATCC, let us come back to the TCC statement and propose a refined version of it. For both conjectures, we consider solutions of effective theories \eqref{actionintro} with a Friedmann-Lemaitre-Robertson-Walker metric, having a scale factor $a(t)$; see Section \ref{sec:gen} for more details. We discuss in particular solutions with expanding ($\dot{a}>0$) or contracting ($\dot{a}<0$) spacetimes. The former is considered in our refined TCC, that we now present.
\vspace{0.1in}
\begin{quote}
\onehalfspacing
{\bf Refined Trans-Planckian Censorship Conjecture:}

\vspace{0.05in}

{\it Consider an effective theory of quantum gravity of the form \eqref{actionintro}, admitting a solution describing an expanding universe with $V>0$. In this universe, let us focus on a relativistic mode having a wavelength shorter than the Planck length at an initial time $t_i$, or equivalently a super-Planckian energy. Then, via the redshift, it cannot reach at a later time $t$ an energy smaller than the typical energy scale of the effective theory, without violating its validity. This gets translated into the following inequality:}
\end{quote}
\beq
\frac{a(t_i)}{a(t)} \ \geq\ \frac{\sqrt{V}}{M_p^2}\ . \label{TCCr}
\eeq

\vspace{0.05in}

\noindent The above inequality is derived considering the most constraining case of an initial Planckian wavelength or energy, and the typical energy scale of the effective theory at a later time $t$ to be given by $\sqrt{V}/M_p$, where $V$ stands for $V(\varphi(t))$. Trading $\sqrt{V}/M_p$ for $H$, one obtains the inequality of the original TCC \cite{Bedroya:2019snp}. Close to (quasi-) de Sitter spacetimes, $H$ and $\sqrt{V}/M_p$ represent the same quantity and trading one for the other does not change anything physically, nor in the general TCC reasoning allowing to derive the characterisation of the potential \eqref{TCCVintro}. Using $\sqrt{V}/M_p$ instead of $H$ is motivated here by the fact that beyond (quasi-) de Sitter spacetimes, $H$ does not always provide a meaningful characterisation of a typical length or a horizon of the universe. This will be even more true when turning to $V<0$. We then prefer to talk in terms of the energy scale of the effective theory, which should also correspond to the macroscopic physics of the universe described by that theory, if not the classical physics. It also allows us to add the question of its validity.

Indeed, a more important difference with the original TCC statement is the notion of validity of the effective theory. In the original statement, solutions violating the inequality \eqref{TCCr} are forbidden as a matter of principle by quantum gravity, while here, they are simply said to lie outside the regime of validity of the effective theory. This refinement is then essentially saying that the effective theory energy cutoff has to be below the Planck scale, which is a minimal, though crucial quantum gravity input. In other words, quantum gravity modes of Planckian energy cannot contribute to the physics of the EFT without violating its validity.\footnote{For instance, one can imagine the following physical process: a quantum gravity mode of super-Planckian mass is created, and transfers its energy to a photon or a graviton. The latter redshifts and ends-up in the energy range where the EFT is valid; it then interacts with EFT degrees of freedom. This new input to the EFT violates its validity. We can imagine the reverse process for the ATCC discussed below.} Applying this idea to the typical energy scale of the effective theory \eqref{actionintro}, namely the scalar potential, we get that it should  remain sub-Planckian in the regime of validity, i.e.~$1 \geq \sqrt{V}/M_p^2$. This is consistent with $1 \geq a(t_i)/a(t)$ in an expanding universe and the inequality \eqref{TCCr}. Note that alternative viewpoints have been proposed in \cite{Burgess:2020nec, Komissarov:2022gax}.

Inspired by this refinement, we now turn to negative potentials, and provide the ATCC statement.
\vspace{0.1in}
\begin{quote}
\onehalfspacing
{\bf Anti-Trans-Planckian Censorship Conjecture:}

\vspace{0.05in}

{\it Consider an effective theory of quantum gravity of the form \eqref{actionintro}, admitting a solution describing a contracting universe with $V<0$. In this universe, let us focus on a relativistic mode having as energy the typical energy scale of the effective theory at an initial time $t_i$, or the corresponding wavelength. Then, via the blueshift, it cannot reach at a later time $t$ an energy higher than the Planck scale, without violating the validity of the effective theory. This gets translated into the following inequality:}
\end{quote}
\beq
\frac{a(t)}{a(t_i)} \ \geq\ \frac{\sqrt{|V_i|}}{M_p^2} \ . \label{ATCCintro}
\eeq

\vspace{0.05in}

\noindent Here, $V_i$ stands for $V(\varphi(t_i))$. As above, one gets $1 \geq a(t)/a(t_i) \geq \sqrt{|V_i|}/M_p^2$, saying again that the typical energy scale of the effective theory, given by the potential, and beyond it the cutoff scale, should be sub-Planckian. We detail this statement and derivation in Section \ref{sec:ATCCstat}. For completeness, let us recall (see Section \ref{sec:gen}) that $V<0$ automatically leads to a decelerating universe.\\

As for the TCC, the ATCC inequality \eqref{ATCCintro} eventually leads to a characterisation of the scalar potential. Before reaching this point, we first show in Section \ref{sec:lifetime} that one can derive a bound on the lifetime. We interpret it as being related to the spacetime contraction, and to the final crunch which should occur in a finite time. We then turn to the characterisation of the potential: contrary to the TCC, reaching this characterisation requires here what we call a second assumption, namely a condition on $V$ and $a(t)$. This condition is automatically satisfied for $V>0$ but not for $V<0$; we discuss it in Section \ref{sec:2ndas}. We test this condition as well as the ATCC inequality \eqref{ATCCintro} on concrete solutions describing contracting universes with $V<0$, starting with the anti-de Sitter solution in Section \ref{sec:AdS} and then two dynamical solutions (with rolling fields) in Appendix \ref{ap:dynsol}. These solutions easily obey all these conditions, in their regime of validity.

\begin{figure}[t]
\begin{center}
\includegraphics[width=0.65\textwidth]{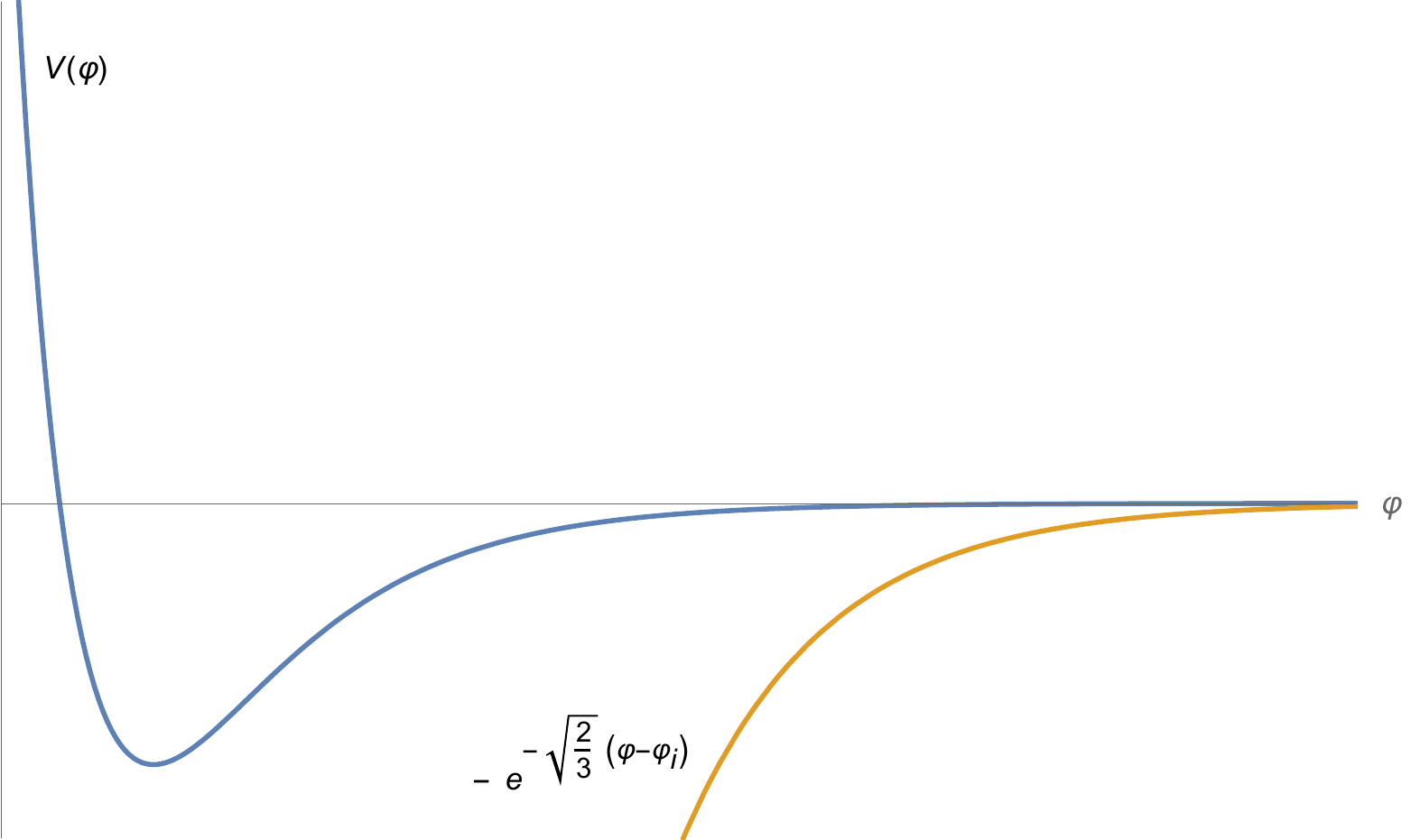}
\caption{Illustration of the ATCC bounds \eqref{boundVintro} and \eqref{boundV'Vintro} for $d=4$. The potential $V(\varphi)$ has a typical shape (see Section \ref{sec:stringpot}), obtained from string theory compactifications, and $|V|$ is small in Planckian units. The negative part of $V$ is bounded by the growing exponential, with the specific rate $c_0$.}\label{fig:intropot}
\end{center}
\end{figure}

This leads us to derive the characterisation of a negative potential for $3 \leq d \leq 10$. Focusing on a single canonically normalized scalar field, we first obtain the following bound on $V<0$ with $V'\geq 0$, in Planckian units
\beq
0 > V(\varphi)\, \geq\,  -\,  e^{ - c_0\, |\varphi - \varphi_i|} \ ,\quad \quad c_0 = \frac{2}{\sqrt{(d-1)(d-2)}} \ ,\label{boundVintro}
\eeq
valid everywhere in field space. We deduce from \eqref{boundVintro}, away from potential extrema (one needs $\dot{\varphi}\neq 0$), the asymptotic condition on the first derivative of the potential\footnote{The condition \eqref{boundV'Vintro} does not strictly forbid anti-de Sitter extrema in the asymptotics, neither did the TCC for de Sitter ones; we make it clear in Section \ref{sec:V}. This condition rather provides a constraint on the asymptotic shape of the potential.}
\beq
\left< -\frac{V'}{V} \right>_{\varphi \to \infty}  \  \geq \ c_0 \ . \label{boundV'Vintro}
\eeq
We illustrate these bounds in Figure \ref{fig:intropot}.
\noindent The similarity of these results with those of the TCC in \eqref{TCCVintro} is clear, as anticipated, even though their derivation contains few subtleties. As is well-known, the condition \eqref{boundV'Vintro} gives a lower bound on exponential rates: $ c \geq c_0$. We refer to Section \ref{sec:V} for more details.

Multifield extensions of these conditions, and related ambiguities, are discussed in Section \ref{sec:multi}. This allows us to test in Section \ref{sec:stringpot} the ATCC and its characterisation of negative potentials on examples obtained from string theory compactifications. The first potential is the semi-universal one, $V(\rho,\tau,\sigma)$, in $d$ dimensions, tested in Section \ref{sec:RhoTauSigmaPot}, and the second one is obtained from compactifications towards so-called DGKT anti-de Sitter solutions in $d=4$, and extension in $d=3$, tested in Section \ref{sec:DGKT}. In those examples, we find no violation of the ATCC bounds in $d\geq4$, providing a confirmation of the above. We find several violations in $d=3$, as already noticed for the TCC in \cite{Andriot:2022xjh}: this can be understood from the peculiarity of gravity in $d=3$.\\

Last but not least, we derive in Section \ref{sec:V''}, for both the TCC and the ATCC, a new condition on the second derivative of the potential. This had not been achieved with the TCC in \cite{Bedroya:2019snp}, and we reach this result here thanks to a minor assumption. The condition we obtain is the following
\bea
\left< \frac{V''}{V} \right>_{\varphi \to \infty}\ \geq\  c_0^2 \ ,\label{boundV''intro}
\eea
in Planckian units, for a single canonically normalized scalar field. We discuss the consequences of this condition for $V>0$ and $V<0$. We then extrapolate this condition, thanks to a possible continuity of the spectrum in field space, towards an anti-de Sitter extremum. We reach the following condition
\beq
{\rm AdS}_{d\,\geq\, 4}:\quad\ {\rm min}\ m^2 \, l^2\ \lesssim \ - 2 \ , \label{boundm2l2intro}
\eeq
where the extrapolation gives us a little flexibility in the bound, expressed by the symbol $\lesssim$. The inequality \eqref{boundm2l2intro} is interpreted as follows
\vspace{0.07in}
\begin{quote}
\onehalfspacing
{\it In a $d$-dimensional anti-de Sitter solution with radius $l$ and $d\geq 4$, the scalar field with lowest squared mass $m^2$ obeys \eqref{boundm2l2intro}.}
\end{quote}
\vspace{0.07in}
Since the BF bound is lower than $-2$ in units of $l$, for $d\geq 4$, this statement is automatically true for perturbatively unstable solutions. We test it upon few perturbatively stable examples found in the literature: {\sl most supersymmetric ones strictly verify this bound}, with the notable exceptions of KKLT \cite{Kachru:2003aw}, LVS \cite{Balasubramanian:2005zx}, and the original DGKT solution \cite{DeWolfe:2005uu}, that we discuss. Non-supersymmetric ones require the mentioned flexibility, with e.g.~${\rm min}\ m^2 \, l^2 \approx -1.5$; we also note two possible exceptions. We however keep in mind that most of these non-supersymmetric examples suffer from non-perturbative instabilities, as conjectured in \cite{Ooguri:2016pdq}. A detailed account of these examples is provided in Section \ref{sec:massbound} with a summary in Table \ref{tab:AdSsusy} and \ref{tab:AdSnonsusy}, as well as a discussion on the holographic consequences of \eqref{boundm2l2intro} for a dual CFT.\\

Similarly to the TCC versus the refined de Sitter conjecture, the ATCC weakens the condition \eqref{LPVcond} of \cite{Gautason:2018gln, Lust:2019zwm} that characterises negative scalar potentials in quantum gravity effective theories. The condition on the first derivative of the potential becomes an asymptotic one, and a value $c_0$ is given for the bound on the exponential rate \eqref{boundV'Vintro}. In addition, the condition on the second derivative is relaxed, especially on anti-de Sitter extrema. Importantly, {\sl relaxing this condition sets aside the on-going debate on scale separation} for $V<0$. The ATCC also brings a physics argument behind those results, related to the regime of validity of a quantum gravity effective theory, thanks to our refinement of the TCC. This argument may explain why the checks of the TCC \cite{Andriot:2020lea, Andriot:2022xjh} have so far been so successful. Last but not least, we found the new condition \eqref{boundV''intro} on the second derivative, whose surprising consequences have just been mentioned. The flexible bound \eqref{boundm2l2intro} on the lowest mass at an anti-de Sitter extremum deserves more study, and we hope to come back to it in future work.

\section{Anti-Trans-Planckian Censorship Conjecture and consequences}\label{sec:ATCCgen}

After recalling in Section \ref{sec:gen} the general cosmological formalism to be used, we introduce the Anti-Trans-Planckian Censorship Conjecture (ATCC) in Section \ref{sec:ATCCstat}, and discuss various consequences, namely a bound on the lifetime in Section \ref{sec:lifetime} and a characterisation of a negative (and climbing) potential in Section \ref{sec:V}, the latter requiring a second mathematical assumption presented in Section \ref{sec:2ndas}. Finally, the well-known anti-de Sitter solution is presented and analysed in this framework in Section \ref{sec:AdS}, while two more dynamical solutions are discussed in Appendix \ref{ap:dynsol}.

\subsection{General framework}\label{sec:gen}

We are interested in $d$-dimensional effective theories of quantum gravity, $3\leq d \leq 10$, whose action is of the form \eqref{actionintro}, that we repeat here for convenience
\beq
{\cal S}= \int \d^d x \sqrt{|g_d|} \left(\frac{M_p^2}{2} \R_d - \frac{1}{2} g_{ij} \del_{\mu}\varphi^i \del^{\mu}\varphi^j - V \right) \ ,\label{action}
\eeq
with scalar fields $\varphi^i$ minimally coupled to gravity. The reduced Planck mass is $M_p$, $g_{ij}$ is the field space metric and $V$ is the scalar potential. Such theories can admit solutions with a $d$-dimensional maximally symmetric spacetime, of cosmological constant $\Lambda_d$. Those solutions correspond to extrema of the potential, $\nabla V= 0$, with no scalar kinetic energy, and the relation $\Lambda_d = \frac{V}{M_p^2} = \frac{d-2}{2d} \R_d $ holds. Here we will also consider more dynamical solutions, with rolling or climbing scalars on gradients of the potential.

To describe all these solutions, it is enough to restrict ourselves to $d$-dimensional spacetimes whose metric is given by Friedmann-Lemaitre-Robertson-Walker (FLRW)
\beq
\d s_d^2 = -\d t^2 + a(t)^2 \left(\frac{\d r^2}{ 1- k\, r^2} + r^2 \d \Omega^2  \right) \ , \label{FLRW}
\eeq
with $k=\pm 1,0$ and $a(t)>0$. We also consider for now a single scalar field $\varphi$, canonically normalized ($g_{ij}=\delta_{ij}$), and set $M_p=1$ in the rest of this subsection. The equations of motion (e.o.m.) are then the two Friedmann equations in $d$ dimensions and the e.o.m. of $\varphi$
\bea
& \frac{(d-1)(d-2)}{2} \left( H^2 + \frac{k}{a^2} \right) = \rho \ , \label{F1eq}\\
& (d-2) \frac{\ddot{a}}{a} + \frac{d-3}{d-1} \rho + p =0  \ \Leftrightarrow \ \dot{H} - \frac{k}{a^2} + \frac{\rho + p}{d-2} = 0 \ , \label{F2eq} \\
& \ddot{\varphi} + (d-1) H \dot{\varphi} + V' = 0 \ , \label{feom}
\eea
with the Hubble parameter, energy density and pressure given by
\beq
H= \frac{\dot{a}}{a} \ ,\quad \rho = \frac{1}{2} \dot{\varphi}^2 + V \ ,\quad p = \frac{1}{2} \dot{\varphi}^2 - V \ ,
\eeq
the dot denoting the derivative with respect to $t$, and we consider a homogeneous scalar field. The equation of state parameter is given by $w = \frac{p}{\rho}$.\\

In most of the paper, we will focus on negative potentials
\beq
V < 0\ . \label{V<0}
\eeq
This has some consequences, as we now recall. To start with, the second Friedmann equation can be rewritten as
\beq
(d-1) \frac{\ddot{a}}{a} = \frac{2}{d-2} V -  \dot{\varphi}^2 \ ,
\eeq
from which we conclude that
\beq
\ddot{a} < 0 \ ,
\eeq
i.e.~we face a {\sl decelerating} universe.

A second observation goes as follows. We will allow ourselves to reach situations without kinetic energy, $\tfrac{1}{2}\dot{\varphi}^2=0$, and therefore cases where $\rho <0$. This includes in particular anti-de Sitter vacua. The first Friedmann equation then imposes to pick
\beq
k=-1 \ .
\eeq
This choice disagrees with cosmological observations ($k=0$), but we do not aim here at a realistic cosmology. However, it also differs from the situation of the TCC \cite{Bedroya:2019snp} which had $k=0$; we will then adapt our analysis to the extra related complications.\\

We will give in Section \ref{sec:AdS} and Appendix \ref{ap:dynsol} explicit solutions to the above equations, starting with the anti-de Sitter solution. Before doing so, we turn in full generality to the ATCC statement and its consequences.

\subsection{ATCC statement}\label{sec:ATCCstat}

As recalled in the Introduction, the TCC discussed in \cite{Bedroya:2019snp} considers a universe in expansion, and makes a statement on the fate of modes which evolve between a sub-Planckian regime and a classical regime (via the growth of their wavelength). A characterisation of positive scalar potentials, $V>0$, is then deduced. We consider here a contracting universe, $\dot{a} <0$, and discuss analogously the fate of modes which would change regime, with the aim of deducing a characterisation of negative scalar potentials, $V<0$.

Let us recall from Section \ref{sec:gen} that $V<0$ automatically gives a decelerating universe, $\ddot{a}<0$. In addition, the solution examples with $V<0$ discussed in Section \ref{sec:AdS} and Appendix \ref{ap:dynsol}, namely the anti-de Sitter solution and more dynamical solutions, all exhibit contracting phases. Given the existence of such solutions, we can safely discuss contracting universes.

A mode having as wavelength the typical length scale of the universe would usually be considered as classical. However, contrary to a de Sitter universe, there is not necessarily a horizon when $V<0$, as we will see for instance in Section \ref{sec:AdS} for an anti-de Sitter universe. So we cannot use the concept of a mode freezing-out and becoming classical by crossing the horizon, as is familiar in cosmology, and implicit for the TCC. We thus trade here the notion of a classical regime for the one of the regime of validity of an effective theory, essentially dictated by an energy cutoff. Degrees of freedom described by the effective theory, meaning those having the relevant length or energy scale, cannot get mixed through a physical process with gravitational quantum modes, as long as the validity of the effective theory is preserved. This is simply because the cutoff scale of an effective theory of quantum gravity is expected to be lower than the Planck scale. This may also be viewed as an extreme version of scale separation. Let us now provide the ATCC statement.
\vspace{0.1in}
\begin{quote}
\onehalfspacing
{\bf Anti-Trans-Planckian Censorship Conjecture:}

\vspace{0.05in}

{\it Consider an effective theory of quantum gravity of the form \eqref{actionintro}, admitting a solution describing a contracting universe with $V<0$. In this universe, let us focus on a relativistic mode having as energy the typical energy scale of the effective theory at an initial time $t_i$, or the corresponding wavelength. Then, via the blueshift, it cannot reach at a later time $t$ an energy higher than the Planck scale, without violating the validity of the effective theory.}

\end{quote}
\vspace{0.1in}

Introducing this notion of validity of the effective theory led us to the above ATCC statement, but also to a refined version of the TCC, as presented in \eqref{TCCr} and further commented there. Another, more technical difference with the original TCC is the trade of $H$ for $\sqrt{|V|}/M_p$. As will be made clear in the solution examples of Section \ref{sec:AdS} and Appendix \ref{ap:dynsol}, the Hubble parameter $H$ or the notion of horizon are not meaningful in defining a ``typical length scale of the universe'' for $V<0$. Rather, the radius $l$ in the anti-de Sitter solution, directly related to the cosmological constant $\Lambda_d=V/M_p^2$, seems more relevant. Therefore, for a general contracting universe with $V<0$, one could propose to take as a typical length scale $M_p/\sqrt{|V|}$. As argued above, we prefer to phrase the statement in terms of energy, given that the potential should provide the typical energy scale in the validity range of the effective theory. The energy scale $\sqrt{|V|}/M_p$ could of course be multiplied by an order one factor, but such a factor will not alter the conclusion on the potential characterisation in Section \ref{sec:V}, so we neglect it here. Then, the ATCC statement gets translated as follows for a contraction between an initial time $t_i$ and a time $t$: in natural units, the energy $1/(a(t)\lambda_0)$ should be smaller than $M_p$, for an initial energy $1/(a(t_i)\lambda_0) \sim \sqrt{|V_i|}/M_p$, with $V_i$ standing for $V(\varphi(t_i))$. In other words, the ATCC condition is
\beq
\text{ATCC:}\quad \boxed{\frac{a(t)}{a(t_i)} \ \geq\ \frac{\sqrt{|V_i|}}{M_p^2}\quad \Leftrightarrow \quad \int_{t_i}^t \d t' \, H(t') \ =\ \ln \frac{a(t)}{a(t_i)} \ \geq \ \ln \frac{\sqrt{|V_i|}}{M_p^2} }  \label{ATCC}
\eeq
The claim is that \eqref{ATCC} should hold in the regime of validity of the effective theory. Consistently with this, we get $\sqrt{|V_i|} < M_p^2$, as it should. Finally, note that when $M_p \to \infty$, the condition becomes trivial as expected from swampland criteria.

It is straightforward to see that \eqref{ATCC} provides a constant lower bound to the scale factor. In particular, as the universe is contracting, the inequality prevents $a(t)$ to reach zero, i.e.~a big crunch. The condition makes $a(t)$ stop before, at a Planckian scale, where the effective theory is not valid anymore. We will illustrate this for instance in the anti-de Sitter solution in Section \ref{sec:AdS}.\\

Let us finally mention that in the TCC analysis, a contracting universe was considered \cite[Foot. 2]{Bedroya:2019snp}, by performing a time reversal on the TCC condition: this led to the condition $\frac{a(t)}{a_i} \geq |H_i|$ in Planckian units. This condition is the same as \eqref{ATCC} up to trading the scale of $H$ for that of $V$. The ATCC however focuses on negative scalar potentials, which imply having a decelerating universe. In that situation, we argued already that $H$ (even $H_i$) is not a relevant parameter to capture a typical length or energy scale of the universe. We then prefer our condition \eqref{ATCC} to that obtained from the TCC for a contracting universe. We now turn to a first consequence of condition \eqref{ATCC}.

\subsection{Bound on the lifetime}\label{sec:lifetime}

Analogously to the TCC, one can derive a bound on the lifetime for a decelerating, contracting universe. Physically, it is conceivable to get such a bound because of the final crunch: having deceleration and contraction forces the crunch to happen in a finite time. Indeed, the function $a(t)$ is concave and positive. Starting at a finite value $a_i=a(t_i)$, one reaches $a(t)=0$ (or any other finite value $a<a_i$) in a finite time. It is however difficult to derive a lifetime bound from this reasoning. The ATCC condition will allow us to get a bound, as we now show.

We consider here as only content the scalar field and its potential, so $\rho + p = \dot{\varphi}^2 \geq 0$. With $k=-1$, we deduce from the second Friedmann equation that $\dot{H}<0$ (related to deceleration), therefore $H_i > H_f$ between an initial time $t_i$ and a final one $t_f$. We deduce $\int_{t_i}^{t_f}  \d t' H(t') < H_i\, (t_f-t_i)$. Using finally the ATCC condition \eqref{ATCC}, we conclude
\beq
\boxed{t_f-t_i\ <\ \frac{1}{|H_i|}\ \ln \frac{M_p^2}{\sqrt{|V_i|}} } \label{boundlifetime}
\eeq
where we recall that $H<0$ (encoding the contraction). This gives us an upper bound on the lifetime of contracting and decelerating phase. More precisely, beyond this time \eqref{boundlifetime}, we reach a Planckian regime so we cannot trust our effective theory anymore. The same interpretation could be given to the TCC lifetime bound \cite{Bedroya:2019snp}, in view the refined TCC \eqref{TCCr}.

It is however unclear how to evaluate this bound in general, given its dependence on $H$. As can be seen in solution examples of Section \ref{sec:AdS} and Appendix \ref{ap:dynsol}, $H(t)$ is not bounded. In these solutions, we see that $|H_i|$ is typically close to zero, and $M_p^2 / \sqrt{|V_i|}$ is supposed to be large, so we get a very high upper bound. If however we push the initial time closer to the crunch, $|H_i|$ would then become very large, leaving a much smaller lifetime. These observations are qualitatively consistent.

\subsection{A second assumption}\label{sec:2ndas}

While a first assumption (in the mathematical sense), the ATCC condition \eqref{ATCC},  was motivated by physics, a second one will be necessary to reach an interesting characterisation of negative scalar potentials. This second assumption is the following
\beq
\boxed{-\frac{k}{a^2}\, \frac{(d-1)(d-2)}{2}  + V \ \geq\ 0 } \label{2ndas}
\eeq
where we set $M_p=1$, as well as in the following. Let us first note that for the TCC, where $k=0$ and $V>0$, \eqref{2ndas} is automatically satisfied; this explains why the assumption is not considered explicitly in the characterisation of positive potentials. Here however with $V<0$, this assumption imposes us to pick $k=-1$, a choice already discussed in Section \ref{sec:gen} and necessary to the anti-de Sitter solution, as well as other contracting and decelerating solutions. This choice leads to various complications w.r.t.~the TCC, and to start with, the non-triviality of this second assumption \eqref{2ndas}.

Restoring an $M_p$ dependence would give $V/M_p^2$, thus making this second assumption trivial in the limit $M_p \to \infty$ with $k=-1$ or $0$. We further check \eqref{2ndas} on the anti-de Sitter solution of Section \ref{sec:AdS} and the two dynamical solutions of Appendix \ref{ap:dynsol}: this condition is always easily verified. It could be automatically satisfied, given appropriate initial conditions, and it would then boil down to a constraint on initial conditions. To be safe, we treat for now \eqref{2ndas} as an independent assumption, but there could be some rationale behind it.

\subsection{A bound on $V<0$ and on $V'/V$}\label{sec:V}

We now have all ingredients to proceed to a characterisation of negative scalar potentials $V<0$ in effective theories of quantum gravity. Combining the ATCC condition \eqref{ATCC} together with the second assumption \eqref{2ndas}, we will proceed analogously to \cite{Bedroya:2019snp} to derive this characterisation. As for the TCC, we will consider a potential slope of definite sign: while TCC was considering rolling down a positive potential, we focus here on a field climbing up a negative potential. Picking one direction, we then take for simplicity $\dot{\varphi} \geq 0$ and $V' \geq 0$; we could equivalently move along the opposite direction with $\dot{\varphi} \leq 0$ and $V' \leq 0$. All we eventually need is $V(\varphi) \geq V_i$.

\begin{figure}[H]
\begin{center}
\includegraphics[width=0.7\textwidth]{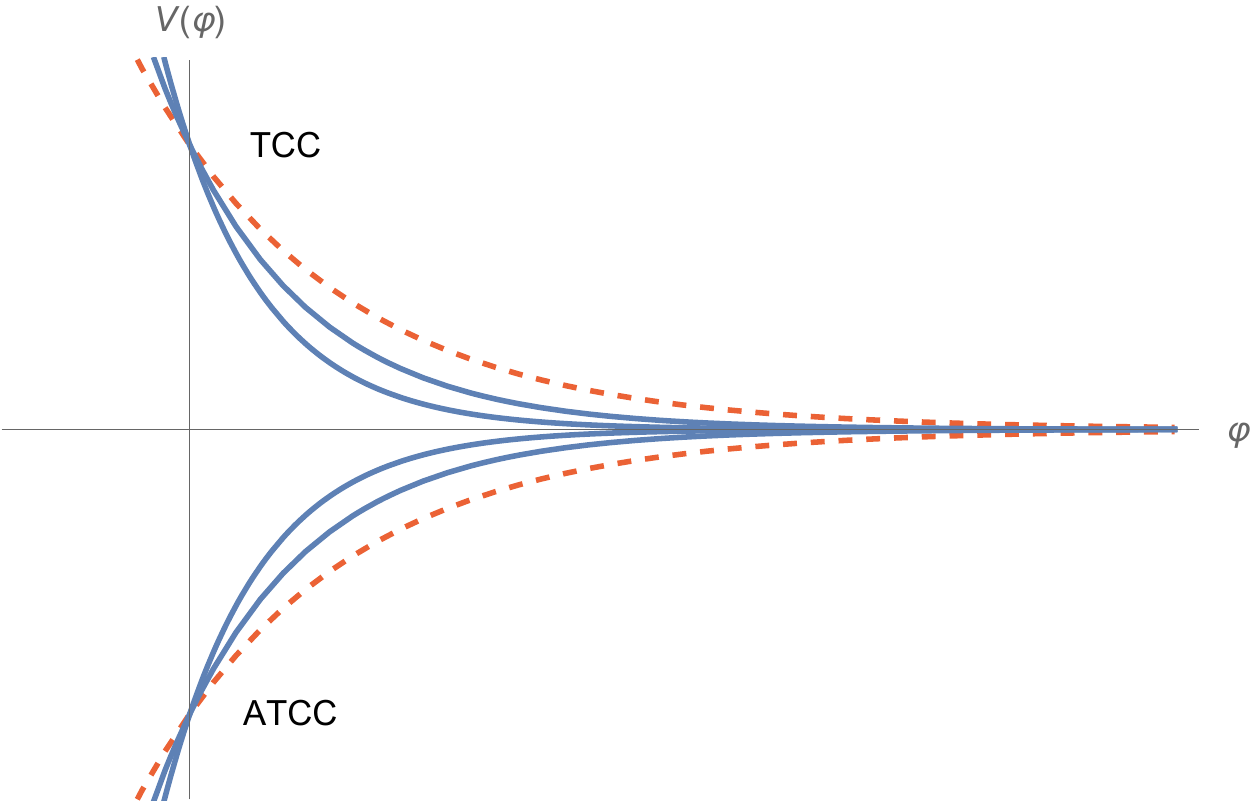}
\caption{Illustration of the TCC and ATCC exponential bounds (in dashed red) on scalar potentials (in blue), as in \eqref{boundV} for the ATCC. We display here exponential potentials $V(\varphi) = \pm e^{-c\, \varphi}$ in Planckian units, bounded by the (A)TCC exponentials for $\varphi\geq \varphi_i=0$.}\label{fig:TCCATCC}
\end{center}
\end{figure}

To start with, the second assumption \eqref{2ndas} can be rewritten thanks to the first Friedmann equation as
\beq
H^2  \geq \frac{\dot{\varphi}^2}{(d-1)(d-2)} \ \Leftrightarrow \ H\, \d t \leq - \frac{\d \varphi}{\sqrt{(d-1)(d-2)}} \ ,
\eeq
where we recall that we consider a contracting universe, i.e.~$H \leq 0$, and that $\dot{\varphi} \geq 0$. We then integrate the above from an initial time $t_i$ to any later time $t$
\beq
\int_{t_i}^t  H\, \d t' \leq - \frac{\varphi - \varphi_i}{\sqrt{(d-1)(d-2)}} \ ,
\eeq
and recall that $\varphi - \varphi_i \geq 0$; in the following, we write $| \varphi - \varphi_i |$ to be more general with the field direction. Using finally the ATCC condition \eqref{ATCC}, we get
\beq
\ln \, \sqrt{|V_i|} \leq - \frac{|\varphi - \varphi_i|}{\sqrt{(d-1)(d-2)}} \ .
\eeq
Since we consider $V' \geq 0$, i.e.~$V(\varphi) \geq V_i$, and $V < 0$, we conclude
\beq
\boxed{V(\varphi)\, \geq\,  -\,  e^{ - \frac{2\, |\varphi - \varphi_i|}{\sqrt{(d-1)(d-2)}} } } \label{boundV}
\eeq
The potential is bounded from below by a growing exponential. The amplitude here is $1$ in Planckian units, and could be adjusted by some order one factor mentioned in Section \ref{sec:ATCCstat}, but this does not change the exponential and its rate. We illustrate \eqref{boundV} as well as its TCC counterpart in Figure \ref{fig:TCCATCC}.

Let us make a few comments on this exponential lower bound \eqref{boundV}. First, \eqref{boundV} is certainly verified initially: for $\varphi = \varphi_i$, this becomes $V_i \geq -1$, which was implicitly considered above to distinguish the initial typical energy scale from the Planck scale (see around \eqref{ATCC}). The bound \eqref{boundV} is then valid at {\sl any further time $t$ and corresponding field value $\varphi$}, as long as the two conditions \eqref{ATCC} and \eqref{2ndas} are verified (in particular as long as we are in the regime of validity of the effective theory); this is remarkable, in contrast to the asymptotic claim to be made below. Note that the same holds for the TCC analogous exponential bound depicted in Figure \ref{fig:TCCATCC}.\\

Proceeding as for the TCC, we deduce a condition on the slope of the potential at large field values. This requires to consider $\Delta \varphi = |\varphi - \varphi_i| \neq 0$; as a consequence, {\sl the following cannot apply to an anti-de Sitter solution}; we come back to this important point below. We consider the following average with $V<0$
\beq
\left<- \frac{V'}{V} \right> = \frac{1}{\Delta \varphi} \int_{\varphi_i}^{\varphi} \frac{V'}{-V} \d \tilde{\varphi} = \frac{\ln |V_i|}{\Delta  \varphi} - \frac{\ln |V|}{\Delta  \varphi} \  \geq \ \frac{\ln |V_i|}{\Delta  \varphi} + \frac{2}{\sqrt{(d-1)(d-2)}} \ ,
\eeq
where the last inequality comes from \eqref{boundV}. We conclude with the large field limit
\beq
\boxed{ \left< -\frac{V'}{V} \right>_{\varphi \to \pm \infty}  \  \geq \ \frac{2}{\sqrt{(d-1)(d-2)}} } \label{boundV'V}
\eeq
where the sign in $\varphi \to \pm \infty$ is the same as that of $V'$ and corresponds to the direction in which the field climbs up the potential.\footnote{Note that in any case, climbing up the potential means $V' \d \tilde{\varphi} \geq 0$, making the average positive, in agreement with the positive bound \eqref{boundV'V}.} Taking for instance $V'\geq 0$ and $\varphi \to \infty$, we can apply this bound to an exponential potential with rate $c>0$, meaning $V(\varphi) \sim V_0 \ e^{-c\, \varphi}$ with $V_0 <0$, as occurs typically in large field limits in string compactifications. Then the bound \eqref{boundV'V}, as well as \eqref{boundV}, become
\beq
V(\varphi) = V_0 \ e^{-c\, \varphi}:\quad c \  \geq \ \frac{2}{\sqrt{(d-1)(d-2)}} \ . \label{rate}
\eeq
This ATCC bound on the rate $c$ is actually the same as the TCC bound obtained in the original paper \cite{Bedroya:2019snp}, and further tested for instance in supergravity potentials in \cite{Andriot:2020lea,Andriot:2022xjh}.

As mentioned above, the bound \eqref{boundV'V} does not apply to an anti-de Sitter solution, which verifies $\dot{\varphi}=0$, while one requires $\Delta \varphi \neq0$ to get the bound. This has the important conceptual consequence that the bound \eqref{boundV'V} on the slope does not forbid anti-de Sitter extrema of the potential. The same is actually true for the TCC and de Sitter: {\sl the bound on the slope does not forbid de Sitter extrema at large field values}, it simply does not apply to them. This should be put in contrast with the swampland de Sitter conjectures. On the other hand, the exponential bound \eqref{boundV}, also present for the TCC, indicates that there is ``less room'' at large field values for (anti-) de Sitter extrema.

\subsection{Example: the anti-de Sitter solution}\label{sec:AdS}

We finally provide a first example of solution in a contracting and decelerating phase, with $V<0$, and test it upon the various conditions discussed previously. The solution is the well-known anti-de Sitter solution, phrased in the formalism introduced in Section \ref{sec:gen} (in particular with a FLRW metric). Two more dynamical solutions are provided and studied in Appendix \ref{ap:dynsol}.

An anti-de Sitter spacetime is best known as a maximally symmetric $d$-dimensional spacetime with cosmological constant $\Lambda_d <0$. Following \cite{Hawking:1973uf}, this spacetime can be viewed as a $d$-dimensional hyperboloid of radius $l$ (related to $\Lambda_d$ as in \eqref{lL}). The corresponding metric is obtained by embedding it in a $(d+1)$-dimensional flat spacetime. This way, one obtains the following anti-de Sitter metric
\beq
\label{eq:AdSFLRW}
\d s_d^2= l^2 \left( -\d \psi^2+ \sin^2(\psi) \left( \d \chi^2 + \sinh^2(\chi)\, \d \Omega^2 \right)\right) \ .
\eeq
This formulation corresponds to a FLRW metric \eqref{FLRW} with
\beq
k=-1 \ ,\quad a(t)= l\ \sin\left(\frac{t}{l}\right) \ , \label{aAdS1}
\eeq
given that $\int \frac{\d r}{\sqrt{1 + r^2}} = {\rm arcsinh}(r)$.

This can be reproduced by solving the equations \eqref{F1eq}-\eqref{feom} of Section \ref{sec:gen}. In that formalism, we consider the anti-de Sitter solution as an extremum of the potential with $\Lambda_d = V <0$ in Planckian units, and without kinetic energy, $\dot{\varphi}=0$. As mentioned in Section \ref{sec:gen}, the latter requires to have $k=-1$. While \eqref{feom} is satisfied, the two Friedmann equations can be rewritten as
\beq
 \dot{a}^2 + \frac{1}{l^2}\, a^2 = 1 \ ,\quad \ddot{a} + \frac{1}{l^2}\, a = 0 \ ,\quad {\rm with}\quad \frac{1}{l} \equiv \sqrt{\frac{-2\Lambda_d}{(d-1)(d-2)} }\ , \label{lL}
\eeq
in terms of the anti-de Sitter radius $l$, and the solution is \eqref{aAdS1}, when imposing the standard initial condition $a(0)=0$.

\begin{figure}[H]
\begin{center}
\begin{subfigure}[H]{0.45\textwidth}
\includegraphics[width=\textwidth]{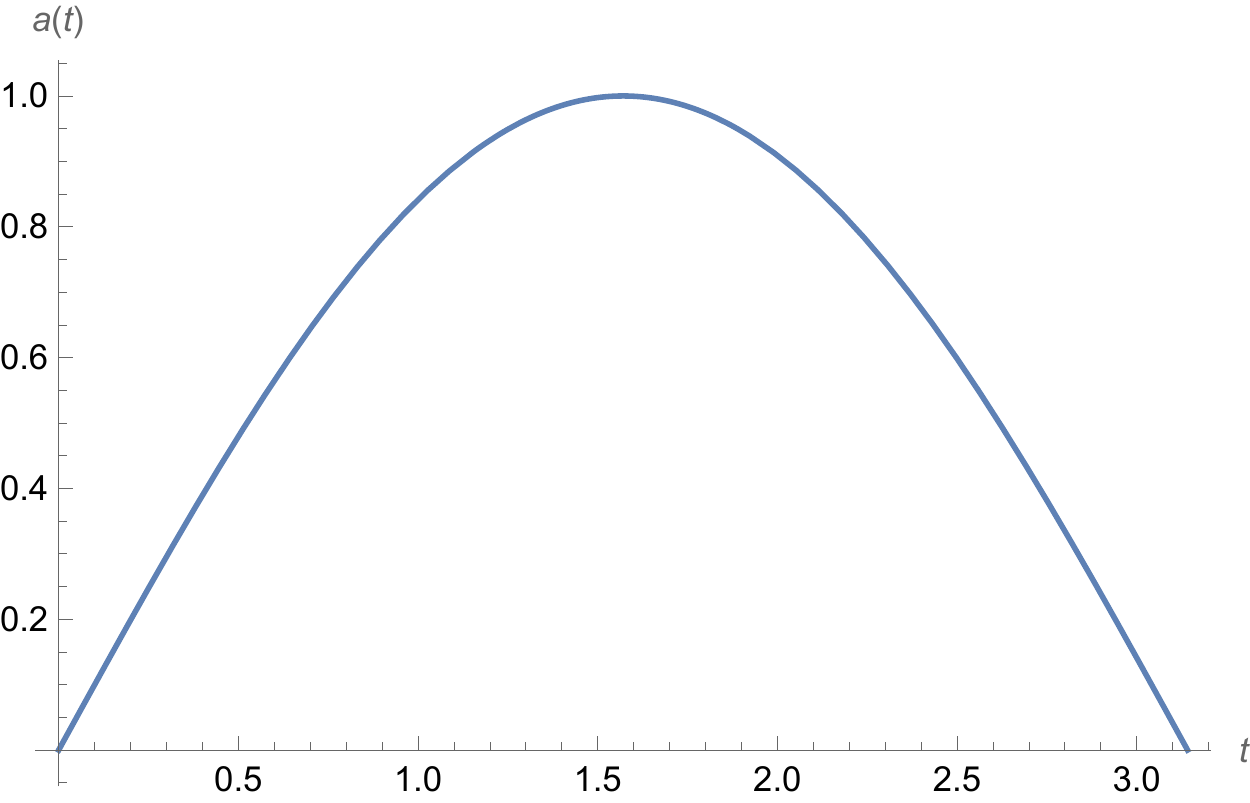}\caption{$a(t)$}\label{fig:AdSat}
\end{subfigure}\qquad \quad
\begin{subfigure}[H]{0.45\textwidth}
\includegraphics[width=\textwidth]{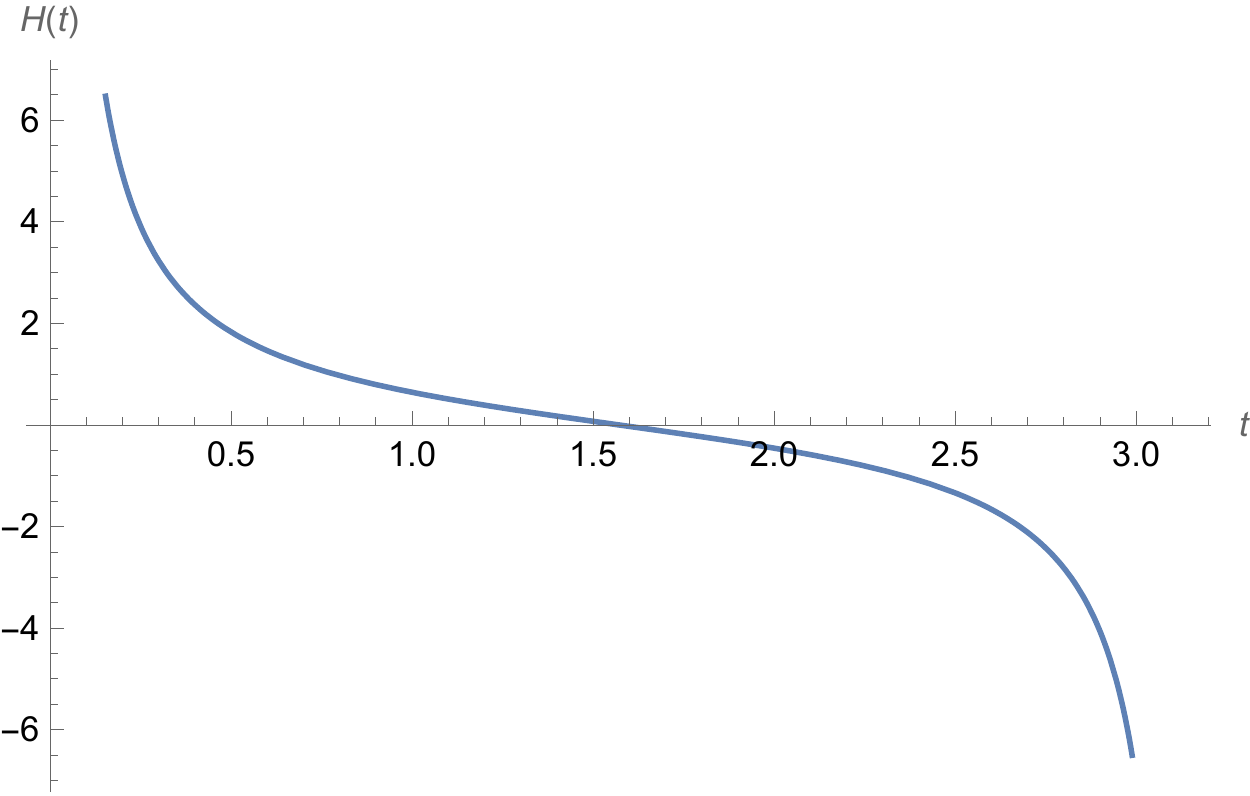}\caption{$H(t)$}\label{fig:AdSHt}
\end{subfigure}
\caption{$a(t)$ and $H(t)$ for the anti-de Sitter solution, with $l=1$.}\label{fig:AdS}
\end{center}
\end{figure}

Let us now depict $a(t)$ and $H(t)$ in Figure \ref{fig:AdS} and comment on them.
\noindent As expected, the anti-de Sitter solution corresponds to a decelerating universe, with an expanding and a contracting phase. The latter starts at $t= \frac{\pi}{2} \, l$, and the big crunch occurs at $t_f= \pi \, l$. The Hubble parameter is given by
\beq
H(t) = \frac{1}{l}\, \cot\left(\frac{t}{l}\right) \ .
\eeq
It is not constant, contrary to a de Sitter solution. In addition, neither $H(t)$ nor $1/H(t)$ is bounded. It is thus very different from the cosmological constant.

Related to this, an anti-de Sitter universe does not admit any horizon: its particle and event horizons are given by
\bea
& d_p=a(t) \int_{t_i=0}^{t} \frac{\d t'}{a(t')}= l\ \sin\left(\frac{t}{l}\right) \times \left. \ln \,\tan \left(\frac{t'}{2\, l}\right) \right|_{t_i=0}^{t} \to \infty \ , \label{eq:particlehorizon}\\
& d_e=a(t) \int_{t}^{t_f=\pi \, l} \frac{\d t'}{a(t')} =  l\ \sin\left(\frac{t}{l}\right) \times \left. \ln \,\tan \left(\frac{t'}{2\, l}\right) \right|_{t}^{t_f=\pi \, l} \to \infty \ . \label{eq:eventhorizon}
\eea
This is again different from a de Sitter universe, admitting an event horizon. We conclude that neither $H(t)$ nor the notion of horizon provides any sensible characteristic length for anti-de Sitter, contrary to de Sitter. This was the motivation for using, in the ATCC as well as the refined TCC \eqref{TCCr}, the scalar potential (here related to $\Lambda_d$ or $l$) to define a typical length or energy scale.\\

Let us now test this solution against the various conditions and quantities discussed for the ATCC.

\begin{itemize}
  \item {\bf ATCC condition} \eqref{ATCC}
\end{itemize}

\noindent Applied to the anti-de Sitter solution, this inequality becomes
\beq
\sin\left(\frac{t}{l}\right) \geq \frac{l_p}{l} \times \sqrt{\frac{(d-1)(d-2)}{2}}\, \sin\left(\frac{t_i}{l}\right) \ ,
\eeq
where $\frac{l_p}{l}$ is a priori a small number. Let us take an initial time close to the maximal size of anti-de Sitter, to be sure to be in a valid initial regime, i.e.~$\sin\left(\frac{t_i}{l}\right) \lesssim 1$: we take for simplicity $\sqrt{\frac{(d-1)(d-2)}{2}}\, \sin\left(\frac{t_i}{l}\right) =1$. We also introduce a time $\tilde{t}$ that measures the difference to the final time $t_f= \pi l$, namely $\tilde{t}= t_f-t$. Close to the big crunch, when $\tilde{t} \to 0$, one obtains
\beq
\sin\left(\pi - \frac{\tilde{t}}{l}\,\right) = \sin\left(\,\frac{\tilde{t}}{l}\,\right) \sim \frac{\tilde{t}}{l} \ \geq\ \frac{l_p}{l} \ .\label{AdSPlancktime}
\eeq
The ATCC condition then indicates that a Planckian time away from the big crunch, we run out of validity of the theory, which is consistent.

\begin{itemize}
  \item {\bf Bound on the lifetime} \eqref{boundlifetime}
\end{itemize}

\noindent For anti-de Sitter in the contracting phase ($t_i > \frac{\pi}{2}\, l$), we get the following lifetime bound
\beq
t_f-t_i\ <\ l \, \left|\tan\left(\frac{t_i}{l}\right)\right| \ \ln \left( \frac{l}{l_p} \times \sqrt{\frac{2}{(d-1)(d-2)} } \right) \ ,
\eeq
where the relation between $\Lambda_d$ and the radius $l$ is given in \eqref{lL}. As discussed below \eqref{boundlifetime}, it is difficult to evaluate in general this bound, depending on the initial time. We also note that it should be contrasted with the complete contraction time, $\frac{\pi}{2}\, l$, of the anti-de Sitter solution.

\begin{itemize}
  \item {\bf Second assumption} \eqref{2ndas}
\end{itemize}

\noindent Remarkably, the second assumption \eqref{2ndas} is perfectly verified for the anti-de Sitter solution (with $a>0$): indeed, this inequality becomes
\beq
\frac{1}{a^2} - \frac{1}{l^2} \geq 0 \quad \Leftrightarrow \quad \sin^2\left(\frac{t}{l}\right) \leq 1 \ ,
\eeq
which certainly holds.

\begin{itemize}
  \item {\bf Potential bound} \eqref{boundV}
\end{itemize}

\noindent Since both the ATCC condition and the second assumption are satisfied, it is without surprise that we get the bound on the potential \eqref{boundV} verified. It remains interesting to see how. Since in the anti-de Sitter solution $\dot{\varphi}=0$, one has $\varphi=\varphi_i$. Therefore, the bound becomes
\beq
\Lambda_d = V \geq -1 \ ,
\eeq
in Planckian units. This is certainly true, given our initial assumption that the typical scale $|\Lambda_d|$ of a solution described by the effective theory should be smaller than $M_p$. We finally recall from Section \ref{sec:ATCCstat} that the bound on the slope \eqref{boundV'V} does not apply to the anti-de Sitter solution.\\

Given that anti-de Sitter vacua are among the best established solutions of string theory, it is rather satisfying to verify that this solution agrees with the ATCC and its consequences.

\section{A bound on $V''/V$ and consequences}\label{sec:V''}

\subsection{Bounding $V''/V$ with the (A)TCC}

In this section, we derive an asymptotic bound involving the second derivative of the scalar potential, $V''$, for $V\neq 0$. It is valid for both the TCC ($V>0, V' \leq 0$, rolling-down) and the ATCC ($V<0, V' \geq 0$, climbing-up), where the specified field direction is chosen for simplicity. Let us emphasize that no bound on $V''$ had been found locally (i.e.~pointlike in field space) in \cite{Bedroya:2019snp}. We achieve this here thanks to an extra minor assumption detailed below; this assumption holds in particular for potentials which are asymptotically exponential, a common situation in string compactifications.

For simplicity of the derivation, we consider $\varphi - \varphi_i \geq 0$, and the large field limit to be $\varphi \to \infty$. We start with the following equalities for $V\neq 0$
\beq
\int_{\varphi_i}^{\varphi} \frac{V''}{V} \d\tilde{\varphi} =  \int_{\varphi_i}^{\varphi}  \frac{V'' V - V' V'}{V^2} \d\tilde{\varphi}+ \int_{\varphi_i}^{\varphi} \left(\frac{V'}{V} \right)^2 \d\tilde{\varphi} = \int_{\varphi_i}^{\varphi}  \left(\frac{V'}{V}\right)' \d\tilde{\varphi}+ \int_{\varphi_i}^{\varphi} \left|\frac{V'}{V} \right|^2 \d\tilde{\varphi} \ .
\eeq
Using the Cauchy–Schwarz inequality for definite Riemann integrals
\beq
\left( \int f(\tilde{\varphi}) g(\tilde{\varphi}) \d\tilde{\varphi} \right)^2 \le \int f(\tilde{\varphi})^2 \d\tilde{\varphi} \int g(\tilde{\varphi})^2 \d\tilde{\varphi} \ ,
\eeq
with $f(\tilde{\varphi})=\left|\frac{V'}{V} \right|,\ g(\tilde{\varphi})=1$, yields
\beq
\int_{\varphi_i}^{\varphi} \frac{V''}{V} \d\tilde{\varphi} \ge  \int_{\varphi_i}^{\varphi}  \left(\frac{V'}{V}\right)' \d\tilde{\varphi}+ \frac{1}{\Delta \varphi} \left( \int_{\varphi_i}^{\varphi} \frac{|V'|}{|V|} \d\tilde{\varphi} \right)^2 = \left.\frac{V'}{V}\right|_{\varphi_i}^{\varphi} + \Delta \varphi\, \left< \frac{|V'|}{|V|} \right>^2 \ ,
\eeq
with the average introduced in Section \ref{sec:V}. We deduce the following inequality
\beq
\left< \frac{V''}{V} \right> \ \ge\ \frac{1}{\Delta \varphi} \, \left.\frac{V'}{V}\right|_{\varphi_i}^{\varphi} + \left< \frac{|V'|}{|V|} \right>^2 \ . \label{condbla}
\eeq
This is where we introduce the extra assumption on the potential: we simply require that
$|V'/V|$ at $\varphi$ is bounded from above in the large field limit $\varphi\to \infty$. This holds in the prototypical example of an (asymptotic) exponential potential. This assumption could even be relaxed, the only requirement being that the first term in the right-hand side of \eqref{condbla} vanishes in this limit.

Provided this minor assumption holds, we are left to use the TCC or ATCC bound on $\left< \frac{|V'|}{|V|} \right>$ when $\varphi \to \infty$. In either case, the absolute value is only needed in the numerator or denominator and amounts to a sign. The resulting bounds are the same and given by \eqref{boundV'V}. We conclude, for a single canonically normalized field,
\beq
\boxed{\left< \frac{V''}{V} \right>_{\varphi \to \infty}\ \geq\  \frac{4}{(d-1)(d-2)} } \label{boundV''}
\eeq
This new bound is actually no surprise: once the potential is bounded by an exponential, as in the TCC and ATCC \eqref{boundV}, one deduces a bound on its (averaged) first derivative but also second derivative. We naturally get in \eqref{boundV''} the square of the exponential rate of \eqref{boundV'V}.\\

The consequences of this new bound are interesting. Let us investigate them while dropping the average, as for instance in the case of an exponential potential. We take $V''$ to be a mass square, $V''=m^2$, which holds for a single canonical field as here. The new bound \eqref{boundV''} is valid in the asymptotics, and to derive it, we have used the (A)TCC bound on $V'/V$, so strictly speaking, \eqref{boundV''} is not meant to be applied at an (anti)-de Sitter critical point. By continuity in field space, one may still wonder about its consequences at such an extremum, as we will discuss.

\subsubsection*{Consequences for $V>0$}

For $V>0$, we get a positive lower bound on $m^2$
\beq
V>0:\quad m^2 \ \geq \ \frac{4}{(d-1)(d-2)} \, V  \ .\label{boundVpos}
\eeq
This should be contrasted with the tachyon commonly observed in de Sitter solutions \cite{Andriot:2019wrs, Andriot:2021rdy} and conjectured within the swampland program \cite{Andriot:2018wzk, Garg:2018reu, Ooguri:2018wrx}. As mentioned, the bound \eqref{boundVpos} does not apply a priori to critical points of the potential, so this observation does not appear inconsistent. But this bound may also be interpreted as the prediction of a state with a positive $m^2$. Looking at the complete mass spectrum obtained by consistent truncation for a database of de Sitter solutions in \cite{Andriot:2022bnb}, we see that most of the states actually obey this bound. So predicting the existence of a state with positive $m^2$, at an extremum or not, seems reasonably true.

Let us now view $m$ as the typical mass scale of a tower of states, in the asymptotics of a positive potential. In that case, the bound indicates the possibility of scale separation. This is reminiscent of the statement that classical de Sitter string backgrounds, if they exist (and then correspond to some large field limit, as here), are very likely to be scale separated \cite[(5.12)]{Andriot:2019wrs}. Note that such a scale separation is then probably local, i.e.~numerical \cite{Andriot:2020vlg, Cicoli:2021fsd}, and not parametrically controlled \cite{Junghans:2018gdb}

\subsubsection*{Consequences for $V<0$}

For $V<0$, we get a negative upper bound on $m^2$
\beq
V<0:\quad m^2 \ \leq \ \frac{4}{(d-1)(d-2)} \, V \ . \label{boundVneg}
\eeq
The new bound is then (as for $V>0$) different than those of the swampland literature \cite{Gautason:2018gln, Lust:2019zwm} reported around \eqref{LPVcond} (see also \cite{Bernardo:2021wnv}), which typically have to do with $m^2 >0$, the question of scale separation or light modes. Here, one may again interpret the new bound \eqref{boundVneg} as the prediction of a state with negative $m^2$. Let us now test this idea in more details.

\subsection{A mass bound for anti-de Sitter and holographic interpretation}\label{sec:massbound}

Using a possible continuity in field space towards a potential extremum, let us rewrite \eqref{boundVneg} at a $d$-dimensional anti-de Sitter critical point for $d\geq 4$.\footnote{Given that $d=3$ provides several examples of violations of the ATCC, as we will show in Section \ref{sec:stringpot}, as well as a violation of the TCC \cite{Andriot:2022xjh}, we do not consider the mass bound \eqref{boundm2l2} for $d=3$. The peculiarity of gravity in $d=3$ may justify this exception. Indeed, gravitational fluctuations, invoked to derive the (A)TCC, are absent in $d=3$.} To that end, we trade the potential for the cosmological constant, and further for the anti-de Sitter radius $l$ \eqref{lL}. The bound \eqref{boundVneg} becomes the mass bound
\beq
\boxed{{\rm AdS}_d:\quad m^2 \, l^2\ \lesssim \ - 2 }\label{boundm2l2}
\eeq
While a strict rewriting of \eqref{boundVneg} would lead to \eqref{boundm2l2} with a symbol $\leq$, we prefer to use here an approximate $\lesssim$, as we now explain. The reason is that the bound \eqref{boundVneg} is meant to be applied away from critical points, in the asymptotics. We therefore draw the bound \eqref{boundm2l2} at an anti-de Sitter extremum invoking a possible continuity of the spectrum from the point in field space where \eqref{boundVneg} applies towards a potential extremum. Moving this way in field space, the spectrum may get a little deformed. Such a deformation is the reason why we allow ourselves for a little flexibility with the symbol $\lesssim$. We will also discuss whether bigger modifications of the spectrum can occur while moving in field space, when considering possible counter-examples to this bound.

The interpretation of the bound \eqref{boundm2l2} is the following
\vspace{0.07in}
\begin{quote}
\onehalfspacing
{\it In a $d$-dimensional anti-de Sitter solution with radius $l$ and $d\geq 4$, there is a scalar field whose mass $m$ obeys \eqref{boundm2l2}.}
\end{quote}
\vspace{0.07in}
Let us recall that the BF bound is given by $m^2\, l^2 \geq -(d-1)^2 / 4$. Our new bound \eqref{boundm2l2} is then greater than the BF bound for $d\geq 4$. This implies that for $d\geq 4$, perturbatively unstable anti-de Sitter solutions automatically satisfy our bound \eqref{boundm2l2}. In the following, we then test \eqref{boundm2l2} for perturbatively stable anti-de Sitter solutions in $d\geq 4$, by looking at their mass spectrum in various examples; we summarize them in Table \ref{tab:AdSsusy} and \ref{tab:AdSnonsusy}, and detail them below. Many other examples could certainly be considered, and we do not aim for completeness here. However, our set of examples already displays interesting features that we now summarize.

\begin{table}[ht!]
  \begin{center}
      \begin{tabular}{|c||c|c|c|c|}
    \hline
\multirow{2}{*}{AdS${}_d$} & \multirow{2}{*}{$\N$} & \multirow{2}{*}{Specification} & Spectrum & Scalar lowest\\
 & & & reference & $m^2\, l^2$ \\
        \hhline{=::====}
    \multirow{25}{*}{$d=4$} & & \multirow{2}{*}{AdS${}_4$, M-th., with:} & & \\
    &  &  &  &  \\
    & 8 & $SO(8)$ & \cite[Tab. 4]{Nilsson:2018lof} & $-9/4$ \\
    & 2 & $SU(3) \times U(1)$ & \multirow{4}{*}{\cite{Comsa:2019rcz}} & $-2.222$ \\
    & 1 & $G_2$ &  & $-2.242$ \\
    & 1 & $U(1)\times U(1)$ &  & $-2.25$ \\
    & 1 & $SO(3)$ &  & $-2.245$ \\[3pt]
    \hhline{~||----}
    &  & \multirow{2}{*}{AdS${}_4 \times$ S${}^6$, IIA, with:} &  & \\
    &  &  &  &  \\
    & 1 & $G_2$  &  & $-2.24158$ \\
    & 2 & $SU(3) \times U(1)$ &  & $-20/9$ \\
    & 3 & $SO(3) \times SO(3) $ & \cite[App. B]{Bobev:2020qev} & $-9/4$ \\
    & 1 & $SU(3)$ & \cite[App. A]{Guarino:2020jwv} & $-20/9$ \\
    & 1 & $U(1)$ &  & $-2.23969$ \\
    & 1 & $\varnothing$  &  & $-2.24943$ \\
    & 1 & $U(1)$ &  & $-2.24908$ \\[3pt]
    \hhline{~||----}
    & & & & \\[-8pt]
    & 1 & DGKT, IIA & \cite{DeWolfe:2005uu, Conlon:2021cjk} & $>0$\\
    & 1 & DGKT-like Branch A1-S1, IIA & \cite[Tab. 2]{Marchesano:2019hfb} & $-2$\\[3pt]
    \hhline{~||----}
    & & & & \\[-8pt]
    & 1 & KKLT, IIB & \cite{Kachru:2003aw, Conlon:2020wmc} & $\geq 0$\\
    & 1 & LVS, IIB & \cite[Sec. 2]{Conlon:2018vov} & $\geq 0$\\[3pt]
    \hhline{~||----}
    &  & \multirow{2}{*}{S-fold, IIB, with:} & \multirow{5}{*}{\cite{Guarino:2020gfe}} & \\
    & & & & \\
    & 1 & $U(1)^2$ &  & $-2$ \\
    & 2 & $U(1)^2$ &  & $-2$ \\
    & 4 & $SO(4)$ &  & $-2$ \\[3pt]
    \hhline{-||----}
    \multirow{4}{*}{$d=5$} & & \multirow{2}{*}{AdS${}_5 \times$ S${}^5$, IIB, with:} & & \\
    & & & & \\
    & 8 & $SO(6)$ & \cite{Kim:1985ez} & $-4$ \\
    & 2 & $SU(2)\times U(1)$ & \cite[Tab. D.4]{Bobev:2020ttg} & $-4$ \\[3pt]
    \hhline{-||----}
    \multirow{2}{*}{$d=7$} & \multirow{2}{*}{1} & \multirow{2}{*}{AdS${}_7 \times$ S${}^3$, IIA} & \multirow{2}{*}{\cite{Apruzzi:2019ecr}} & \multirow{2}{*}{$-8$} \\
    & & & & \\
    \hline
    \end{tabular}
     \caption{Sample of $d$-dimensional supersymmetric anti-de Sitter solutions with a quantum gravity uplift, together with their number of preserved supersymmetries $\N$, and further specifications including symmetry groups. In order to compare to the bound \eqref{boundm2l2}, we give the lowest mass squared for a scalar in units of the radius $l$, as can be found in the reference indicated. Details are provided in the main text.}\label{tab:AdSsusy}
  \end{center}
\end{table}

\begin{table}[ht!]
  \begin{center}
      \begin{tabular}{|c||c|c|c|c|}
    \hline
\multirow{2}{*}{AdS${}_d$} & \multirow{2}{*}{Specification} & Spectrum & Scalar lowest & Non-pert.\\
 & & reference & $m^2\, l^2$ & instability ref. \\
        \hhline{=::====}
\multirow{17}{*}{$d=4$} & \multirow{2}{*}{M-th., $SO(3) \times SO(3)$} & \multirow{2}{*}{\cite[Tab. 2]{Fischbacher:2010ec}} & \multirow{2}{*}{$-12/7$} & \multirow{2}{*}{\cite{Bena:2020xxb}} \\
    & & & & \\
    \hhline{~||----}
 & \multirow{2}{*}{AdS${}_4 \times$ S${}^6$, IIA, with:} &  & & \\
& & & & \\
    & $G_2$ &  & $-1$ & \cite{Bomans:2021ara} \\
    & $SU(3)$ &  & $-1.58174$ & \\
    & $SO(3) \times U(1) $ &  &  $-1.71379$ & \\
    & $SO(3)$ & \cite[App. B]{Bobev:2020qev} & $-1.71663$ & \\
    & $SU(3)$ & \cite[App. A]{Guarino:2020jwv} & $-1.70679$ & \\
    & $SO(3) \times U(1)$  &  & $-1.70677$ & \\
    & $SO(3) \times SO(3)$ &  & $-1.96422$ & \\
    & $U(1)$& & $-2.18141$ & \\
    & $\varnothing$ & & $-2.24727$ & \\[3pt]
    \hhline{~||----}
    & \multirow{2}{*}{DGKT-like, IIA, with:} & & & \\
    & & & & \\
    & Branch A1-S1 & \cite[Tab. 2]{Marchesano:2019hfb} & $-2$ & $\, $\cite{Marchesano:2022rpr}? \\
    & Branch A2-S1 & & $0$ & \cite{Marchesano:2022rpr} \\[3pt]
    \hhline{-||----}
\multirow{4}{*}{$d=7$} & \multirow{2}{*}{AdS${}_7 \times$ S${}^3$, IIA, with:} &  &  & \\
 & & & & \\
    & d = 2 & \cite[Tab. 1]{Apruzzi:2019ecr} & $-9$ & \cite{Apruzzi:2019ecr} \\[3pt]
    \hline
    \end{tabular}
     \caption{Sample of $d$-dimensional non-supersymmetric perturbatively stable solutions with a quantum gravity uplift, together with further specifications including symmetry groups. In order to compare to the bound \eqref{boundm2l2}, we give the lowest mass squared for a scalar in units of the radius $l$, as can be found in the reference indicated. We also provide a reference where a non-perturbative instability of the solution has been pointed-out. Details are provided in the main text.}\label{tab:AdSnonsusy}
  \end{center}
\end{table}

After testing the bound \eqref{boundm2l2} on examples of perturbatively stable anti-de Sitter solutions in $d\geq 4$ (with quantum gravity uplifts), the result is twofold. First, we get that {\sl most supersymmetric solutions verify our bound}, the only exceptions being KKLT, LVS, and DGKT detailed below. Second, {\sl most non-supersymmetric solutions also verify the bound \eqref{boundm2l2} in the flexible sense}, meaning that the lowest $m^2 \, l^2$ is slightly above $-2$; we find two exceptions that admit no negative $m^2$. Interestingly, many non-supersymmetric solutions, including the two exceptions, are non-perturbatively unstable as conjectured in \cite{Ooguri:2016pdq}. Depending on the interpretation of that conjecture, this may put them in the swampland. In case the (non)-supersymmetric counter-examples to our bound rather fall in the landscape, another interpretation is that the mass spectrum has been drastically modified when moving in field space from the point where \eqref{boundV''} or \eqref{boundVneg} applies to the anti-de Sitter critical point. It would be interesting to study such an evolution of the spectrum in field space. Given our tests on solutions, it seems as well that the spectrum is better preserved in supersymmetric cases, and more deformed in non-supersymmetric ones. We hope to come back to these questions in future work.

We summarize our set of examples in Table \ref{tab:AdSsusy} and \ref{tab:AdSnonsusy}, and detail them in the following.

\subsubsection*{Examples of anti-de Sitter solutions and their spectrum, compared to \eqref{boundm2l2}}

\begin{itemize}

\item We start with the AdS${}_4 \times$ S${}^7$ solution in M-theory. Its Kaluza--Klein spectrum can be found in \cite[Tab. 4]{Nilsson:2018lof} (see also \cite{Duff:1986hr} or \cite[Tab. 1]{DHoker:2000pvz}).\footnote{The spectrum in \cite[Tab. 4]{Nilsson:2018lof} should be understood as follows, using the relation \eqref{Deltam}: one has $\Delta = E_{0\, {\rm (there)}}$ and also $m^2 \, l^2 = -2 + ({\rm Mass})^2_{{\rm (there)}}/4$, where ``there'' refers to the notations of \cite{Nilsson:2018lof}. We thank N.~Bobev and H.~Samtleben for related discussions.} The lowest scalar squared mass is obtained at level $n=1$, and is given by $m^2 \, l^2 = -9/4$ (the BF bound). The level $n=0$ gives $m^2 \, l^2 = -2$, and corresponds to a gauged supergravity mode \cite{Comsa:2019rcz}. These two lowest squared masses are found within the states $0^{\pm(1)}$, and verify our mass bound \eqref{boundm2l2}.\footnote{In \cite[Tab. 7]{Cassani:2012pj}, one can find AdS${}_4 \times$ S${}^7$ supersymmetric solutions with a spectrum verifying $m^2 \geq 0$. This is because of a specific truncation considered in that work, that differs from the one leading to $SO(8)$ 4d supergravity. The previously mentioned Kaluza--Klein spectrum remains valid, thus providing once again a mode verifying our bound.}

\item A systematic search for AdS${}_4$ solutions from M-theory was performed in \cite{Comsa:2019rcz}, using maximal gauged supergravity. In addition to the above solution, 4 supersymmetric solutions are indicated, one of which is new; we refer to \cite{Comsa:2019rcz} for the original references. All these solutions satisfy our bound \eqref{boundm2l2}.

\item Many non-supersymmetric solutions are also found in \cite{Comsa:2019rcz}. All of them are perturbatively unstable (see also \cite{Bena:2020xxb} and references therein) except one AdS${}_4$ solution \cite{Warner:1983du, Warner:1983vz} with $G=SO(3) \times SO(3)$. The scalar spectrum of that solution can be found in \cite[Tab. 2]{Fischbacher:2010ec}: the lowest mass is $m^2 = -12/7 \approx -  1.714$, which appears compatible with the flexible bound \eqref{boundm2l2}. That solution was also found to have a non-perturbative brane-jet instability \cite{Bena:2020xxb}.

\item Perturbatively stable AdS${}_4 \times$ S${}^6$ solutions of massive type IIA supergravity offer an interesting set of examples, summarized in \cite[Tab. 4.1]{Bobev:2020qev} (see also the older \cite[Tab. 2]{Guarino:2020jwv}). These solutions can be found as critical points of 4d ${\cal N}=8$ dyonic ISO(7) supergravity, as first shown in \cite{Guarino:2015jca}. They differ from one another by the number of preserved supersymmetry and a residual symmetry group $G$. Let us first focus on supersymmetric solutions: there are 7 of them. One can verify in \cite[App. B]{Bobev:2020qev} that all of them admit a scalar satisfying $m^2 \, l^2 \leq -2$.

\item Let us now consider the perturbatively stable non-supersymmetric AdS${}_4 \times$ S${}^6$ solutions: there are 9 of them \cite[Tab. 4.1]{Bobev:2020qev}. Their spectrum is given in \cite[App. C]{Bobev:2020qev} and we can read from there that their lowest mass is close to $-2$, if not lower. The furthest away from $-2$ is the solution with $G=G_2$ that has $m^2 = -1$, and then $G=SU(3)$ with $m^2=-1.582$. One may wonder whether Kaluza-Klein modes at higher levels could have lower masses as e.g.~in \cite{Malek:2020mlk}. The study of the Kaluza-Klein spectrum for these solutions shows however that it is not the case \cite{Guarino:2020flh} (see Figure 1 there for all but $G=G_2$). It also confirms the perturbative stability of these solutions. We may then consider that $m^2 = -1$ or $-1.582$ are in agreement with the flexible bound \eqref{boundm2l2}.

One may also express doubts on these solutions because they are not supersymmetric, following the conjecture of \cite{Ooguri:2016pdq}. While these solutions are perturbatively stable and do not have brane-jet instabilities \cite{Guarino:2020jwv}, the solution with $G=G_2$ admits another kind of non-perturbative instability, in the form of a bubble of nothing \cite{Bomans:2021ara}, in line with \cite{Ooguri:2016pdq}.

\item Three well-known supersymmetric AdS${}_4$ solutions seem to provide counter-examples to our bound: the one from KKLT \cite{Kachru:2003aw}, from LVS \cite{Balasubramanian:2005zx}, and the original DGKT solution \cite{DeWolfe:2005uu}. Indeed, the three solutions have light scalars whose masses squared verify $m^2 \geq 0$. We start with KKLT, a construction which requires non-perturvative contributions. The Calabi-Yau complex structure moduli and the dilaton are stabilized there by the tree-level potential. Thanks to its no-scale structure and supersymmetry,  one verifies that their masses satisfy $m^2 \geq 0$ \cite{Kachru:2003aw}. The spectrum of the K\"ahler moduli can be found e.g.~in \cite[Sec. 4.3.3]{Conlon:2020wmc}, where one verifies that the corresponding conformal weights are such that $\Delta > 3$, i.e.~$m^2 > 0$ with \eqref{Deltam}. We then turn to LVS, which requires perturbative corrections. Its light spectrum can be found in \cite[Sec. 2]{Conlon:2018vov}, from which we read again $m^2 \geq 0$, the axion being massless. Finally, we turn to the DGKT solution: as can be read in \cite[Sec. 3.4]{DeWolfe:2005uu}, the metric and dilaton fluctuations have positive definite mass squared. Regarding the other, axionic fields, the spectrum depends on flux signs, related to supersymmetry. Picking the supersymmetric choice (corresponding to case 4 in \cite[App. C.2.1]{Conlon:2021cjk}), we get again $m^2 > 0$ for those fields. Last but not least, the blow-up modes coming from the resolution of the orbifold singularity are there K\"ahler moduli, and those are also stabilized with $m^2 >0$. Let us also mention that a dual version of this original DGKT solution and its spectrum has been discussed recently in \cite{Bardzell:2022jfh}: this type IIB solution is obtained from a Landau-Ginzburg model without K\"ahler moduli, the mirror situation of the original DGKT solution which has no complex structure moduli. The spectrum obtained is the same.

These three solutions and their quantum gravity origin are highly debated in the literature. One question is the control on the (non)-perturbative contributions just mentioned, as well as the smearing of the sources. We refrain from entering this debate here. In case these solutions are in the landscape, a possible explanation for a violation of our bound \eqref{boundm2l2} is a drastic modification of the spectrum when moving in field space; we refer to the above summary for a discussion of this idea.

\item Let us also mention DGKT-like solutions in $d=4$ classified in \cite{Marchesano:2019hfb}, generalizing the original ones \cite{DeWolfe:2005uu, Camara:2005dc}. The spectrum of the various solution branches is discussed there: the modes with lowest mass can be found in \cite[Tab. 2]{Marchesano:2019hfb}. We see that for two of the three branches, one gets precisely a state with $m^2 \, l^2 = -2$, obeying \eqref{boundm2l2}. The last branch (A2-S1) however admits no state with negative $m^2$. Those solutions could then provide again a counter-example to our mass bound \eqref{boundm2l2}. Note however that they are non-supersymmetric, and suffer from a non-perturbative instability \cite{Marchesano:2022rpr}.

\item Finally, some AdS${}_4$ solutions were shown to uplift to type IIB string theory as S-folds: see \cite{Guarino:2020gfe, Bobev:2021yya, Berman:2021ynm, Bobev:2021rtg}  and references therein. In \cite{Guarino:2020gfe}, one can find 3 supersymmetric solutions and their spectra. The lowest scalar mass verifies $m^2 \, l^2 = -2$, thus satisfying our bound. There is also one perturbatively unstable non-supersymmetric solution.

\item We turn to $d=5$ and start with the mass spectrum of AdS${}_5 \times$ S${}^5$ \cite{Kim:1985ez} (see also \cite{Bigazzi:2002gyi}). The lowest masses of the scalars are $m^2 \, l^2=-4,-3$, at and above the BF bound. Our mass bound \eqref{boundm2l2} is then satisfied.

\item A systematic search for AdS${}_5$ solutions of ${\cal N}=8$ SO(6) gauged supergravity (consistent truncation of type
IIB string theory on S${}^5$) was conducted in \cite{Bobev:2020ttg}. All non-supersymmetric solutions that were found are perturbatively unstable. The only two stable, supersymmetric solutions are the previous one ($SO(6)$, $\N=8$), and the one of \cite{Khavaev:1998fb} ($SU(2)\times U(1)$, $\N=2$) whose spectrum was first given in \cite{Freedman:1999gp} (see also \cite[Tab. D.4]{Bobev:2020ttg}). Its lowest mass squared is $m^2 \, l^2 =-4$.

\item Last but not least, AdS${}_7$ solutions of type IIA string theory and their stability are discussed in \cite{Apruzzi:2019ecr}. The supersymmetric solutions have been classified, and they are considered in \cite{Apruzzi:2019ecr} together with non-supersymmetric counterparts. The BF bound in $d=7$ is given by $m_{{\rm BF}}^2\, l^2 = -9$. Supersymmetric solutions have a dilaton of mass $m^2 \, l^2 = -8$, so they always satisfy our bound. For non-supersymmetric solutions, the dilaton mass is $m^2 \, l^2 = 12$. In addition to the dilaton, the spectrum for some sets of scalars was studied for these solutions. Those sets are related to representations of an SU(2) R-symmetry (see \cite[Tab. 1]{Apruzzi:2019ecr}), as well as to a number of $D_8$-branes present in the solution. Within those sets of scalars, the non-supersymmetric solutions are either found perturbatively unstable, or they admit scalars with $m^2 \, l^2 = -9$ (representation ${\rm d}=2$, shown in Table \ref{tab:AdSnonsusy}), or $m^2 \, l^2 \geq 0$ (representation ${\rm d}=3$). We cannot conclude from the latter on a violation of our bound, because many other scalars and their spectrum are not studied. We also note that these seemingly perturbatively stable non-supersymmetric solutions are shown to exhibit a non-perturbative instability, related to $N\!S_5$-brane bubbles. Finally, let us also mention an analogous situation for AdS${}_6$ solutions \cite{Apruzzi:2021nle}.

\end{itemize}

\subsubsection*{Holographic consequences of \eqref{boundm2l2}}

\begin{figure}[t]
\begin{center}
\includegraphics[width=0.7\textwidth]{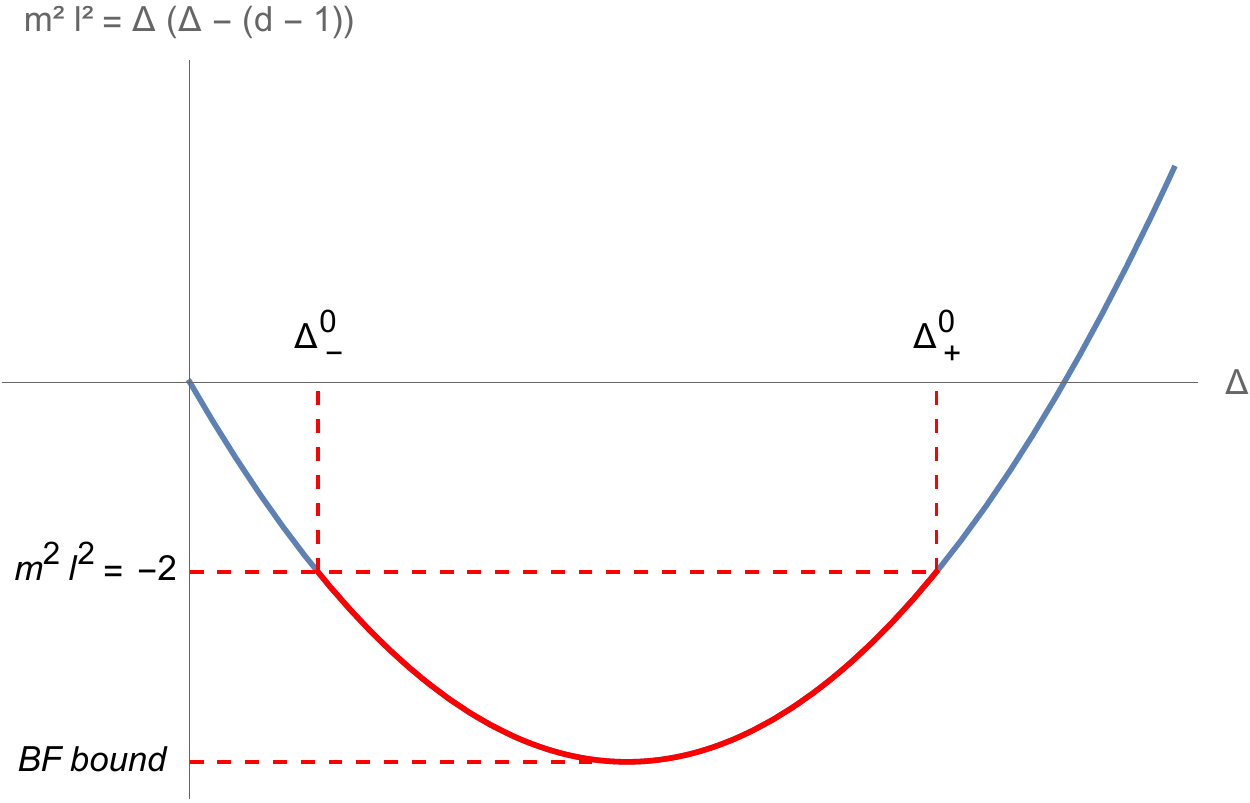}
\caption{$m^2\, l^2$ as a function of the conformal dimension $\Delta$, with the mass bound $m^2\, l^2 \leq -2$ \eqref{boundm2l2} and the corresponding dimensions $\Delta_{\pm}^0$. The masses allowed by this upper bound and above the BF bound, can be found in the red region at the bottom of the curve. The graph is displayed here for $d=5$, but it is qualitatively similar for any $d \geq 4$.}\label{fig:Deltaplot}
\end{center}
\end{figure}

Let us now investigate the consequences of the mass bound \eqref{boundm2l2} for holography with $d \geq 4$. Given the standard relation to the conformal dimension $\Delta$ of an operator in a dual CFT
\beq
\Delta(\Delta -(d-1)) = m^2 l^2 \ , \quad \Delta_{\pm} = \frac{d-1}{2} \pm \frac{1}{2} \sqrt{(d-1)^2 + 4 m^2 l^2} \ ,\label{Deltam}
\eeq
the mass bound \eqref{boundm2l2} gets translated into
\beq
\Delta_-^0  \leq \Delta \leq \Delta_+^0 \ ,\quad \Delta_{\pm}^0 = \frac{d-1}{2} \pm \frac{1}{2} \sqrt{(d-1)^2 -8 } \ ,\label{Deltaineq}
\eeq
as illustrated in Figure \ref{fig:Deltaplot}. Predicting the existence of a state with such a mass correspond to having a systematic relevant operator whose dimension obeys \eqref{Deltaineq}.

One may also compare \eqref{Deltaineq} to the unitary bound for CFT scalar fields: $\Delta \geq  (d-3)/2$. For $d > 4$, we get $\Delta_-^0  \leq (d-3)/2 \leq \Delta_+^0$. For $d=4$ however, we get $\Delta_-^0  \geq (d-3)/2$ so the mass bound guarantees there the scalar unitarity bound.\\

Last but not least, for $d=4$, the bound becomes $1 \leq \Delta \leq 2$. This is the only dimension for which one gets integer values for $\Delta_{\pm}^0$. In other words, saturating the bound $m^2\, l^2 = -2$ gives $\Delta=1$ or $2$. Getting these integers in this way is remarkable enough to be mentioned. Indeed, these integers were argued to play an important role in scale separated anti-de Sitter solutions and their holographic dual CFT \cite{Conlon:2021cjk, Apers:2022zjx, Apers:2022tfm, Quirant:2022fpn, Plauschinn:2022ztd}, where $m^2\, l^2 = -2$ was shown to correspond to saxions \cite{Marchesano:2019hfb, Conlon:2021cjk}. The peculiarity of having a CFT with integer conformal dimensions was there associated to the peculiarity of having scale separation. The mass bound \eqref{boundm2l2} may then also play a role in these arguments.

\section{Multifield extension}\label{sec:multi}

The ATCC and its consequences, in particular the characterisation of the scalar potential $V$, have so far been studied for a single canonical scalar field. Typical string compactifications however lead to multifield effective theories \eqref{actionintro}. This situation requires a multifield extension of Section \ref{sec:ATCCgen} and \ref{sec:V''}. We briefly discuss in this section several multifield extensions, leaving a more thorough study to future work.\\

A first natural extension is to trade the single canonical field for a one-dimensional path in field space with canonical parameter $\hat{s}$; we will also refer to it in Section \ref{sec:stringpot} as a canonical field direction $\hat{t}$. As for the TCC in \cite[Sec. 3.2]{Bedroya:2019snp}, the reasoning pursued here in Section \ref{sec:V} to characterise the scalar potential can be adapted, using the path in field space. Then, one reaches similar bounds: schematically, one gets
\beq
V<0: \quad\quad V(\varphi^i(\hat{s})) \geq\,  -\,  e^{ - \frac{2\, |\hat{s} - \hat{s}_i|}{\sqrt{(d-1)(d-2)}} } \ ,\quad\quad  \left< \frac{\nabla_{\hat{s}} V}{|V|} \right>_{\hat{s} \to \pm \infty}  \geq \ \frac{2}{\sqrt{(d-1)(d-2)}}  \ ,\label{boundVs}
\eeq
considering $V<0$, $\nabla_{\hat{s}} V \geq 0$, and where the gradient along the path and corresponding average are defined as
\beq
\nabla_{\hat{s}} V = \frac{\del \varphi^i}{\del \hat{s}} \del_i V \ , \quad \quad \left< \frac{\nabla_{\hat{s}} V}{V} \right> = \frac{1}{\Delta \hat{s}} \int_{\hat{s}_i}^{\hat{s}} \frac{\nabla_{s} V}{V} \, \d s \ .
\eeq
Note that the distance along the path $|\hat{s} - \hat{s}_i| = \Delta \hat{s}$ could be replaced, in the lower bound on $V$, by a shorter distance: the geodesic distance between the two points considered. This brief derivation shows that considering the gradient of the potential along one field direction, $\nabla_{\hat{s}} V$, the ATCC bound on the first derivative of the potential \eqref{boundV'V} and the corresponding rate are unchanged, as given in \eqref{boundVs}.

A different gradient has however been considered in the literature: $\nabla V = \sqrt{g^{ij} \del_i V \del_j V}$. It captures derivatives of the potential along all fields at any point in field space. One necessarily has $\nabla V \geq |\nabla_{\hat{s}} V|$, as can be seen with a canonical field basis. This gradient played a crucial role in the Strong de Sitter Conjecture \cite{Rudelius:2021oaz, Rudelius:2021azq}, stating that the following holds in asymptotics of field space
\beq
V>0: \quad \left. \frac{\nabla V}{V}\right|_{\infty} \geq \frac{2}{\sqrt{d-2}} \ .\label{StrongdSC}
\eeq
The main point is that the rate appearing in \eqref{StrongdSC} is greater than the TCC one, but both can be compatible since $\nabla V \geq |\nabla_{\hat{s}} V|$. Let us recall that the precise value of the bound rate has been obtained by requiring it to be preserved under dimensional reduction. Here with $V<0$, a second possible multifield extension for the ATCC is then
\beq
V<0: \quad \left. \frac{\nabla V}{|V|}\right|_{\infty} \geq \frac{2}{\sqrt{d-2}} \ ,\label{StrongVneg}
\eeq
also possibly compatible with the first one \eqref{boundVs}.\\

The first extension above, involving a single field direction, raises several questions that we now discuss. We will face these questions again in Section \ref{sec:stringpot} when considering explicit potentials from string compactifications. To start with, we see that different paths or field directions can a priori be chosen. In order to test the ATCC, the ideal situation is to have a field direction along which the potential is a negative, growing exponential, but it is not always the case. Put differently, finding a specific field direction with such an asymptotic behaviour that allows comparing the exponential rate to the ATCC bound \eqref{boundVs}, is actually non-trivial. The freedom in field directions and asymptotic behaviours complicates the analysis.

A second question is whether it is meaningful to focus on a single field direction and its asymptotics in a (multi)field space. In Section \ref{sec:stringpot}, we will do so and simply ``freeze'' the other fields, by setting them to a finite value. But one could ask for a mechanism realising this, such as having them stabilized. Note that stabilizing the other fields, or at least placing them at a potential extremum, brings $\nabla V$ closer, if not equal, to $|\nabla_{\hat{s}} V|$, therefore confusing the two extensions \eqref{boundVs} and \eqref{StrongVneg}.

An interesting example found recently in \cite{Calderon-Infante:2022nxb} provides a good illustration of these various points for $V>0$ and the TCC. Certain Calabi-Yau compactifications of string theory to $d=4$ were found there to provide a field direction, in the large complex structure and weak string coupling limit, having an exponential rate lower than the TCC one, namely $\sqrt{2/7} < \sqrt{2/3}$. This illustrates that while the TCC has been tested among many field directions, especially combinations of (sub)volumes and dilaton, \cite{Andriot:2020lea,Andriot:2022xjh}, one may still find other field directions having a different potential asymptotic behaviour and rate. In addition, these compactifications exhibit further fields, the K\"ahler moduli, appearing in the scalar potential through the K\"ahler potential, and those cannot be stabilized perturbatively. That situation illustrates the question of the appropriate gradient, or more generally for us, whether the multifield extension \eqref{boundVs} or \eqref{StrongVneg} is preferred.\\

Last but not least, one may wonder about the multifield extension of the new bound \eqref{boundV''} on the second derivative of the potential. While this would deserve a more thorough analysis, we propose for now the following natural extensions, valid in the asymptotics of field space
\bea
& V>0:\quad \quad \left. \frac{{\rm max}\,\nabla \del V}{V} \right|_{\infty}  \ \geq \ \ \frac{4}{(d-1)(d-2)} \ ,\label{condetaVpos}\\[0.1in]
& V < 0: \quad\quad \left. \frac{{\rm min}\, \nabla \del V}{|V|} \right|_{\infty}  \ \leq \ -\frac{4}{(d-1)(d-2)} \ , \label{condetaV}
\eea
where we use the same notations as around \eqref{LPVcond}, $\nabla \del V$ standing for the mass matrix. The difference between \eqref{condetaV}, and \eqref{LPVcond} from \cite{Lust:2019zwm} for $V<0$ is interesting. The condition \eqref{condetaV} is actually the same as the one of the refined de Sitter conjecture \cite{Palti:2019pca} for $V>0$, up to the constant in the right-hand side that we specify here. While the condition \eqref{LPVcond} was meant in \cite{Lust:2019zwm} to extend the refined de Sitter conjecture to negative potentials, we rather obtain that the condition does not need to be modified for negative potentials.

However, we recall that \eqref{condetaVpos} and \eqref{condetaV} are not meant to be applied at potential extrema. They may though be extrapolated there by continuity in field space, as already discussed and tested in Section \ref{sec:massbound}. For $V<0$, this leads to the anti-de Sitter mass bound
\beq
{\rm AdS}_d:\quad {\rm min}\ m^2 \, l^2\ \lesssim \ - 2 \ , \label{boundm2l2multi}
\eeq
predicting the existence of at least one scalar whose mass obeys this bound. We refer to the discussion around \eqref{boundm2l2} for more details.

\section{Scalar potentials from string theory}\label{sec:stringpot}

In this section, we consider examples of scalar potentials obtained from string theory compactifications, and compare their behaviour to the one predicted by the ATCC, namely that a negative potential is bounded from below by a growing exponential \eqref{boundV}, and that it obeys the asymptotic bound \eqref{boundV'V} (see also \eqref{boundVs}).

In the asymptotics, potentials of string effective theories typically have three possible behaviours: either the potential diverges positively (the field can then be stabilized), or it converges to zero positively, or negatively. Indeed, having a potential diverging to $-\infty$ would be suspicious in terms of physics, and having an asymptotic positive or negative constant potential would also be unusual, since cosmological constants are typically rather obtained as local extrema of the potential than actual constant potentials. Therefore, sticking to the first three asymptotic behaviours, only the potential converging to zero negatively is relevant to test the ATCC.

When expressed in terms of canonically normalized fields, the potentials considered here are given as sums of exponentials. Relevant potentials are then those which are asymptotically negative, but growing exponentially to zero. In this situation, the comparison to the ATCC predictions for one field becomes straightforward. If for $V(\varphi) < 0$ at $\varphi \to \infty$ the dominant term is $V_0 \ e^{-c\, \varphi}$ ($V_0 <0$, $c>0$), which means that $c$ is the minimum rate among the different terms, then one checks \eqref{boundV} and \eqref{boundV'V} (or \eqref{boundVs}) essentially by verifying
\beq
\text{ATCC bound:}\quad \quad c \geq \ \frac{2}{\sqrt{(d-1)(d-2)}} \ , \label{cbound}
\eeq
as in \eqref{rate}. For this reason, we mostly focus in the following on the rates of the exponentials.

The potentials to be considered are multifield. As discussed in Section \ref{sec:multi}, this complicates the analysis on several aspects. In particular, there is no guarantee that the potentials exhibit the desired behaviour (negative and growing exponential) along the fields on which the potential depends. Most of the time, such a behaviour only appears along one specific field direction, which is a combination of the appearing fields. For each of the following potentials, we will make such specific field directions appear, in various manners.

We will also ignore the orthonormal field directions; in other words we just freeze the other fields to some finite value. This point was discussed in Section \ref{sec:multi}, where we explained that including other fields could increase the gradient $|\nabla V|$. Here we only test single field direction gradients \eqref{boundV'V} or \eqref{boundVs}, with the lower bound \eqref{cbound}, so we allow ourselves to ignore the other, orthonormal fields.

We start in Section \ref{sec:RhoTauSigmaPot} with a semi-universal potential $V(\rho, \tau, \sigma)$ in classical string compactifications to $d$ dimensions, $3 \leq d \leq 10$, and turn in Section \ref{sec:DGKT} to potentials obtained by compactifications to $d=3,4$ DGKT-like anti-de Sitter solutions \cite{DeWolfe:2005uu, Camara:2005dc, Farakos:2020phe}.

\subsection{Scalar potential for $(\rho, \tau, \sigma)$} \label{sec:RhoTauSigmaPot}

We first consider the scalar potential $V(\rho, \tau, \sigma)$, motivated in \cite{Hertzberg:2007wc, Danielsson:2012et} and derived (as well as the kinetic terms) in arbitrary dimension $d$ in \cite{VanRiet:2011yc, Andriot:2022xjh}. The fields $\rho$ (the $(10-d)$-dimensional volume) and $\tau$ (the $d$-dimensional dilaton) are universal fields in classical string compactifications. We consider such compactifications here, using 10d supergravity. The field $\sigma$ depends on a set of parallel $D_p$-branes and orientifold $O_p$-planes: it is related to the internal volume wrapped by these $O_p/D_p$ sources, or equivalently, to the volume of their transverse space. We consider for now only one such set of sources. In the case of several intersecting sets, the potential (and kinetic terms) have been derived in \cite{Andriot:2019wrs, Andriot:2020wpp, Andriot:2022yyj}; see also one example below in \eqref{pot5fields}. Note that we do not consider other types of sources, in particular no $\overline{D}_p$-brane.

The potential $V(\rho, \tau, \sigma)$ in arbitrary dimension $d$ is given in \cite[(4.20)]{Andriot:2022xjh}. We consider in the following the simplified notation where we drop the volume integral, and set $M_p=1$. Trading the fields for their canonically normalized counterparts
\beq
\tau=e^{\frac{1}{\sqrt{d-2}}\hat{\tau}}\, , \qquad \rho=e^{\sqrt{\frac{4}{10-d}}\hat{\rho}}\, ,\qquad \sigma=e^{\sqrt{\frac{-4}{AB(B-A)}}\hat{\sigma}} \ , \label{canfields}
\eeq
with $A=p-9$, $B=p+1-d$, we rewrite the potential as follows
\bea
2\ & V(\hat{\rho}, \hat{\tau}, \hat{\sigma}) \label{Vrts}\\
= &\  e^{\frac{-2}{\sqrt{d-2}}\hat{\tau}} \Bigg(-e^{-\sqrt{\frac{4}{10-d}}\hat{\rho}}\, \R_{10-d}(\hat{\sigma}) + \frac{1}{2} e^{-3\sqrt{\frac{4}{10-d}}\hat{\rho}} \sum_n e^{(-An-B(3-n))\sqrt{\frac{-4}{AB(B-A)}}\hat{\sigma}} |H^{(n)}|^2 \Bigg) \nn \\&
+ \frac{1}{2} e^{\frac{2-2d}{\sqrt{d-2}}\hat{\tau}} e^{(3-d)\sqrt{\frac{4}{10-d}}\hat{\rho}} \sum_n e^{(-An-B(7-n))\sqrt{\frac{-4}{AB(B-A)}}\hat{\sigma}} |H_7^{(n)}|^2 \nn\\&
- e^{-\frac{d+2}{2\sqrt{d-2}}\hat{\tau}}\, e^{\frac{2p-8-d}{4} \sqrt{\frac{4}{10-d}}\hat{\rho}}\, e^{\frac{1}{2}B(p-9)\sqrt{\frac{-4}{AB(B-A)}}\hat{\sigma}} \, g_s \frac{T_{10}}{p+1} \nn \\&
+ \frac{1}{2} g_s^2\, e^{\frac{-d}{\sqrt{d-2}}\hat{\tau}} \sum_{q=0}^{10-d} e^{\frac{10-d-2q}{2} \sqrt{\frac{4}{10-d}}\hat{\rho}} \sum_n e^{(-An-B(q-n))\sqrt{\frac{-4}{AB(B-A)}}\hat{\sigma}} |F_q^{(n)}|^2 \ ,\nn
\eea
and we refer to \cite{Andriot:2022xjh} for more details on notations. In the case where the (internal) $10-d$ dimensions form a group manifold, then one has
\bea
{\cal R}_{10-d} (\hat{\sigma}) =& -e^{-\sqrt{\frac{-4B}{A(B-A)}}\hat{\sigma}}\ (\delta^{cd}   f^{b_{\bot}}{}_{a_{||}  c_{\bot}} f^{a_{||}}{}_{ b_{\bot} d_{\bot}})^0 + e^{\sqrt{\frac{-4A}{B(B-A)}}\hat{\sigma}} \left( {\cal R}_{||} +  {\cal R}_{||}^{\bot} \right)^0 \nn \\&
- \frac{1}{2} e^{(-2B+A)\sqrt{\frac{-4}{AB(B-A)}}\hat{\sigma}} |f^{{}_{||} 0}{}_{{}_{\bot} {}_{\bot}}|^2  \ . \label{curvV}
\eea
Let us make use of this semi-universal scalar potential to test the ATCC bound \eqref{cbound}, first along $(\hat{\rho}, \hat{\tau}, \hat{\sigma})$ and then along specific field directions.

\subsubsection{Analysis of the exponential rates}\label{sec:rhorate}

In the potential $V(\hat{\rho}, \hat{\tau}, \hat{\sigma})$ given in \eqref{Vrts}, all flux terms are positive. The only terms that have a chance to be negative are the source term depending on $T_{10}$, and some of the curvature terms in \eqref{curvV}. In order to test the ATCC, especially the bound \eqref{cbound}, let us then look at the exponential rates there. As for the asymptotics, we consider $\hat{\tau}, \hat{\rho} \to \infty$ which correspond to the large volume and small string coupling limits; the opposite limits would invalidate the supergravity approximation of the classical string regime. Regarding the last field, we can however consider both $\hat{\sigma} \to \pm \infty$.

In the source and curvature term, it is straightforward to verify for $d\geq 3$ that the rates for $\hat{\tau}$ verify the ATCC bound \eqref{cbound}. Turning to $\hat{\rho}$, we first consider the curvature term: we obtain that the ATCC bound \eqref{cbound} is satisfied for $d\geq 4$, and is violated for $d=3$. This is not a surprise: as discussed already in \cite{Andriot:2022xjh}, a violation of the TCC has also been noticed for $d=3$ and can be interpreted as being due to the peculiarity of gravity in $d=3$.\footnote{It remains legitimate to ask whether a $d=3$ compactification can indeed be found with a background where the dominant contribution comes from the curvature term: for instance, is it possible to satisfy the Einstein equations with internal curvature but without flux? Since we will get further violations in $d=3$, we leave this question aside.} We now turn to the source term, which is more subtle. We focus on the case where $2p-8-d<0$ to get the right asymptotics, and then consider the square of the rate minus the ATCC bound squared, giving the quantity
\beq
\frac{(2p-8-d)^2}{4(10-d)} - \frac{4}{(d-1)(d-2)} \ .
\eeq
Up to $d=3$ cases ($p=2,...,5$) and two more exceptions, we get this difference to be positive or zero whenever $2p-8-d<0$. This means that the ATCC bound \eqref{cbound} is again satisfied.

The two exceptions are $d=7, p=7$ and $d=4, p=5$, and their analysis is interesting. It is worth noting that it is not possible to get a non-zero $T_{10}$ as considered here without flux, in a meaningful compactification, because of the 10d Bianchi identities (or tadpole cancellation). For $d=7, p=7$, one must have $F_1 \neq 0$, magnetically sourced by $O_7/D_7$. In turn, an $F_1$ in $d=7$ leads to a positive and growing exponential in the potential, thus dominating the source term in the asymptotics, so this first exception is not relevant. In the case of $d=4, p=5$, an $F_3$ flux is magnetically sourced, but the tadpole cancellation can also be verified with a combination of $H$ and $F_1$: the Bianchi identity is $\d F_3 - H\w F_1 \propto T_{10}$. Having an $F_1$ would lead to the same conclusion as for the previous exception. However, in absence of $F_1$, one must have a non-zero $F_3$. Surprisingly, there is then no contribution to the $\hat{\rho}$ dependence from the $F_3$ flux term. Still, as far as $\hat{\rho}$ is concerned, this flux term adds a constant positive contribution to the potential, so in the asymptotics the potential stops being negative; then again, this exception is not relevant. It is remarkable that the two possible exceptions, beyond $d=3$, are removed thanks to the requirement of Bianchi identities or tadpole cancellation, which from the point of view the effective theory come as a quantum gravity input. To conclude, there is no violation of the ATCC bound for $\hat{\rho}$ with $d\geq 4$.

We finally turn to $\hat{\sigma}$. Considering this field is only meaningful for $AB\neq0$, which means $d \leq p \leq 8$. We start with the source term and the limit $\hat{\sigma} \to \infty$, given the negative sign in the exponential. We consider the square of the rate minus the ATCC bound squared, giving the quantity
\beq
\frac{(9-p)(p+1-d)}{10-d} - \frac{4}{(d-1)(d-2)} \ .
\eeq
We verify that this is strictly positive for $d \leq p \leq 8$ and $d\geq 4$, and strictly negative for $d=3$. This shows again a verification of the ATCC bound \eqref{cbound} for $d\geq 4$, and a violation at $d=3$. The curvature terms are more tricky. The last term in \eqref{curvV} gives a positive contribution to the potential, so we are rather interested in the first two terms: those have indefinite signs. In those two terms, one has competing exponentials: one grows when the other one diminishes. Therefore, while one might get a violation of the ATCC bound from one of the exponential, the other one could diverge and thus dominate. Without a more thorough analysis of the possible signs of these terms and details of the compactification, it is difficult to conclude.

This ends our general analysis of the exponential rates appearing explicitly in $V(\hat{\rho}, \hat{\tau}, \hat{\sigma})$ \eqref{Vrts}, seeing no violation of the ATCC bound \eqref{cbound} for $d\geq 4$ but several at $d=3$. This provides an interesting confirmation of the ATCC for $d\geq 4$. We now turn to specific field directions, encoded in no-go theorems.

\subsubsection{No-go theorems for anti-de Sitter and specific field directions}

Standard no-go theorems on (quasi) de Sitter solutions can be recast as follows, with $a >0$
\beq
a\, V+\sum_i b_i\, \partial_{\hat{\varphi}^i} V \leq 0 \quad \Leftrightarrow \quad V + \frac{1}{c}\, \del_{\hat{t}^{\,1}} V \leq 0 \ ,\quad \quad {\rm with}\ \ c = \frac{a}{\sqrt{\sum_i b_i^2}} \ , \label{nogogen}
\eeq
as pointed-out in \cite{Andriot:2019wrs, Andriot:2020lea} and formally clarified in \cite{Andriot:2021rdy}. This allows to rewrite the obstruction carried by the no-go theorem into a swampland format, and then compare the no-go theorem to corresponding conjectures, testing for instance the TCC bound on $c$ \cite{Bedroya:2019snp}. In addition, this rewriting indicates a (canonically normalized) field direction $\hat{t}^{\,1}$ through
\beq
\sum_i b_i\, \partial_{\hat{\varphi}^i} = \sqrt{\sum_i b_i^2} \ \del_{\hat{t}^{\,1}} \ . \label{relfields}
\eeq
The reformulation \eqref{nogogen} then shows that $c$ is not a random constant, but it can correspond to the lowest exponential rate for $\hat{t}^{\,1}$ in the scalar potential \cite{Andriot:2020lea, Andriot:2021rdy}, provided the assumptions of the no-go theorem hold. This interpretation will be made explicit in an example below. This point of view is interesting as it relates no-go theorems to specific field directions, having specific exponential rates; those can then be compared to the TCC bound \cite{Andriot:2020lea, Andriot:2022xjh}. In the following, we make use of this idea on (rare) no-go theorems against (quasi) anti-de Sitter solutions in $d=4$. We will compare this way the corresponding rates for specific field directions to our ATCC bound \eqref{cbound}. To reach this result, a first step is to reproduce the 10d no-go theorems in terms of the potential of a 4d effective theory.\\

We start with the no-go theorem obtained in \cite[(3.15)]{Andriot:2022way} for the solution class $s_{555}$. Specifying to this class amounts to an assumption for the no-go, since it determines the orientifolds and projects-out various fields; in particular, one has $F_1=F_5=H=0$. To rewrite it in a 4d fashion, we simply follow \cite[(4.39)]{Andriot:2022xjh}; the 10d derivation via \cite[(3.14)]{Andriot:2022way}, matching \cite[(2.19)]{Andriot:2022xjh}, indicates that the 4d version is indeed given by \cite[(4.39)]{Andriot:2022xjh}. We use the latter with $d=4, p=5$ and obtain
\beq
2\, V(\hat{\rho}, \hat{\tau}) + \frac{1}{\sqrt{2}}\, \partial_{\hat{\tau}} V + \sqrt{\frac{3}{2}}\, \partial_{\hat{\rho}}V=0 \, , \label{nogos555}
\eeq
where the right-hand side vanishes because of $s_{555}$; the no-go theorem actually forbids both de Sitter and anti-de Sitter solutions. Note that the potential $V(\hat{\rho}, \hat{\tau})$ is the same as \eqref{Vrts} with $\hat{\sigma}=0$, even with multiple sets of $O_5/D_5$. From \eqref{nogos555} and \eqref{nogogen}, we deduce the value
\beq
c=\sqrt{2} \ ,
\eeq
which in $d=4$ is in agreement with the ATCC bound \eqref{cbound}.

Let us now illustrate the fact that this no-go theorem defines a specific canonically normalized field direction with a corresponding rate $c$. Following \cite{Andriot:2021rdy}, we can introduce the field space diffeomorphism as a matrix $P$
\beq
P= \left( \frac{\del \hat{\varphi}}{\del \hat{t}} \right) \ ,\ P^T \left(\del_{\hat{\varphi}} \right) = \left(\del_{\hat{t}} \right) \ , \ P^{-1} (\d \hat{\varphi})^T = (\d \hat{t})^T  \ , \label{Pdel}
\eeq
where we gave the usual diffeomorphism relations for vectors (here $\left(\del_{\hat{\varphi}} \right)$ is a column matrix) and for one-forms (here $(\d \hat{\varphi})$ is a row matrix). Having here $\hat{\varphi}^i$ and $\hat{t}^i$ canonically normalized imposes that $P$ is an orthonormal matrix; one can verify that $(\d \hat{\varphi}) \delta (\d \hat{\varphi})^T$ is then preserved. The relation \eqref{relfields} provides us with the first line of $P^T$. Since $P^T=P^{-1}$, we deduce the first line of $P^{-1}$, and that the same relation as \eqref{relfields} holds for the one-forms. We infer, up to a constant, a relation between the fields: for two fields, we obtain
\beq
\hat{t}^{\,1} = \frac{b_1}{\sqrt{b_1^2 + b_2^2}} \hat{\varphi}^1 + \frac{b_2}{\sqrt{b_1^2 + b_2^2}} \hat{\varphi}^2 \, , \quad \hat{t}^{\,2} = -\frac{b_2}{\sqrt{b_1^2 + b_2^2}} \hat{\varphi}^1 + \frac{b_1}{\sqrt{b_1^2 + b_2^2}} \hat{\varphi}^2 \ ,
\eeq
as also mentioned in \cite{Andriot:2020lea}.

Applying this formalism to $\hat{\varphi}^1=\hat{\rho},\ \hat{\varphi}^2=\hat{\tau}$, with $\hat{t}^{\,1}=\hat{t},\ \hat{t}^{\,2}=\hat{t}_{\perp}$, we obtain from the no-go theorem \eqref{nogos555} the fields
\beq
\hat{t} = \frac{\sqrt{3}}{2} \hat{\rho} + \frac{1}{2} \hat{\tau} \, , \quad \hat{t}_\perp = -\frac{1}{2} \hat{\rho} + \frac{\sqrt{3}}{2} \hat{\tau} \ .
\eeq
We now rewrite the potential $V(\hat{\rho}, \hat{\tau})$ from \eqref{Vrts} in terms of the new specific fields, for $s_{555}$, and get
\beq
2\, V(\hat{t}, \hat{t}_\perp)= e^{-\sqrt{2}\, \hat{t}} \left(- e^{-\sqrt{\frac{2}{3}}\, \hat{t}_\perp} {\cal R}_6 - g_s \frac{T_{10}}{6} e^{-\sqrt{\frac{8}{3}}\, \hat{t}_\perp} + \frac{1}{2} g^2_s |F_3|^2 e^{-\sqrt{6}\, \hat{t}_\perp}\right) \ .
\eeq
As predicted, we see that $c=\sqrt{2}$ appears as the exponential rate of $\hat{t}$. A fair comparison to the ATCC bound would require the overall sign of the terms in the brackets to be negative; this however depends on the details of the compactification. We also verify that the exponential rates for $\hat{t}_\perp$ satisfy the ATCC bound in $d=4$.\\

We turn to the no-go theorem obtained in \cite[Sec. 3.3.3]{Andriot:2022way} for the solution class $m_{466}$ in $d=4$. Although it seems to correspond to the T-dual situation of the previous no-go theorem (and we thus expect the same final $c$ value), its 10d derivation requires different equations. As a consequence, a 4d derivation has not been provided. We work it out here, using a different scalar potential, $V(\rho, \tau, \sigma_1, \sigma_2, \sigma_3)$, where $\sigma_{1}$ corresponds to the set of $O_4/D_4$ along internal direction 4 and $\sigma_{I=2,3}$ to sets of $O_6/D_6$ along directions $123$ and $156$. Explicitly, the scalar potential is obtained thanks to the code {\tt MSSSp} \cite{Andriot:2022yyj} and reads
\bea
2\, V =& - \tau^{-2} \rho^{-1}\, {\cal R}_6(\sigma_1, \sigma_2, \sigma_3) + \frac{1}{2} \tau^{-2} \rho^{-3} \sigma_1^{-3} \sigma_2^{3} \sigma_3^{3}\, |H|^2 + \frac{1}{2} \tau^{-4} \rho \sigma_1^{-2}\, |F_2|^2 + \frac{1}{2} \tau^{-4} \rho^{-1} \sigma_1^{2}\, |F_4|^2 \nn \\&
- \tau^{-3} \rho^{-1} \sigma_1^{\frac{-5}{2}}
\sigma_2^{\frac{3}{2}} \sigma_3^{\frac{3}{2}}\,  \frac{T_{10}^{(4)}}{5}  - \tau^{-3} \sigma_1^{\frac{3}{2}}
\sigma_2^{-\frac{9}{2}} \sigma_3^{-\frac{9}{2}} \left(\sigma_3^6\, \frac{T_{10}^{(6)_2}}{7} + \sigma_2^6 \, \frac{T_{10}^{(6)_3}}{7}\right) \ ,\label{pot5fields}
\eea
where
\bea
{\cal R}_6(\sigma_1, \sigma_2, \sigma_3) & = R_1 \, \sigma_1^{5} \sigma_2^{-9} \sigma_3^{3} +R_2\, \sigma_1^{-7} \sigma_2^{3} \sigma_3^{3} + R_3\, \sigma_1^{5} \sigma_2^{3} \sigma_3^{-9} \\
& +R_4\, \sigma_1^{-1} \sigma_2^{-3} \sigma_3^{3} +R_5\, \sigma_1^{5} \sigma_2^{-3} \sigma_3^{-3} + R_6\, \sigma_1^{-1} \sigma_2^{3} \sigma_3^{-3} \ ,\nn
\eea
with
\bea
-2 R_1 & = {f^{2}{}_{45}}^2 + {f^{2}{}_{46}}^2 + {f^{3}{}_{45}}^2 + {f^{3}{}_{46}}^2 \ ,\nn\\
-2 R_2 &= {f^{4}{}_{25}}^2 + {f^{4}{}_{26}}^2 + {f^{4}{}_{35}}^2 + {f^{4}{}_{36}}^2 \ ,\nn\\
-2 R_3 &= {f^{5}{}_{24}}^2 + {f^{5}{}_{34}}^2 + {f^{6}{}_{24}}^2 + {f^{6}{}_{34}}^2 \ ,\\
-R_4 &= f^{2}{}_{45} f^{4}{}_{25} + f^{2}{}_{46} f^{4}{}_{26} + f^{3}{}_{45} f^{4}{}_{35} + f^{3}{}_{46} f^{4}{}_{36} \ ,\nn\\
R_5 &= f^{2}{}_{45} f^{5}{}_{24} + f^{3}{}_{45} f^{5}{}_{34} + f^{2}{}_{46} f^{6}{}_{24} + f^{3}{}_{46} f^{6}{}_{34} \ , \nn\\
-R_6 &= f^{4}{}_{25} f^{5}{}_{24} + f^{4}{}_{35} f^{5}{}_{34} + f^{4}{}_{26} f^{6}{}_{24} + f^{4}{}_{36} f^{6}{}_{34} \ . \nn
\eea
We then reproduce the 10d no-go theorem by finding the following combination
\beq
V(\rho, \tau, \sigma_1, \sigma_2, \sigma_3) + \frac{1}{3} \rho \del_{\rho} V + \frac{1}{4} \tau \del_{\tau} V + \frac{1}{6} \sum_{I=1}^3 \sigma_I \del_{\sigma_I} V = 0 \ , \label{nogom4660}
\eeq
which is vanishing for $m_{466}$. This no-go theorem, as the previous one, is excluding both de Sitter and anti-de Sitter solutions.

To extract the corresponding value for $c$, we need to express \eqref{nogom4660} in terms of canonical fields. While we have them for $\rho,\tau$ in \eqref{canfields}, we need to determine them for $\sigma_{I=1,2,3}$. To that end, we recall the field space metric \cite[(3.15)]{Andriot:2022yyj}
\beq
g_{ij} =
\begin{pmatrix}
\mathlarger{\frac{3}{2 \rho^2}} & 0 & 0 &0 & 0  \\[10pt]
0 & \mathlarger{\frac{2}{\tau^2}} & 0 &0 &0 \\[10pt]
0 & 0 & \mathlarger{\frac{15}{2 \sigma_1^2}} & \mathlarger{-\frac{9}{2\sigma_1 \sigma_2}} & \mathlarger{-\frac{9}{2\sigma_1 \sigma_3}} \\[10pt]
0 & 0 &\mathlarger{-\frac{9}{2\sigma_1 \sigma_2}} & \mathlarger{\frac{27}{2\sigma_2^2}} & \mathlarger{-\frac{9}{2\sigma_2 \sigma_3}}  \\[10pt]
0 & 0 &\mathlarger{-\frac{9}{2\sigma_1 \sigma_3}} & \mathlarger{-\frac{9}{2\sigma_2 \sigma_3}} & \mathlarger{\frac{27}{2\sigma_3^2}}
\end{pmatrix} \ . \label{gijm461}
\eeq
Focusing on the metric block of sigma fields, one determines a basis of 3 orthonormal eigenvectors. Scaling them with the square root of the eigenvalues, it is straightforward to find the canonical field directions $\hat{\sigma}_I$ such that $g_{IJ}\del \sigma_I \del \sigma_J=\sum_I (\del \hat{\sigma}_I)^2$. We obtain
\bea
& \hat{\sigma}_1 = 3\, \ln \left(\frac{\sigma_2}{\sigma_3}\right)\, , \quad \hat{\sigma}_2 = \sqrt{\beta_2}\, \ln \left(\sigma_1^{\alpha_2} \sigma_2 \sigma_3 \right) \, , \quad \hat{\sigma}_3 = \sqrt{\beta_3}\, \ln \left(\sigma_1^{\alpha_3} \sigma_2 \sigma_3 \right) \ , \\
& \alpha_2= \frac{1}{6}(1-\sqrt{73})\, ,\, \alpha_3=\frac{1}{6}(1+\sqrt{73})\, ,\, \beta_2= \frac{9}{4}\left(1 + \frac{7}{\sqrt{73}}\right) \, ,\, \beta_3= \frac{9}{4}\left(1 - \frac{7}{\sqrt{73}}\right) \ .\nn
\eea
We then determine the field space diffeomorphism, as a matrix $\left(\frac{\del \hat{\sigma}}{\del \sigma} \right)$ of elements $(i,j)$ given by $\frac{\del \hat{\sigma}_i}{\del \sigma_j}$
\beq
\left(\frac{\del \hat{\sigma}}{\del \sigma} \right) =
\begin{pmatrix}
  0 & \frac{3}{\sigma_2} & -\frac{3}{\sigma_3}\\
  \frac{\alpha_2 \sqrt{\beta_2}}{\sigma_1} & \frac{\sqrt{\beta_2}}{\sigma_2} & \frac{\sqrt{\beta_2}}{\sigma_3} \\
  \frac{\alpha_3 \sqrt{\beta_3}}{\sigma_1} & \frac{\sqrt{\beta_3}}{\sigma_2} & \frac{\sqrt{\beta_3}}{\sigma_3} \\
\end{pmatrix} \ .
\eeq
As explained in \eqref{Pdel} and in \cite{Andriot:2021rdy}, this allows us to compute the relation
\beq
\begin{pmatrix}
\del_{\sigma_1} \\
\del_{\sigma_2} \\
\del_{\sigma_3} \\
\end{pmatrix}= \left(\frac{\del \hat{\sigma}}{\del \sigma} \right)^T
\begin{pmatrix}
\del_{\hat{\sigma}_1} \\
\del_{\hat{\sigma}_2} \\
\del_{\hat{\sigma}_3} \\
\end{pmatrix} = \begin{pmatrix}
\frac{1}{\sigma_1} ( \alpha_2 \sqrt{\beta_2} \, \del_{\hat{\sigma}_2} + \alpha_3 \sqrt{\beta_3} \, \del_{\hat{\sigma}_3} ) \\
\frac{1}{\sigma_2} (3\, \del_{\hat{\sigma}_1}  + \sqrt{\beta_2}\, \del_{\hat{\sigma}_2} + \sqrt{\beta_3}\,  \del_{\hat{\sigma}_3}) \\
\frac{1}{\sigma_3} (-3\, \del_{\hat{\sigma}_1}  + \sqrt{\beta_2}\, \del_{\hat{\sigma}_2} + \sqrt{\beta_3}\, \del_{\hat{\sigma}_3})
\end{pmatrix} \ ,
\eeq
from which we finally rewrite the no-go theorem \eqref{nogom4660} with canonical fields
\bea
&  V + \frac{1}{4} \tau \del_\tau V + \frac{1}{3} \rho \del_\rho V + \frac{1}{6} \sum_{I=1}^3 \sigma_I \del_{\sigma_I} V  = 0 \\
\Rightarrow\quad &  V + \frac{1}{2\sqrt{2}}\, \del_{\hat{\tau}} V + \frac{1}{\sqrt{6}}\, \del_{\hat{\rho}} V +  \frac{2 + \alpha_2}{6} \sqrt{\beta_2}\, \del_{\hat{\sigma}_2} V + \frac{2 + \alpha_3}{6} \sqrt{\beta_3}\, \del_{\hat{\sigma}_3} V  = 0 \nn \ .
\eea
It is then straightforward with \eqref{nogogen} to compute for this no-go theorem the value
\beq
c=\sqrt{2} \ ,
\eeq
which as expected, though non-trivially, matches the one obtained for the previous no-go theorem on the class $s_{555}$. This value for $c$ once again satisfies the ATCC bound \eqref{cbound}.

\subsection{DGKT-inspired scalar potentials}\label{sec:DGKT}

In this section, we study string effective theories \eqref{actionintro} and their scalar potentials, obtained by a compactification towards a $d=4$ anti-de Sitter solution \cite{DeWolfe:2005uu, Camara:2005dc}, the so-called DGKT solution. Such a solution is part of a larger $d=4$ family \cite{Marchesano:2019hfb} leading to various 4d theories, and similar solutions have been obtained in $d=3$ \cite{Farakos:2020phe, VanHemelryck:2022ynr}. We first consider in Section \ref{sec:DGKT4d} the simplest $d=4$ solution and theory, and turn to $d=3$ in Section \ref{sec:DGKT3d}.

\subsubsection{4d isotropic solution, potential and field directions}\label{sec:DGKT4d}

We study here a theory obtained by a compactification of 10d type IIA supergravity with orientifold $O_6$-planes on a torus orbifold $T^6/\mathbb{Z}^2_3$. It leads to a $d=4$ classical anti-de Sitter solution with parametric control and scale separation \cite{DeWolfe:2005uu}. The 4d effective theory is an ${\cal N}=1$ supergravity. The discrete symmetries of the internal space reduce supersymmetry, but also truncate degrees of freedom, eventually leading to only few moduli that get all stabilized thanks to fluxes.

A first task is to obtain the kinetic terms of the scalar fields, in order to reach a canonical basis. To that end, we use the ${\cal N}=1$ supergravity formalism, deduced from a Calabi-Yau compactification in type IIA \cite{Grimm:2004ua}. We also follow \cite{Conlon:2021cjk,Apers:2022tfm}. The moduli content of the effective theory is expressed by the following superfields
\begin{align}
    t_i=b_i+\i\, \upsilon_i\,,
    \quad S=e^{-D}+\i\, \frac{\xi}{\sqrt{2}}\,,\quad i=1,2,3 \ ,
\end{align}
where $b_i$ stand for the moduli of the $B$-field for each $T^2$, $\upsilon_i$ are the K\"ahler moduli related to the volumes of the $T^2$, $D$ is the 4d dilaton and $\xi$ is the axion arising from $C_3$. These scalar fields are also related to the internal volume ${\rm vol}$ and the 10d dilaton $\phi$ as follows
\begin{align}
    {\rm vol}=\kappa\, \upsilon_1\upsilon_2\upsilon_3\,,\quad e^D=\frac{e^{\phi}}{\sqrt{{\rm vol}}}\,,
\end{align}
with the constant $\kappa=\kappa_{123}$ corresponding to an intersection number. Complex structure moduli are projected out by the internal space discrete symmetries, and $\xi, b_i$ are stabilized \cite{DeWolfe:2005uu}. The kinetic terms for the remaining fields, $D$ and $\upsilon_i$, can be obtained from the K\"ahler potential
\begin{align}
K =-\ln\left(\frac{4}{3}\times 6\, {\rm vol} \right)-4\,\ln\left(S+\overline{S}\right)\,.
\end{align}
They are given as follows for superfields $\Phi^I$, standing here for $S,t_i$
\begin{align}
    e^{-1} \mathcal{L}_{kin}&=K_{I\bar{J}}\,\partial\Phi^I\partial\Phi^{\bar{J}}
    =(\partial D)^2
    +\frac{1}{4}\sum^3_{i=1}\frac{1}{\upsilon_i^2}(\partial\upsilon_i)^2\,,
\end{align}
with $e$ the 4d volume and $K_{I\bar{J}}= \frac{\del^2 K}{\partial\Phi^I\partial\Phi^{\bar{J}} }$ the K\"ahler metric; we also recall that $\frac{\del t_i}{\del \upsilon_i} = 2 \i$.

The setting considered here is isotropic, meaning actually that the three geometric moduli are related $\upsilon_i=\upsilon/\vert e_i\vert$ up to a constant $e_i$, leaving us effectively with only one K\"ahler modulus $\upsilon$. For future convenience in the scalar potential, the two scalar fields are redefined in \cite{DeWolfe:2005uu} towards $g, r$ as follows
\begin{align}\label{DilatonUpsilon}
e^D=\vert p\vert\sqrt{\frac{\vert m_0\vert}{E}}\,g\,,\quad
\upsilon=\sqrt{\frac{E}{\vert m_0\vert}}\,r^2\,.
\end{align}
The constants $m_0$, $p$ and $e_i$ are related to flux quantas of $F_0$, $H_3$ and $F_4$ respectively, while $E=\upsilon^3/{\rm vol}= |e_1e_2e_3|/\kappa $. In terms of the new fields, the kinetic terms are given by
\begin{align}
   e^{-1} \mathcal{L}_{kin}&=\frac{1}{2} \left(\frac{2}{g^2}(\partial g)^2 +\frac{6}{r^2}(\partial r)^2 \right)\,. \label{kingr}
\end{align}
It is then straightforward to obtain the canonical fields $\hat{r}, \hat{g}$ as
\begin{align}
r=e^{\frac{1}{\sqrt{6}}\,\hat{r}}\,,\quad  g=e^{\frac{1}{\sqrt{2}}\,\hat{g}}\,.
\end{align}

Next, we turn to the scalar potential. It has been derived in \cite{DeWolfe:2005uu} starting from 10d supergravity,\footnote{This scalar potential has three terms due to fluxes, and one due to the $O_6$-planes. The latter tension $\mu_6$ has been traded for flux numbers after using the tadpole cancellation condition phrased as
\begin{align}
    \int\text{d}F_2=m_0p+2\sqrt{2}\kappa^2_{10}\mu_6=0\,,
\end{align}
where the integral is performed over the imaginary component of the holomorphic Calabi-Yau three-form. This way, coefficients in the potential are only flux dependent, and those got absorbed in the redefined fields, up to the overall factor $\lambda$. Let us also note a slight difference in conventions of 10d supergravity with ours, possibly leading small changes in the numerical coefficients in the potential. This does not affect our analysis since we only care about the field powers or the exponential rates.} and is given by
\begin{align}\label{Vgr}
\frac{1}{\lambda}V(g,r)&=\frac{1}{2}g^4r^6-\sqrt{2}\,g^3+\frac{1}{4}g^2r^{-6}+\frac{3}{2}g^4r^{-2}\,,
\end{align}
where $\lambda=p^4\vert m_0\vert^{5/2}E^{-3/2}$. In terms of the canonical fields, the potential gets rewritten as
\begin{align}
\frac{1}{\lambda}V(\hat{g},
\hat{r})
&=\frac{1}{2}e^{2\sqrt{2}\, \hat{g}}e^{\sqrt{6}\, \hat{r}} -\sqrt{2}\,e^{\frac{3}{\sqrt{2}}\, \hat{g}}+\frac{1}{4}e^{\sqrt{2}\, \hat{g}}e^{-\sqrt{6}\, \hat{r}} +\frac{3}{2}e^{2\sqrt{2}\, \hat{g}}e^{-\sqrt{\frac{2}{3}}\, \hat{r}}\,.\label{Vgrhat}
\end{align}
Note that a comparison of \eqref{Vgr} to the potential \eqref{Vrts}, or that of the kinetic terms \eqref{kingr} to the canonical fields \eqref{canfields}, allows us to identify the following relations between fields
\beq
\rho \propto r^2, \upsilon \ ,\quad \tau \propto g^{-1}, e^{-D}\ ,
\eeq
up to proportionality constants. Given our analysis of the exponential rates for $\hat{\rho},\hat{\tau}$ made in Section \ref{sec:rhorate}, it is without surprise that we see here in the DGKT potential \eqref{Vgrhat} no violation of the ATCC bound \eqref{cbound}.\\

\begin{figure}[H]
\begin{center}
\begin{subfigure}[H]{0.45\textwidth}
\includegraphics[width=\textwidth]{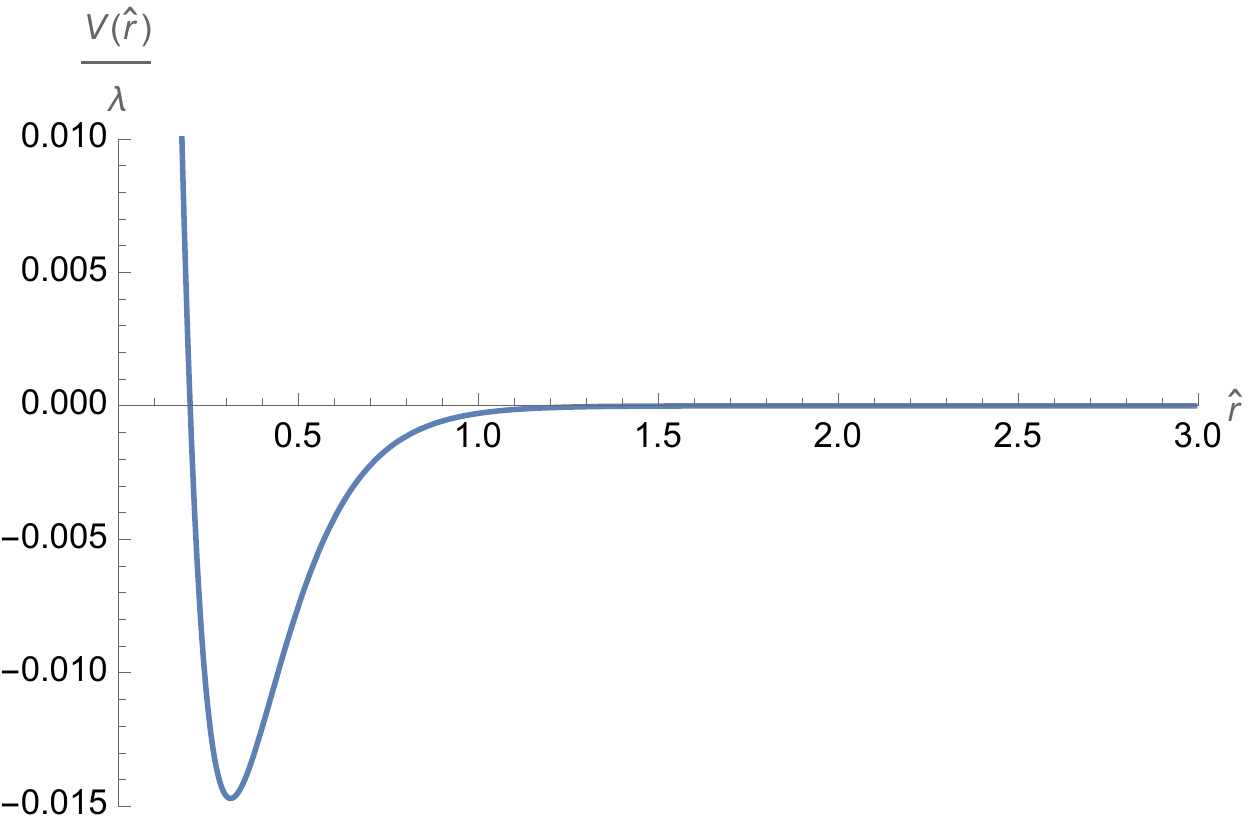}
\caption{$V(\hat{r})/\lambda$}
\label{fig:DGKTVr}
\end{subfigure}\quad  \quad
\begin{subfigure}[H]{0.45\textwidth}
\includegraphics[width=\textwidth]{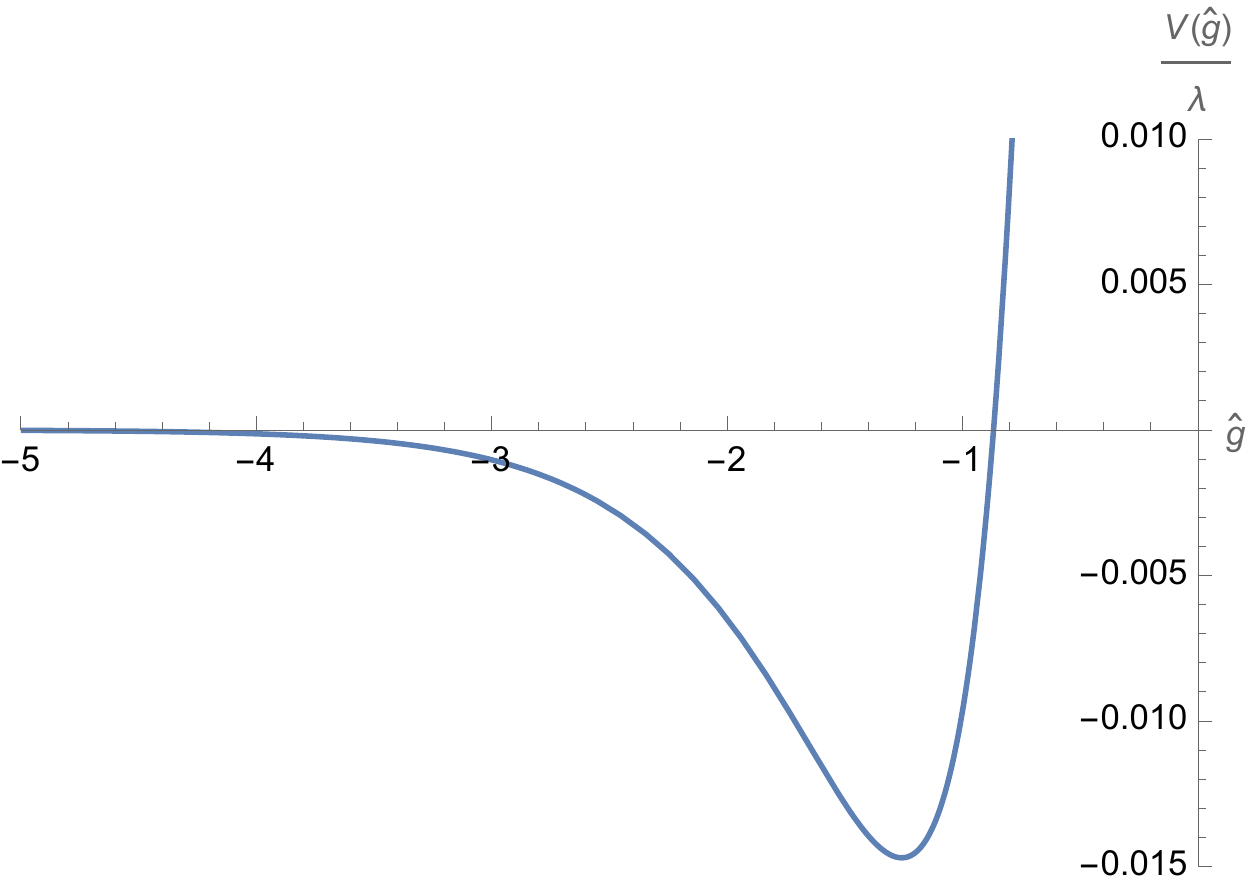}\caption{$V(\hat{g})/\lambda$}
\label{fig:DGKTVg}
\end{subfigure}
\caption{Potentials $V(\hat{r})$ of \eqref{potrhat} and $V(\hat{g})$ of \eqref{potghat}}\label{fig:Ludwig}
\end{center}
\end{figure}

In \cite{DeWolfe:2005uu}, however, a different field direction is exhibited which possesses the desired asymptotic behaviour to test the ATCC, as can be seen in Figure \ref{fig:Ludwig}.
\noindent This field direction is obtained by considering an extremization condition, $g\partial_gV+2r\partial_rV=0$, corresponding to the direction $g\,r^6=5/(4\sqrt{2})$. The potential along this direction is given by either of the following expressions, in terms of the field $\hat{r}$ or $\hat{g}$
\begin{align}
\frac{1}{\lambda}V(\hat{r})&= \frac{1875}{2048}e^{-13\sqrt{\frac{2}{3}}\, \hat{r}} -\frac{975}{2048}e^{-3\sqrt{6}\, \hat{r}}\,\label{potrhat}\,,\\
\frac{1}{\lambda}V(\hat{g})&= \frac{3}{2^{\frac{1}{6}}\times 5^{\frac{1}{3}}}e^{\frac{13}{3\sqrt{2}}\, \hat{g}}
-\frac{39}{40\sqrt{2}}e^{\frac{3}{\sqrt{2}}\, \hat{g}}\label{potghat}\,,
\end{align}
and those are displayed in Figure \ref{fig:Ludwig}. However, $\hat{r}$ and $\hat{g}$ are now related along this field direction, so we need to introduce a new canonically normalized field for this direction. The relation $g\,r^6=5/(4\sqrt{2})$ leads to $\del_{\mu} \hat{g} = - 2\sqrt{3}\, \del_{\mu} \hat{r}$. We then get for the kinetic terms $(\del \hat{g})^2 + (\del \hat{r})^2 = 13 (\del \hat{r})^2$. We then introduce a new canonically normalized field $\hat{t}$, such that
\beq
-\frac{ \sqrt{13}}{2\sqrt{3}}\, \del_{\mu} \hat{g} = \sqrt{13}\, \del_{\mu} \hat{r} =  \del_{\mu} \hat{t} \ . \label{relgrt}
\eeq
Note that proceeding this way, we simply ignore an orthonormal direction $\hat{t}_{\perp}$, meaning we consider $\del_{\mu} \hat{t}_{\perp}=0$. The relation \eqref{relgrt} does not fix the relative constant between the fields. Choosing for simplicity $\sqrt{13}\, \hat{r} =  \hat{t}$, we then get from $g\,r^6=5/(4\sqrt{2})$ the following relation
\beq
-\frac{ \sqrt{13}}{2\sqrt{3}}\, \hat{g} +  \sqrt{\frac{13}{6}}\, \ln \left( \frac{5}{4\sqrt{2}} \right) = \hat{t} = \sqrt{13}\, \hat{r} \ .
\eeq
Both expressions \eqref{potrhat} and \eqref{potghat} of the potential finally get rewritten as a single one, given by
\beq
\frac{1}{\lambda}V(\hat{t}) = \frac{1875}{2048}e^{-\sqrt{\frac{26}{3}}\, \hat{t}} -\frac{975}{2048}e^{-3\sqrt{\frac{6}{13}}\, \hat{t}}\,\label{potthat} \ .
\eeq
Its graph is analogous to that of $V(\hat{r})$. To test the ATCC, the relevant asymptotics is $\hat{t} \to \infty$, where the potential is negative. The exponential rate for this specific field direction is then higher than the ATCC bound \eqref{cbound} in $d=4$
\beq
3\sqrt{\frac{6}{13}} > \sqrt{\frac{2}{3}} \ .
\eeq
The ATCC is then again satisfied.

\subsubsection{3d solution, potential and field directions}\label{sec:DGKT3d}

We study here an analogous solution and theory to the one of $d=4$ DGKT, obtained in $d=3$ in \cite{Farakos:2020phe}. The setup is a compactification of 10d type IIA supergravity on a $G_2$-holonomy internal space, leading to a $d=3$ effective theory with (non)-supersymmetric anti-de Sitter vacua. The presence of $O_2$ and $O_6$-planes, together with $\mathbb{Z}_2$ orbifold involutions on a torus, $T^7/\mathbb{Z}_2^3$, leads to supergravity with minimal supersymmetry in $d=3$. As for DGKT in $d=4$, the anti-de Sitter solution has here interesting features such as parametric scale separation, full moduli stabilization, and a parametric control on the classical regime.

The effective theory contains 8 scalar fields. The first 2 are the universal ones, $x$ and $y$, related to the 10d dilaton $\phi$ and the internal volume $\upsilon$. The internal part of the 10d metric in Einstein frame \cite[(3.22)]{Farakos:2020phe} scales as $e^{2 \beta \upsilon}$ with $\beta=-1/(4\sqrt{7})$. From there one defines
\beq
\frac{x}{\sqrt{7}}= -\frac{3}{8}\phi+\frac{\beta}{2}\upsilon\,, \quad
y= -\frac{1}{4}\phi -\frac{21\beta}{2}\upsilon \ .
\eeq
In addition, one introduces 7 volumes of 3-cycles, $s^{i=1,...,7}$. The 7d internal volume then scales as
\beq
{\rm vol}_7 \sim \left( \prod_{i=1}^7 s^i \right)^{\frac{1}{3}} \sim \left(e^{\beta \upsilon} \right)^7 \ .
\eeq
One defines the corresponding unit volume fluctuations $\tilde{s}^i$, such that
\beq
s^i = \tilde{s}^i\, e^{3\beta \upsilon} \, , \quad \prod_{i=1}^7 \tilde{s}^i = 1 \ . \label{tildesdef}
\eeq
In the following, we will only consider the 6 independent fields $\tilde{s}^{a=1,...,6}$.

Let us discuss the kinetic terms and the canonically normalized fields. The kinetic terms are given in \cite{Farakos:2020phe} as follows
\begin{align}
    e^{-1}\mathcal{L}_{\text{kin}}=
    -\frac{1}{4} (\del x)^2
    -\frac{1}{4} (\del y)^2
    - \frac{\delta_{ij}}{4 \tilde{s}^i \tilde{s}^j} \del_{\mu} \tilde{s}^i \del^{\mu} \tilde{s}^j  \,,
\end{align}
with $e$ the 3d volume, together with a $\tfrac{1}{2} {\cal R}_3$ term in the effective action. Expressing $\tilde{s}^7$ in terms of the 6 independent fields $\tilde{s}^{a}$ thanks to \eqref{tildesdef}, one rewrites the kinetic terms as
\beq
    e^{-1}\mathcal{L}_{\text{kin}}=
    -\frac{1}{4} (\del x)^2
    -\frac{1}{4} (\del y)^2
    - \tilde{G}_{ab}\, \partial_{\mu}\tilde{s}^a\partial^{\mu}\tilde{s}^b \,,\qquad \tilde{G}_{ab} = \frac{\delta_{ab} + 1}{4 \tilde{s}^a \tilde{s}^b } \ ,
\eeq
where this field space metric also appeared in \cite{Emelin:2021gzx}. Let us now work-out the canonically normalized fields. For the first 2 fields, we consider
\beq
\hat{x} =\frac{1}{\sqrt{2}}\, x \ ,\ \hat{y} =\frac{1}{\sqrt{2}}\, y \ .
\eeq
The 6 other fields require more work. For them, one should first find an orthonormal basis of 6 eigenvectors of the field space metric $2\, \tilde{G}_{ab}$. Rescaling those with the square root of the corresponding eigenvalues, one then obtains the canonical fields $\hat{s}^a$. One verifies that $2\,  \tilde{G}_{ab}\, \partial_{\mu}\tilde{s}^a\partial^{\mu}\tilde{s}^b = \delta_{ab}\, \partial_{\mu}\hat{s}^a\partial^{\mu}\hat{s}^b $. The fields are related as follows
\begin{align}
    \ln\, \tilde{s}^1&=\frac{1}{\sqrt{21}}\hat{s}^1
    -\hat{s}^2
    -\frac{1}{\sqrt{3}}\hat{s}^3
    -\frac{1}{\sqrt{6}}\hat{s}^4
    -\frac{1}{\sqrt{10}}\hat{s}^5
    -\frac{1}{\sqrt{15}}\hat{s}^6\,,\label{Lns1}\\
    \ln\, \tilde{s}^2&=\frac{1}{\sqrt{21}}\hat{s}^1
    +\sqrt{\frac{5}{3}}\hat{s}^6\,,\\
    \ln\, \tilde{s}^3&=\frac{1}{\sqrt{21}}\hat{s}^1
    +2\sqrt{\frac{2}{5}}\hat{s}^5
    -\frac{1}{\sqrt{15}}\hat{s}^6\,,\\
    \ln\, \tilde{s}^4&=\frac{1}{\sqrt{21}}\hat{s}^1
    +\sqrt{\frac{3}{2}}\hat{s}^4
    -\frac{1}{\sqrt{10}}\hat{s}^5
    -\frac{1}{\sqrt{15}}\hat{s}^6\,,\\
    \ln\, \tilde{s}^5&=\frac{1}{\sqrt{21}}\hat{s}^1
    +\frac{2}{\sqrt{3}}\hat{s}^3
    -\frac{1}{\sqrt{6}}\hat{s}^4
    -\frac{1}{\sqrt{10}}\hat{s}^5
    -\frac{1}{\sqrt{15}}\hat{s}^6\,,\\
    \ln\, \tilde{s}^6&=\frac{1}{\sqrt{21}}\hat{s}^1
    +\hat{s}^2
    -\frac{1}{\sqrt{3}}\hat{s}^3
    -\frac{1}{\sqrt{6}}\hat{s}^4
    -\frac{1}{\sqrt{10}}\hat{s}^5
    -\frac{1}{\sqrt{15}}\hat{s}^6\,.\label{Lns6}
\end{align}

We now turn to the scalar potential. It is given as follows \cite{Farakos:2020phe}, in terms of $\hat{x},\hat{y}$ and $\tilde{s}^a$
\begin{align}\label{xypotential}
    V(\hat{x},\hat{y},\tilde{s}^a)= &\, F(\tilde{s}^a)\, e^{2\sqrt{2}\, \hat{y}-2\sqrt{\frac{2}{7}}\, \hat{x}}
    +H(\tilde{s}^a)\, e^{2\sqrt{2}\, \hat{y}+2\sqrt{\frac{2}{7}}\, \hat{x}}\\
    + &\,C\,e^{\sqrt{2}\, \hat{y}-\sqrt{14}\, \hat{x}}
    -T(\tilde{s}^a)\, e^{\frac{3}{2}\sqrt{2}\, \hat{y}-\frac{5}{2}\sqrt{\frac{2}{7}}\, \hat{x}}\,,\nn
\end{align}
with
\begin{align}
    F(\tilde{s}^a)&=\frac{f^2}{16}\left(\sum_a (\tilde{s}^a)^2+36\prod_a (\tilde{s}^a)^{-2}\right)\,,\ H(\tilde{s}^a) =\frac{h^2}{16}\left(\sum_a (\tilde{s}^a)^{-2} +\prod_a (\tilde{s}^a)^{2}\right)\,,\\
    C&=\frac{m^2}{16}\,,\ T(\hat{s}^a)=\frac{hm}{8}\left(\sum_a (\tilde{s}^a)^{-1}+\prod_a \tilde{s}^a\right)\,.\nn
\end{align}
The first term in the potential is the contribution from $F_4$, the second one from $H_3$, while the Romans mass is $C$. The last term is the $O_6$-plane contribution, where the tension has been expressed in terms of fluxes using the tadpole cancellation. The $O_2$ contribution is cancelled by an appropriate number of $D_2$, and the corresponding magnetic $F_6$ flux vanishes in the smeared limit. A comparison of this potential to $V(\hat{\rho},\hat{\tau})$ in \eqref{Vrts} for $d=3$ gives the relations
\beq
\hat{\rho}= -\frac{1}{4\sqrt{2}} \left( \sqrt{7} \hat{y} + 5 \hat{x} \right) \ ,\quad \hat{\tau} = -\frac{1}{4\sqrt{2}} \left( 5 \hat{y} - \sqrt{7} \hat{x} \right) \ ,\label{rtxy}
\eeq
up to possible constants. The canonical normalization is also cross-checked from \eqref{rtxy}.

\begin{figure}[t]
\begin{center}
\begin{subfigure}[H]{0.45\textwidth}
\includegraphics[width=\textwidth]{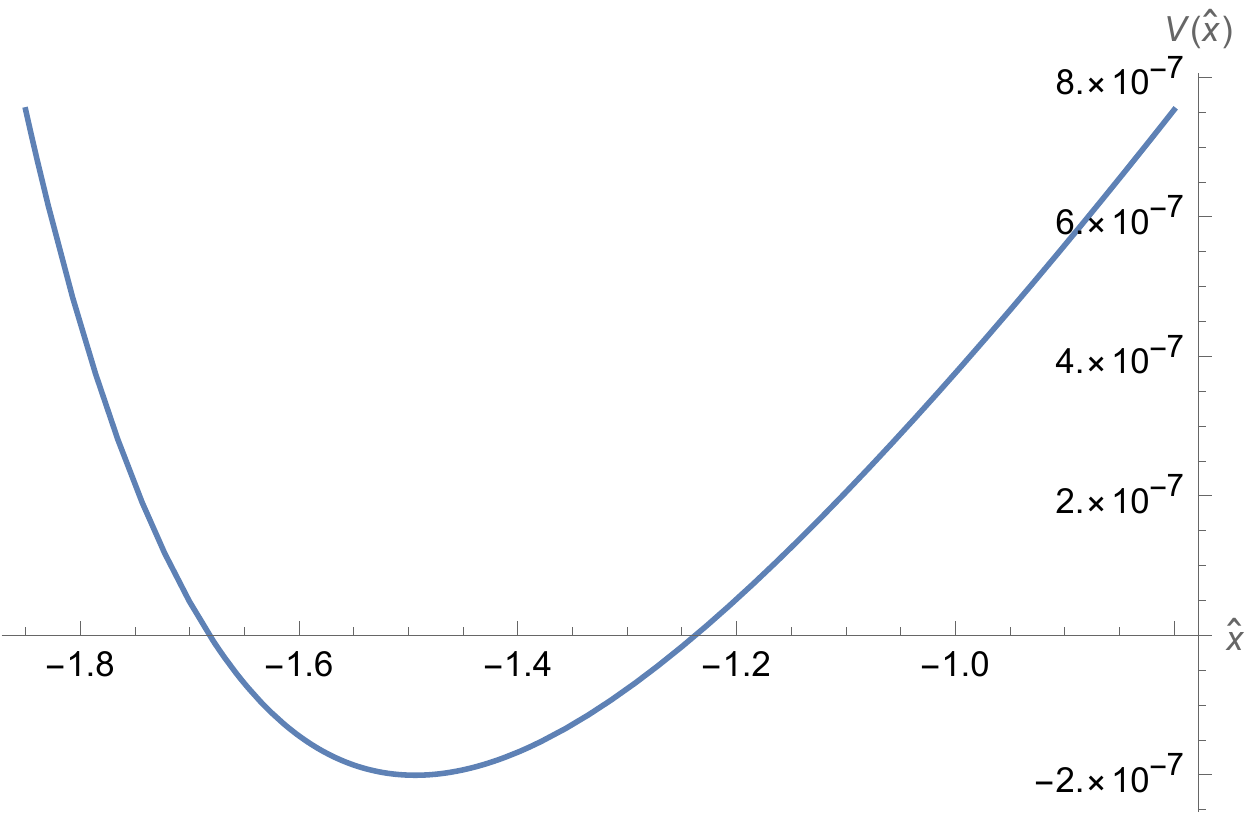}
\caption{$V(\hat{x},\hat{y}_0,\tilde{s}^a_0)$}
\label{fig:V(tildeX)}
\end{subfigure}\quad  \quad
\begin{subfigure}[H]{0.45\textwidth}
\includegraphics[width=\textwidth]{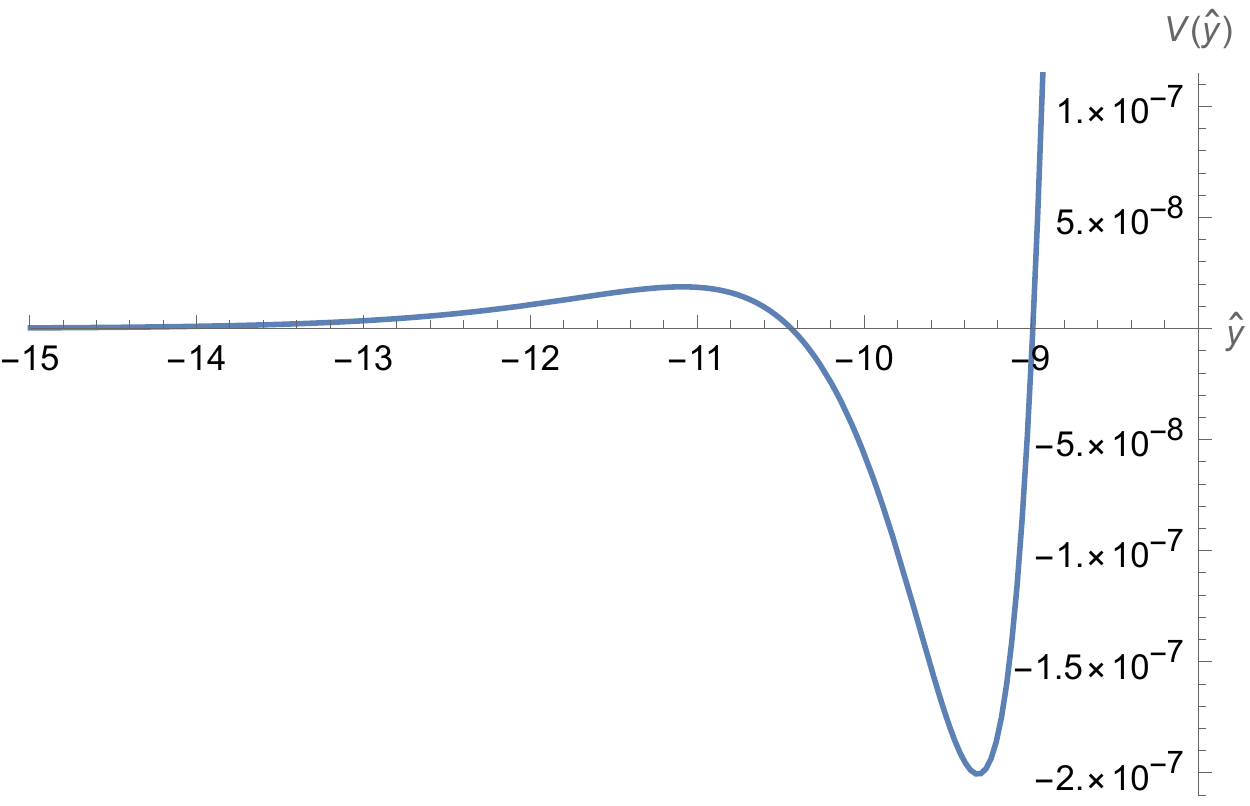}\caption{$V(\hat{x}_0,\hat{y},\tilde{s}^a_0)$}
\label{fig:V(tildeY)}
\end{subfigure}
\begin{subfigure}[H]{0.6\textwidth}
\includegraphics[width=\textwidth]{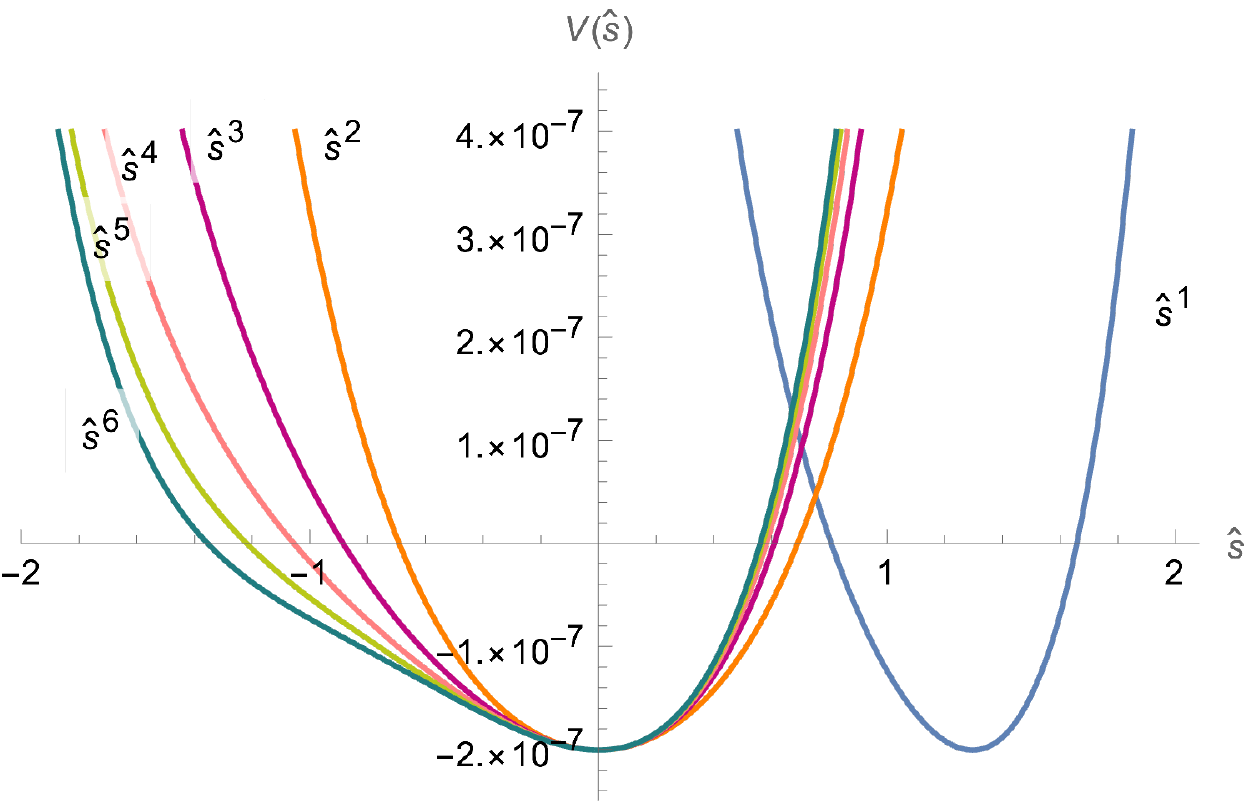}\caption{$V(\hat{x}_0,\hat{y}_0,\hat{s}^a)$}
\label{fig:V(tildeS)}
\end{subfigure}
\caption{Potential \eqref{xypotential} depicted along one canonical field direction, $\hat{x}$, $\hat{y}$ or $\hat{s}^a$ respectively, while fixing the other fields to their vacuum value. To get the dependence on $\hat{s}^a$ in Figure \ref{fig:V(tildeS)}, we use the relations \eqref{Lns1}-\eqref{Lns6}, from which we also deduce the vacuum values of the canonical fields: $\hat{s}_0^1=1.29619$, $\hat{s}^{a=2,...,6}_0=0$.}\label{fig:Vxys}
\end{center}
\end{figure}

In order to study the potential, and depict it in Figure \ref{fig:Vxys}, let us fix the remaining constants, namely the flux numbers, following \cite{Farakos:2020phe}. Flux quantization gives
\begin{align}
    h=(2\pi)^2K\,,~~~~ m=(2\pi)^{-1}M\,,~~~~
    f=(2\pi)^3N\,,~~~~
    KM=16\,,
\end{align}
where $N,K,M$ are integers, $N$ is unbounded, and the last condition comes from the tadpole cancellation. In the following, we choose $K=16,M=1, N=1$. The potential then admits an anti-de Sitter vacuum. It can be determined by extremizing the superpotential. The fields are then stabilized at values that can be read from the corresponding conditions
\begin{align}\label{SUSYpoints}
    0.515696=\frac{h}{f}e^{2\sqrt{\frac{2}{7}}\hat{x}_0}\,,~~~~~~
    3.43111=\frac{m}{f}e^{-\frac{\sqrt{2}}{2}\hat{y}_0-\frac{5}{2}\sqrt{\frac{2}{7}}\hat{x}_0}\,,
    ~~~~~~
    \tilde{s}^a_0=1.32691\,.
\end{align}
We obtain the values $\hat{x}_0=-1.49381$ and $\hat{y}_0=-9.31713$.\\

We now have all needed elements to test the ATCC. As can be seen in Figure \ref{fig:Vxys}, the potential does not have the appropriate behaviour among these field directions to challenge the ATCC. This is confirmed by a close look at the rates in \eqref{xypotential}. The only term that has a chance to be negative is the $O_6$ term. The exponential rates there are however higher in the asymptotics than that of the $F_4$ term for $\hat{x}$, and that of $C$ for $\hat{y}$. The negative term is then dominated by positive terms in what would have been the appropriate asymptotics to challenge the ATCC, so the potential turns positive. The analysis for $\hat{s}^a$ is more involved but we verify graphically that the potential is asymptotically positive along those directions. We conclude that we do not observe here any violation of the ATCC, despite being in $d=3$.\\

In Section \ref{sec:DGKT4d}, we did not observe either a violation of the ATCC bound along the 2 universal moduli, but a specific direction among them eventually offered a chance to challenge it, by providing the appropriate asymptotic behaviour. A property of this specific direction is that it led to the same field dependence (in these 2 moduli) in the $H_3,F_0$ and $O_6$ term, while the $F_4$ term behaved differently. With the same requirement here, we identify a direction obeying this property, up to a constant $A>0$: it is given by the relation
\beq
\hat{x}= - \frac{\sqrt{7}}{9}\, \hat{y} + \frac{\sqrt{14}}{9}\, \ln\, A \ .\label{specdir}
\eeq
The potential \eqref{Vrts} then gets rewritten along this direction as
\beq
V(\hat{y})= F(\tilde{s}^a_0)\, A^{-\frac{4}{9}}\, e^{\frac{20\sqrt{2}}{9}\, \hat{y}} +  A^{-\frac{14}{9}} \Big( H(\tilde{s}^a_0)\, A^{2}  -T(\tilde{s}^a_0)\, A + C \Big) e^{\frac{16\sqrt{2}}{9}\, \hat{y}}\,, \label{V(y)}
\eeq
where we also fix $\tilde{s}^a=\tilde{s}^a_0$ to focus on this single direction. As in $d=4$ with \eqref{potrhat} and \eqref{potghat}, the term with the lowest rate is the one without $F_4$. This term would dominate in the asymptotics $\hat{y}\to -\infty$. This term is found negative, as required to test the ATCC, in the following tight range
\beq
H(\tilde{s}^a_0)\, A^{2}  -T(\tilde{s}^a_0)\, A + C \leq 0 \ \Leftrightarrow 1.38988 \leq 10^5 \times A \leq 13.7585 \ .\label{coefneg}
\eeq
Mimicking the $d=4$ situation, we can obtain the field direction \eqref{specdir} by considering the following extremization condition on the potential $V(\hat{x},\hat{y},\tilde{s}^a_0)$ given in \eqref{Vrts}
\beq
\del_{\hat{x}} V + \frac{1}{\sqrt{7}} \del_{\hat{y}} V = 0 \ \Leftrightarrow \ \eqref{specdir} \ \ {\rm and} \ \ A= \frac{-T(\tilde{s}^a_0) + \sqrt{96\, C\, H(\tilde{s}^a_0) + T(\tilde{s}^a_0)^2}}{8\, H(\tilde{s}^a_0)} \ ,
\eeq
where the latter solves $4 A^2 H(\tilde{s}^a_0) + A T(\tilde{s}^a_0) -6 C = 0$. Requiring an extremization condition then fixes the constant $A$. Remarkably, the corresponding value is $A = 3.78707\, .\, 10^{-5}$, which allows to get the second term in the potential negative \eqref{coefneg} .

\begin{figure}[H]
\begin{center}
\includegraphics[width=0.6\textwidth]{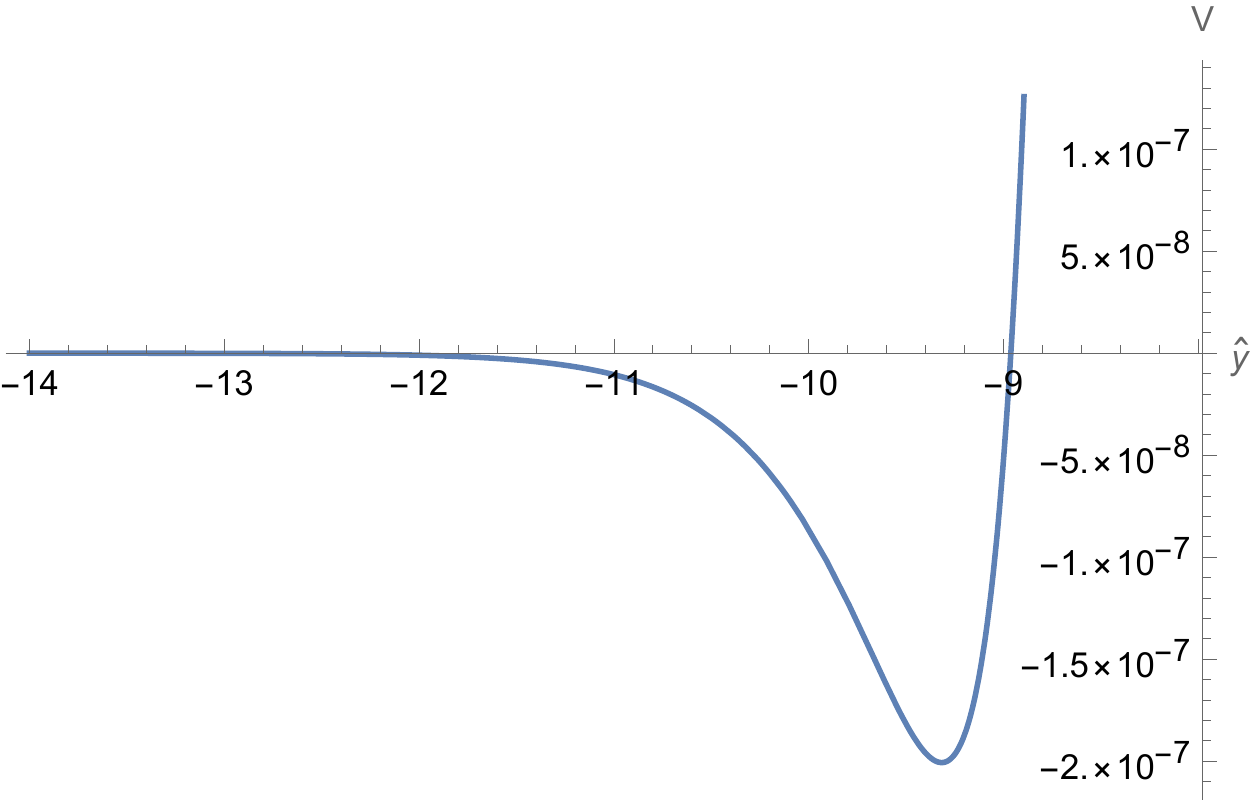}
\caption{Potential $V(\hat{y})$ given in \eqref{V(y)} with $A = 3.78707\, .\, 10^{-5}$, along the specific direction \eqref{specdir}.}\label{fig:V(y)}
\end{center}
\end{figure}

As long as \eqref{coefneg} is obeyed, the potential has the desired asymptotic behaviour, as can be seen in Figure \ref{fig:V(y)}.
\noindent As in $d=4$, we note however that this specific field direction is not canonically normalized anymore. Proceeding as before, we find the normalized direction to be along $\hat{t}$ with
\beq
\del_{\mu} \hat{t} = \frac{2 \sqrt{22}}{9} \del_{\mu} \hat{y} \ .
\eeq
This eventually gives an exponential rate along this specific, normalized direction to be
\beq
\frac{16\sqrt{2}}{9} \times \frac{9}{2 \sqrt{22}} = \frac{8}{\sqrt{11}} > \sqrt{2} \ ,
\eeq
where $\sqrt{2}$ corresponds to the ATCC bound \eqref{cbound} in $d=3$. Therefore, we see once again no violation of the ATCC.

\vfill

\subsection*{Acknowledgements}

We thank F.~Apruzzi, N.~Bobev, F.~Farakos, F.~Gautason, S.~L\"ust, H.~Samtleben, T.~Van Riet and M.~Walters for helpful exchanges during the completion of this work. L.~H.~acknowledges support from the Austrian Science Fund (FWF): project number P34562-N, doctoral program W1252-N27.

\newpage

\begin{appendix}

\section{Examples: dynamical solutions}\label{ap:dynsol}

We provide in this appendix two solutions to equations \eqref{F1eq}-\eqref{feom} for negative potentials, $V<0$, thus decelerating, and with $|V|<1$ in Planckian units. Those solutions are ``more'' dynamical than the anti-de Sitter solution discussed in Section \ref{sec:AdS}, in the sense that both $a(t)$ and $\varphi(t)$ now vary. These solutions are contracting, $\dot{a}<0$, and we find it easy to get such solutions with $V<0$, justifying the relevance of the ATCC. We then test (successfully) these solutions upon the various conditions and quantities introduced in Section \ref{sec:ATCCgen}.\\

Let us first indicate a point related to the resolution of equations \eqref{F1eq}-\eqref{feom}. The latter can be rewritten as $F_1=F_2=E=0$, with the quantities
\bea
& F_1= \frac{(d-1)(d-2)}{2} \left( H^2 + \frac{k}{a^2} \right) - \frac{1}{2} \dot{\varphi}^2 - V  \ ,\ F_2 = \frac{\ddot{a}}{a} - \frac{2}{(d-1)(d-2)} V + \frac{1}{d-1} \dot{\varphi}^2 \ ,\nn\\
& E= \ddot{\varphi} + (d-1) H \dot{\varphi} + V' \ ,
\eea
where we set $M_p=1$. One can show in full generality that the following relation holds
\beq
\dot{F_1} + 2 H \, F_1 = - \dot{\varphi}\, E +(d-1)(d-2)\, H\, F_2 \ .
\eeq
We deduce in particular
\beq
F_1 = 0 \ , \ E =0 \ \Rightarrow \ F_2=0 \ \text{for}\ H \neq 0 \ .\label{F1E}
\eeq
In other words, we do not have to solve the second Friedmann equation as long as we have $H \neq 0$, i.e.~sticking to an expansion or contracting phase. We will stick to this resolution scheme. When solving the first Friedmann equation, one is typically led to choose between an expanding or a contracting solution, and we pick the latter.

Let us now say a word on initial conditions. The potentials considered often have a positive part. We however only want to probe the physics of $V<0$, so we choose initial conditions accordingly: we take in particular an initial $\rho_i <0$. The field $\varphi$ then first rolls-down or climbs-up the potential, depending on the sign of the initial $\dot{\varphi}_i$. The two solutions below only display the first option, but as far as we could test, no result changes with the second option.

Finally, we set $d=4$, $k=-1$, and obtain the solutions numerically. As argued in Section \ref{sec:lifetime}, a contracting and decelerating solution can only exist for a finite time: it starts with $a_i=a(t_i)$ until a final crunch where $a(t_c)=0$. In practice, the numerics stop slightly before due to related divergences, at a final time $t_f$ where $a(t_f)$ is very close to zero as we will indicate. We now present and analyse each solution in Appendix \ref{ap:exppot} and \ref{ap:kkltpot}.

\subsection{Exponential potential}\label{ap:exppot}

\begin{figure}[H]
\begin{center}
\begin{subfigure}[H]{0.45\textwidth}
\includegraphics[width=\textwidth]{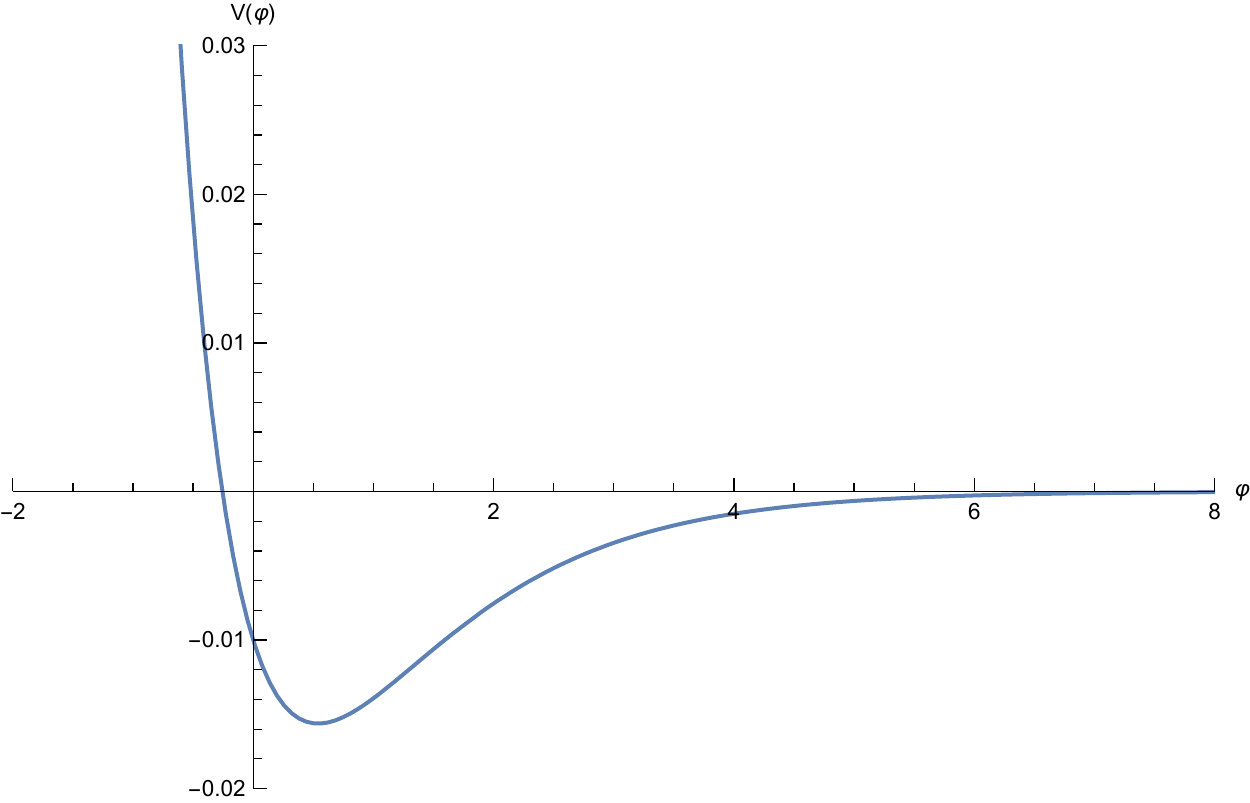}\caption{Potential $V(\varphi)$}\label{fig:dynsol_exp_pot}
\end{subfigure}\qquad \quad
\begin{subfigure}[H]{0.45\textwidth}
\includegraphics[width=\textwidth]{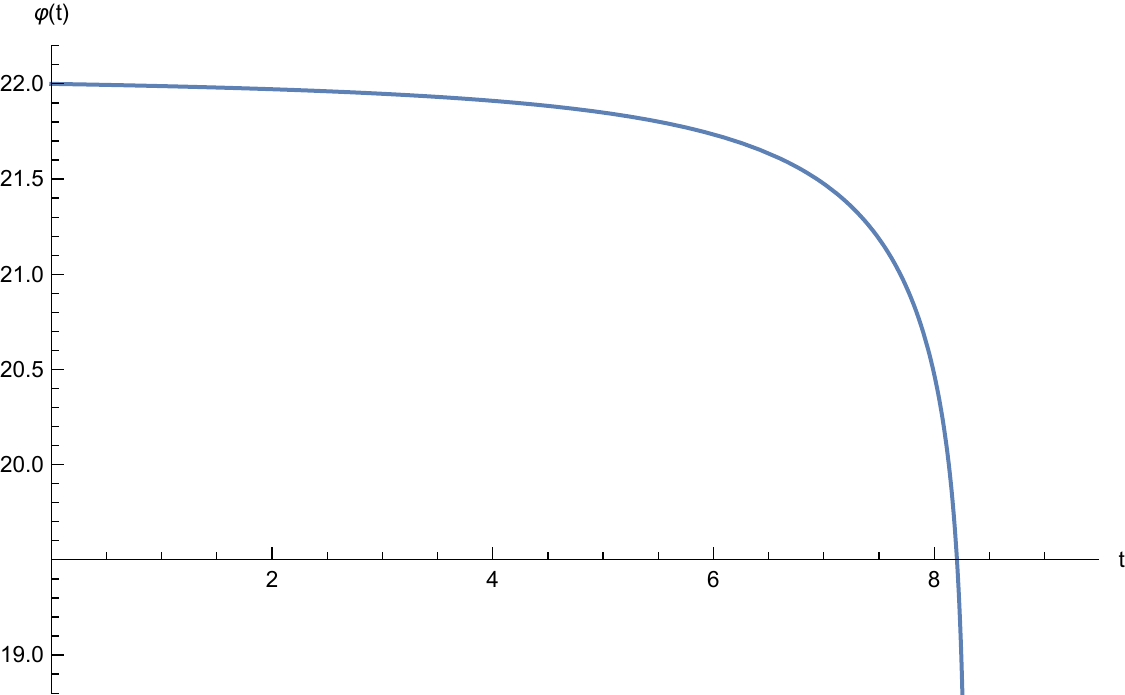}\caption{$\varphi(t)$}\label{fig:dynsol_exp_phi}
\end{subfigure}\\
\begin{subfigure}[H]{0.45\textwidth}
\includegraphics[width=\textwidth]{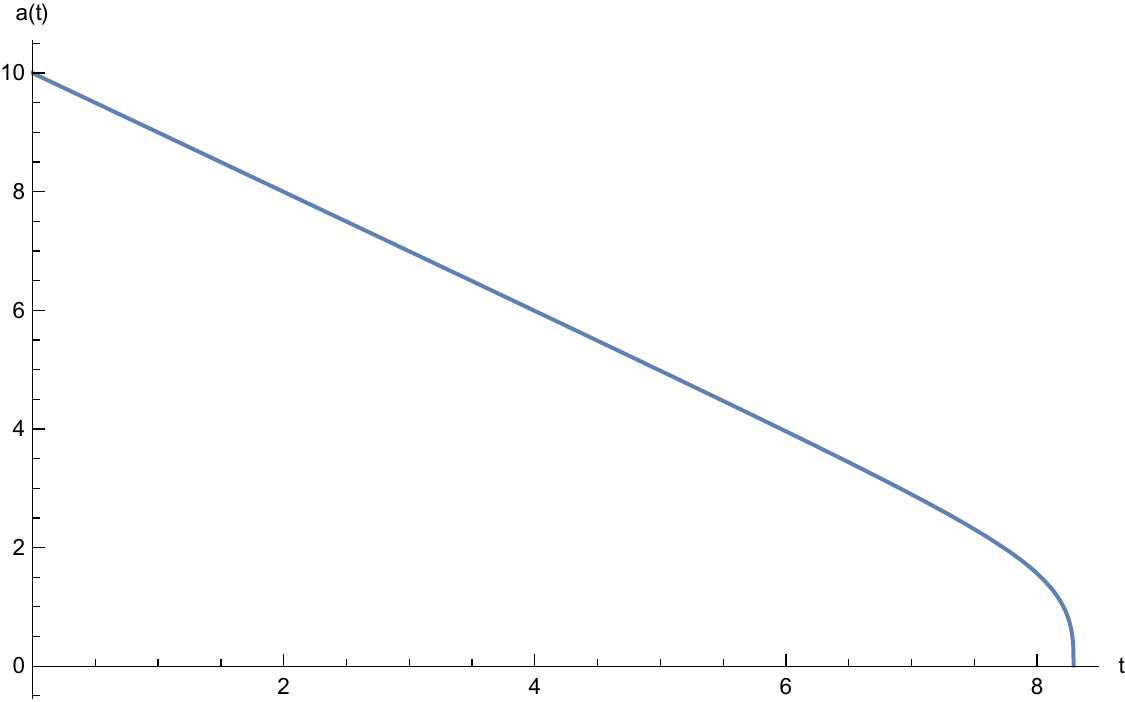}\caption{$a(t)$}\label{fig:dynsol_exp_a}
\end{subfigure}\qquad \quad
\begin{subfigure}[H]{0.45\textwidth}
\includegraphics[width=\textwidth]{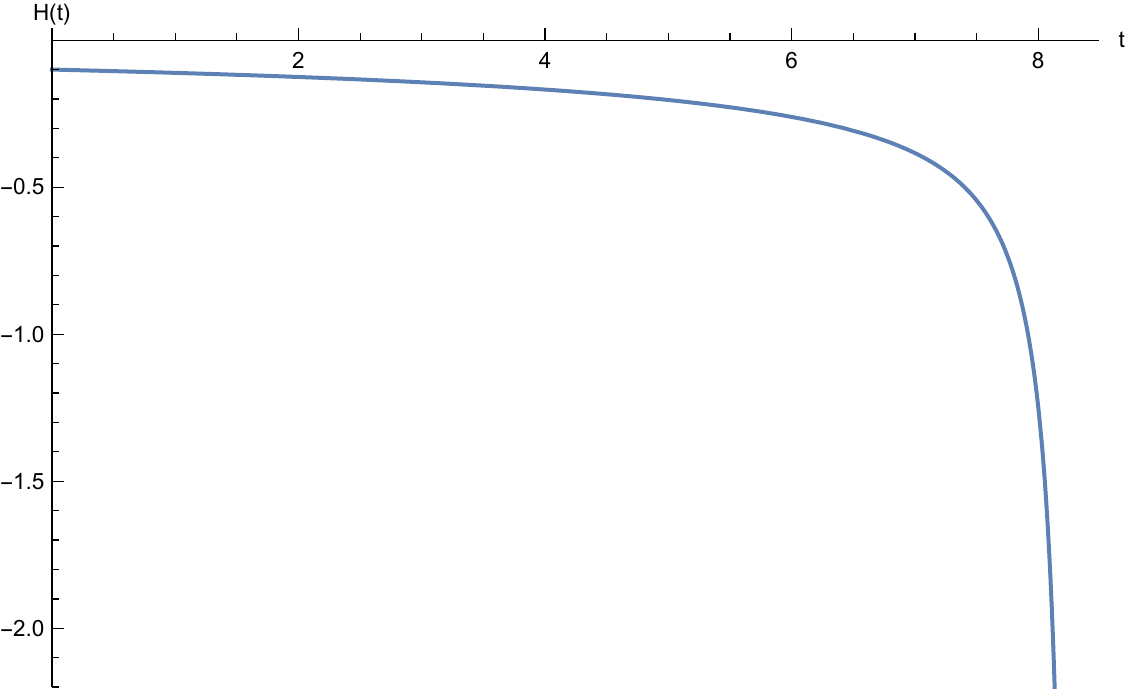}\caption{$H(t)$}\label{fig:dynsol_exp_H}
\end{subfigure}
\caption{Dynamical contracting solution $\varphi(t), \, a(t)$, as well as $H(t)$, in the negative part of the exponential potential \eqref{dynsol_exp_pot}.}\label{fig:dynsol_exp}
\end{center}
\end{figure}

We consider the following potential
\beq
V(\varphi)=0.04\, e^{-1.74\, \varphi} -0.05\, e^{-0.87\, \varphi}\, , \label{dynsol_exp_pot}
\eeq
displayed in Figure \ref{fig:dynsol_exp_pot}. Its minimum is at $\varphi_{\text{min}}=0.540234$, $V_{\text{min}}=-0.015625$. We choose the following initial conditions at $t_i=0$
\beq
\varphi(0)=22\, ,\qquad \dot{\varphi}(0)=-0.01\, ,\qquad a(0)=10\, .
\eeq
The numerical solution $\varphi(t), \, a(t)$, as well as $H(t)$, are given in Figure \ref{fig:dynsol_exp}. We obtain the final time $t_f\approx 8.29096$ for which $a(t_f)=0.00035871$.
\noindent The final values are given by
\beq
\varphi(t_{f})=0.0463959\, , \qquad V(\varphi(t_{f}))=-0.0111242 \ ,
\eeq
where $\varphi(t_{f})$ is not displayed in Figure \ref{fig:dynsol_exp_phi}. These values indicate that the field has rolled-down to the minimum and passed it, climbing-up a little to the left. This did not seem to affect $a(t)$. We note that for this solution, as well as the next one and the anti-de Sitter solution, $H(t)$ displayed in Figure \ref{fig:dynsol_exp_H} varies a lot, contrary to a de Sitter universe.\\

We now test the solution upon the various conditions discussed in Section \ref{sec:ATCCgen}.

\begin{itemize}
  \item {\bf ATCC condition} \eqref{ATCC}
\end{itemize}

\noindent This condition, rephrased as
\beq
M_p^2\, a(t)- \sqrt{|V_i|}\, a(t_i) \ \geq \ 0 \, , \label{ATCC2}
\eeq
is verified here at all times by the solution, until the final time $t_f$ when numerics stop. We obtain there $a(t_{f})-\sqrt{|V_i|} a_i=0.000202651$.

\begin{itemize}
  \item {\bf Second assumption} \eqref{2ndas}
\end{itemize}

\begin{figure}[H]
\begin{center}
\includegraphics[width=0.7\textwidth]{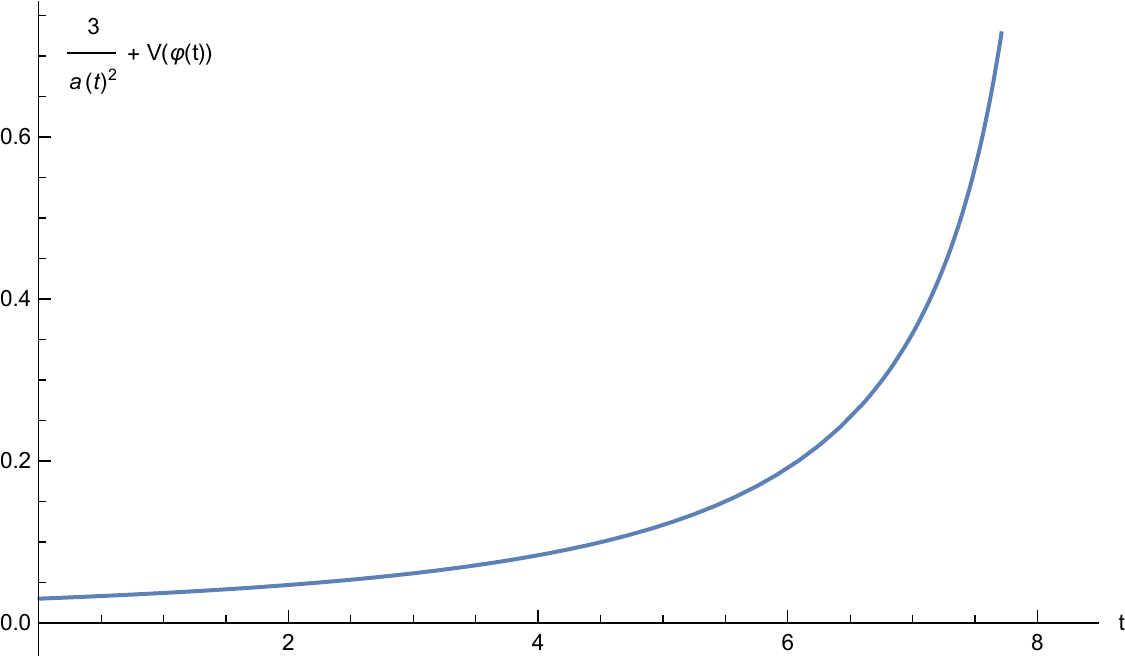}
\caption{$\frac{3}{a(t)^2}+V(\varphi(t))$ with the exponential potential \eqref{dynsol_exp_pot}.}\label{fig:dynsol_exp_check2}
\end{center}
\end{figure}

\noindent The second assumption is easily verified by this solution, as displayed in Figure \ref{fig:dynsol_exp_check2}.

\begin{itemize}
  \item {\bf Potential bound} \eqref{boundV}
\end{itemize}

Since both conditions above are satisfied by the solution, it is without surprise that the potential bound \eqref{boundV} is also satisfied.\footnote{One may complain that the two dynamical solutions presented in this Appendix are not climbing potentials but rolling down (with a little climbing for the case of the exponential potential). We believe this still allows to probe the physics in the negative potential, and does not affect our results. In addition, we obtained as well climbing solutions by simply changing the sign of $\dot{\varphi}(0)$: those exhibited exactly the same behaviour, in terms of the contraction and the verification of the two mathematical assumptions.} Since the potential \eqref{dynsol_exp_pot} is made only of exponential terms, satisfying this bound essentially amounts to the bound on the rate \eqref{rate}, given here in $d=4$ by $\sqrt{\frac{2}{3}} \approx 0.82$. This is indeed verified by the dominant term of the potential \eqref{dynsol_exp_pot} in the large field limit, whose rate is $0.87$.

\subsection{KKLT-inspired potential}\label{ap:kkltpot}

\begin{figure}[H]
\begin{center}
\begin{subfigure}[H]{0.45\textwidth}
\includegraphics[width=\textwidth]{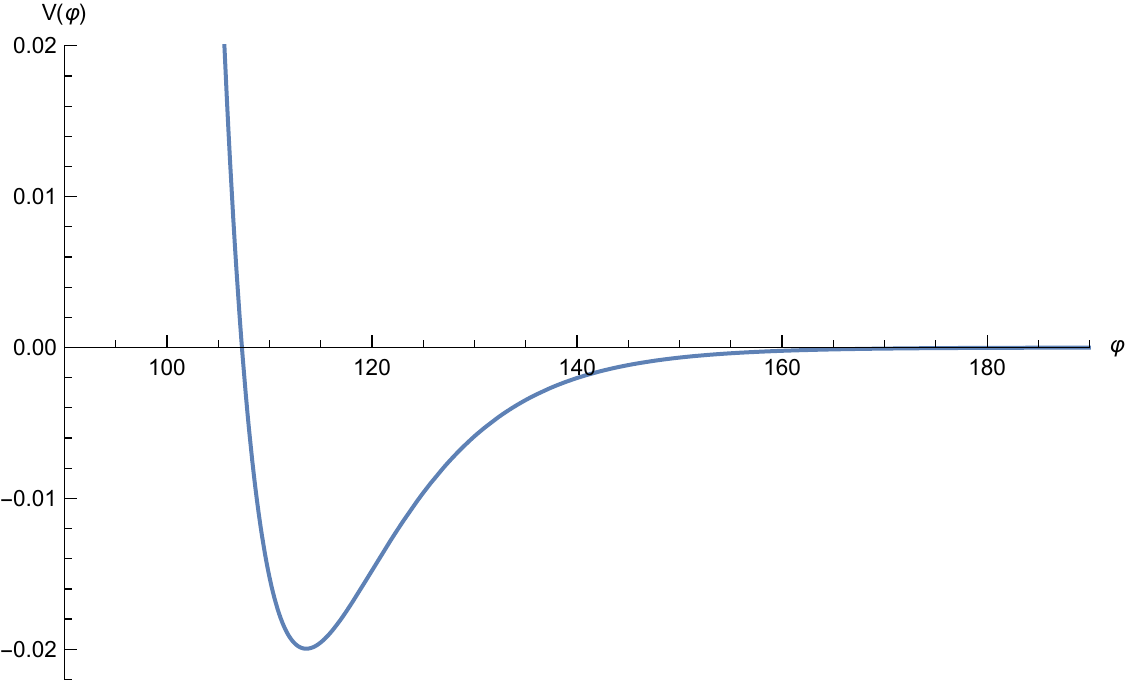}\caption{Potential $V(\varphi)$}\label{fig:dynsol_kklt_pot}
\end{subfigure}\qquad \quad
\begin{subfigure}[H]{0.45\textwidth}
\includegraphics[width=\textwidth]{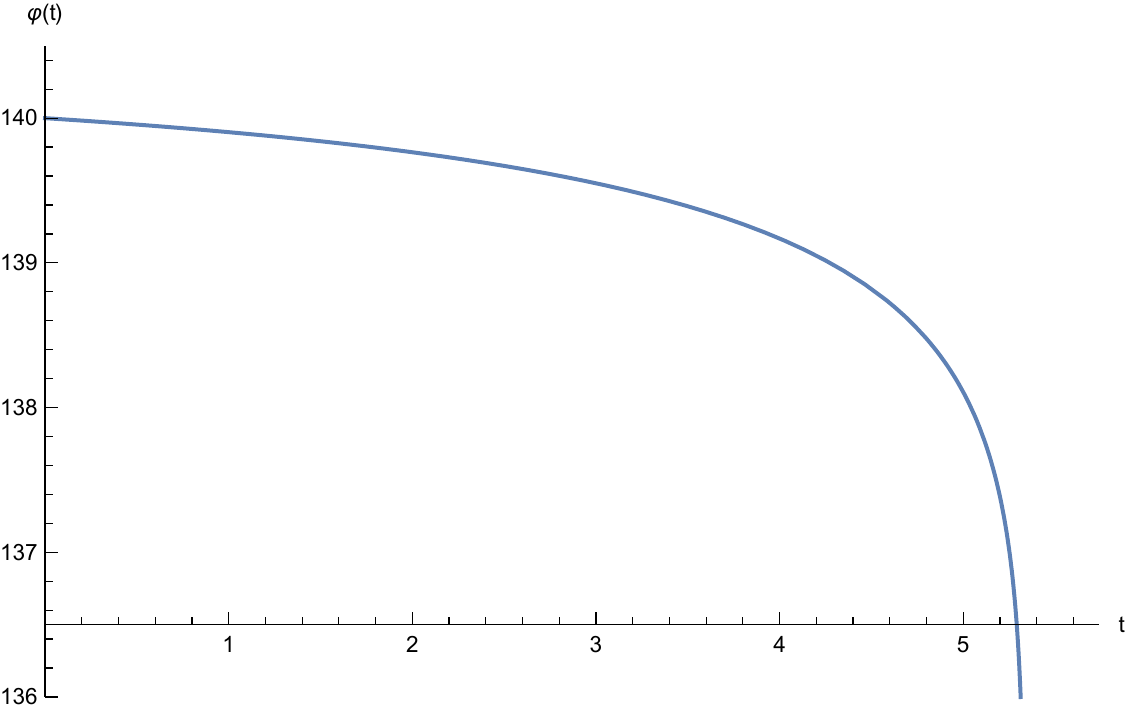}\caption{$\varphi(t)$}\label{fig:dynsol_kklt_phi}
\end{subfigure}\\
\begin{subfigure}[H]{0.45\textwidth}
\includegraphics[width=\textwidth]{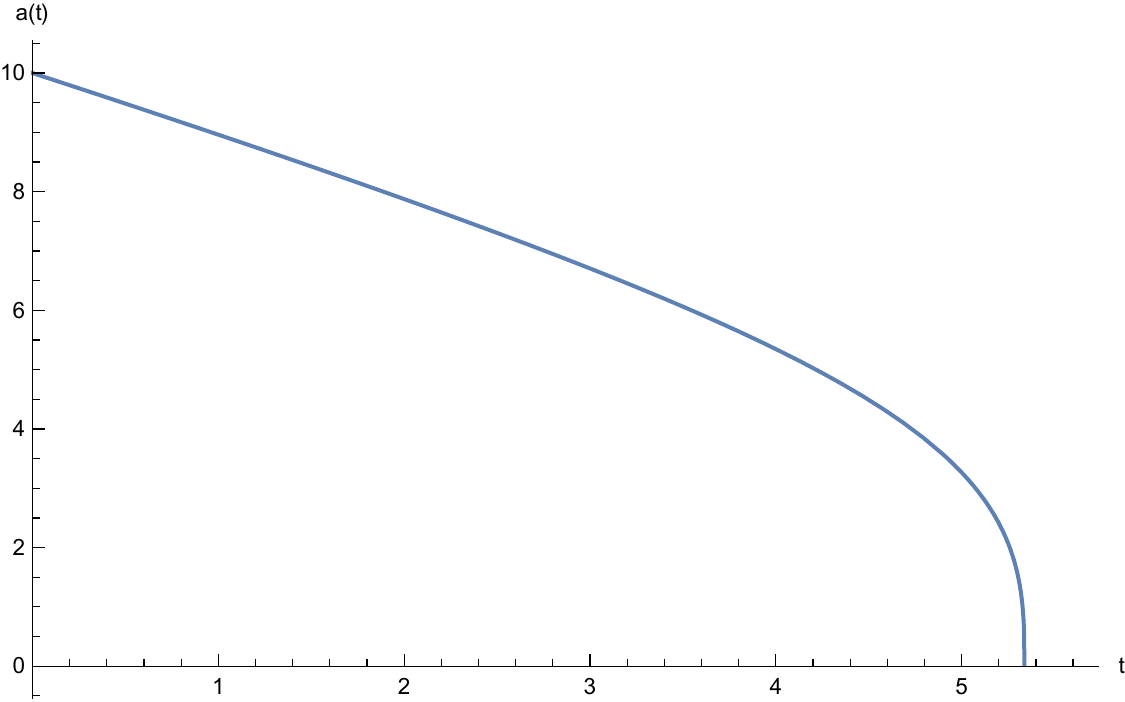}\caption{$a(t)$}\label{fig:dynsol_kklt_a}
\end{subfigure}\qquad \quad
\begin{subfigure}[H]{0.45\textwidth}
\includegraphics[width=\textwidth]{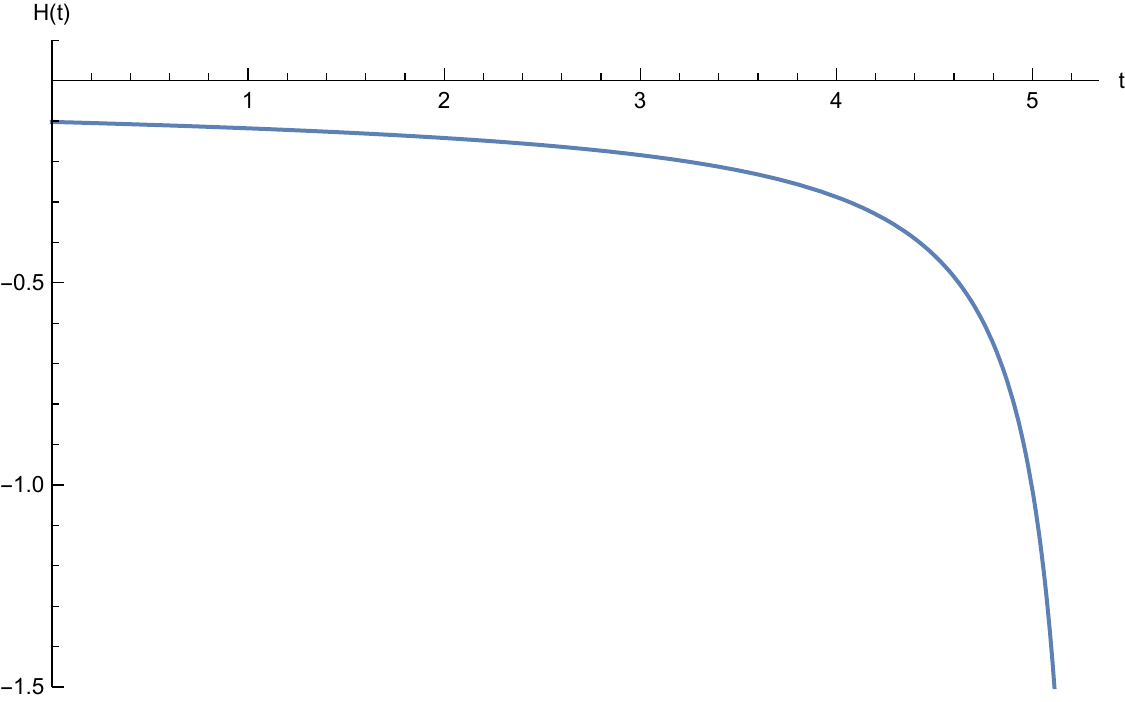}\caption{$H(t)$}\label{fig:dynsol_kklt_H}
\end{subfigure}
\caption{Dynamical contracting solution $\varphi(t), \, a(t)$, as well as $H(t)$, in the negative part of the KKLT-inspired potential \eqref{dynsol_kklt_pot}.}\label{fig:dynsol_kklt}
\end{center}
\end{figure}

We turn to the following ``KKLT-inspired'' potential, whose expression is close to the one of \cite{Kachru:2003aw}
\beq
V(\varphi)= \frac{10^{12} \, e^{-0.2\, \varphi} \left(-3 \times 10^{-4}\, e^{0.1\, \varphi} + 3 +  0.1\, \varphi\right)}{6\, \varphi^2} \, . \label{dynsol_kklt_pot}
\eeq
Its graph is displayed in Figure \ref{fig:dynsol_kklt_pot}. Its minimum is at $\varphi_{\text{min}}=113.589$, $V_{\text{min}}=-0.0199658$. We choose the following initial conditions at $t_i=0$
\beq
\varphi(0)=140\, , \qquad \dot{\varphi}(0)=-0.082\, , \qquad a(0)=10\, .
\eeq
The numerical solution $\varphi(t), \, a(t)$, as well as $H(t)$, are given in Figure \ref{fig:dynsol_kklt}. We obtain the final time $t_f\approx 5.3379$ for which $a(t_f)=0.000627402$.
\noindent The final values are given by
\beq
\varphi(t_{f})=117.175\, , \qquad V(\varphi(t_{f}))=-0.0178131 \ ,
\eeq
where $\varphi(t_{f})$ is not displayed in Figure \ref{fig:dynsol_kklt_phi}. This time, the field rolls-down but does not pass the minimum.\\

We now test the solution upon the various conditions discussed in Section \ref{sec:ATCCgen}.

\begin{itemize}
  \item {\bf ATCC condition} \eqref{ATCC}
\end{itemize}

This condition, rephrased as \eqref{ATCC2}, is verified by the solution at all times, except very close to the final time $t_f$, itself close to the final crunch. This seems consistent with the regime of validity of the effective theory, as already discussed around \eqref{AdSPlancktime} for the anti-de Sitter solution.

\begin{itemize}
  \item {\bf Second assumption} \eqref{2ndas}
\end{itemize}

\begin{figure}[H]
\begin{center}
\includegraphics[width=0.7\textwidth]{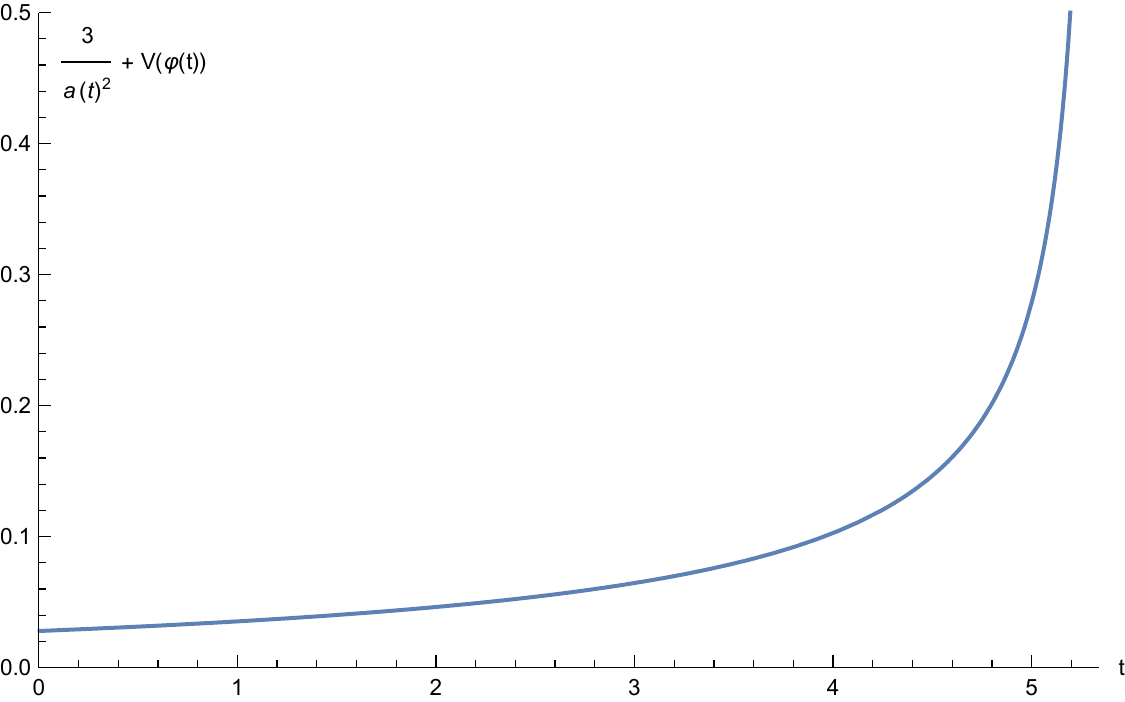}
\caption{$\frac{3}{a(t)^2}+V(\varphi(t))$ with the KKLT-inspired potential \eqref{dynsol_kklt_pot}.}\label{fig:dynsol_kklt_check2}
\end{center}
\end{figure}

The second assumption \eqref{2ndas} is easily verified by the solution, as displayed in Figure \ref{fig:dynsol_kklt_check2}.

\begin{itemize}
  \item {\bf Potential bound} \eqref{boundV}
\end{itemize}

\begin{figure}[H]
\begin{center}
\includegraphics[width=0.7\textwidth]{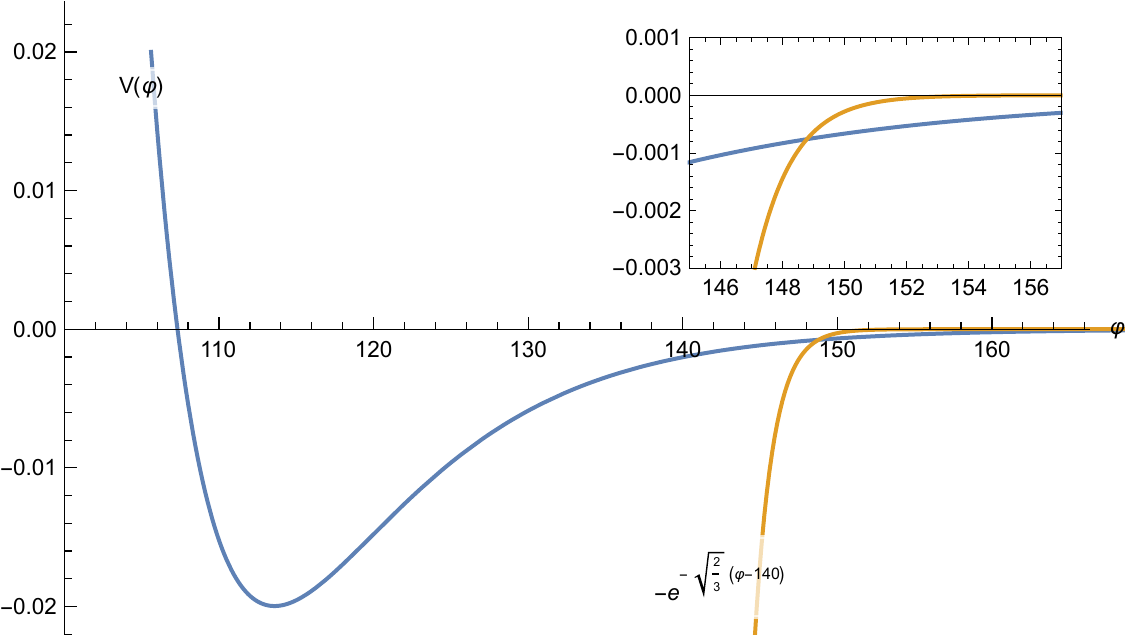}
\caption{Comparison of the KKLT-inspired potential \eqref{dynsol_kklt_pot} in blue and the growing exponential of the bound \eqref{boundV} in orange, with $\varphi_i=140$. The two curves intersect at $\varphi = 148.79$.}\label{fig:KKLTintersect}
\end{center}
\end{figure}

Contrary to the previous solution and potential, the KKLT-inspired potential \eqref{dynsol_kklt_pot} does not satisfy the potential bound \eqref{boundV}, as can be seen on Figure \ref{fig:KKLTintersect}.
\noindent More precisely, the bound \eqref{boundV} is verified as long as $\varphi \leq 148.79$. At $\varphi_{\rm min}$, the exponential bound gets evaluated to be  $-2.31897 \times 10^9$, much lower than the minimum of the potential $V(\varphi_{\rm min})$, while at $\varphi_i=140$, it gets evaluated at $-1$ which is again lower than $V_i$.

Given that the two above conditions are satisfied by our solution up to times very close to the crunch, it may look surprising to get a violation of the potential bound \eqref{boundV}. The resolution of this puzzle is simply that our solution did not probe $\varphi \geq 148.79$, but only lower values; this is why we did not notice any significant violation of the two mathematical assumptions. In the probed region of field space, everything is then consistent and satisfied.

The violation of the potential bound in the large field region can also be seen through the asymptotic condition \eqref{boundV'V}, which also gets violated: we obtain here $\left< -V'/V \right>_{\varphi \to \infty} = 0.1$ which smaller than the bound rate $0.82$. Our dynamical solution does not indicate any issue because it does not probe large enough field values. One may also interpret this by saying that this potential at low field values is admissible in a quantum gravity effective theory, but not its part at large field values.

\end{appendix}

\providecommand{\href}[2]{#2}\begingroup\raggedright\endgroup

\end{document}